\documentclass{article}

\usepackage{jheppub}
\usepackage[utf8]{inputenc}
\usepackage{amsmath}
\usepackage{amssymb}
\usepackage{amsfonts}
\usepackage{amsthm}
\usepackage{appendix}
\usepackage{physics}
\usepackage{xcolor}
\usepackage{upgreek}
\usepackage{tikz}
\usetikzlibrary{arrows.meta} 
\usepackage{tensor}
\usepackage[most]{tcolorbox}
\usetikzlibrary{positioning}
\tcbset{colback=yellow!10!white,colframe=red!50!black}
\usepackage{tikz-cd} 
\usepackage{xcolor}

\usepackage{dsfont} 

 \makeatletter
\def\moverlay{\mathpalette\mov@rlay}
\def\mov@rlay#1#2{\leavevmode\vtop{%
   \baselineskip\z@skip \lineskiplimit-\maxdimen
   \ialign{\hfil$\m@th#1##$\hfil\cr#2\crcr}}}
\newcommand{\charfusion}[3][\mathord]{
    #1{\ifx#1\mathop\vphantom{#2}\fi
        \mathpalette\mov@rlay{#2\cr#3}
      }
    \ifx#1\mathop\expandafter\displaylimits\fi}
\makeatother

\def\LCL{\textit{LCL}}

\def \s= {\!\!\!=\!\!\!}

\def \dg {\delta}
\def \ds {\partial}

\def \ag {\alpha}
\def \bg {\beta}
\def \cg {\gamma}

\def \si {\sigma}

\def \ep {\epsilon}
\def \om {\omega}
\def \Om {\Omega}

\def \la {\lambda}

\def \xx {\mathcal{X}}

\def \oo {\mathcal{O}}

\def \hh {\mathcal{H}}

\def \Zs {\mathbb{Z}}
\def \Rs {\mathbb{R}}

\def \Cs {\mathbb{C}}

\def \vt {\vartheta}

\def \sl {\mathfrak{sl}}

\def \su {\mathfrak{su}}

\def \Dg {\Delta}

\def \det{\mathrm{det}}

\def \orm{\dutchcal{O}}

\def \DC{\mathrm{DC}}
\def \CC{\mathrm{CC}}
\def \du{\dutchcal{u}}
\def \mU{\mathcal{U}}

\def \pgff{p}
\def \cgff{c}

\DeclareMathAlphabet{\dutchcal}{U}{dutchcal}{m}{n}
\SetMathAlphabet{\dutchcal}{bold}{U}{dutchcal}{b}{n}
\DeclareMathAlphabet{\dutchbcal} {U}{dutchcal}{b}{n}

\newcommand{\hypg}[3] {{}_2 F_1\!\begin{bmatrix} #1,#2 \\ #3  \end{bmatrix}\hspace*{-3pt}}

\newcommand{\ba}{\begin{align}}

\newcommand{\be}{\begin{equation}}
\newcommand{\ee}{\end{equation}}
\def\bd{\begin{tikzpicture}}
\def\ed{\end{tikzpicture}}

\title{Comb channel lightcone bootstrap II. Triple-twist anomalous dimensions}

\subheader{\normalsize
        \vspace*{-3.7em}
        \begin{flushright}
	DESY 24-012
        \end{flushright}
}

\author{Sebastian Harris$^1$,}
\author{Apratim Kaviraj$^{1,2}$,}
\author{Jeremy A. Mann$^3$,}
\author{Lorenzo Quintavalle$^{4}$,}
\author{Volker Schomerus$^{1,5}$}
\affiliation{$^1$ Deutsches Elektronen-Synchrotron DESY, Notkestr. 85, 22607 Hamburg, Germany }
\affiliation{$^2$ Department of Physics, Indian Institute of Technology - Kanpur, Kanpur 208016,  India}
\affiliation{$^3$ Department of Mathematics, King’s College London, Strand, London, WC2R 2LS, UK}
\affiliation{$^4$ D\'epartement de Physique, de G\'enie Physique et d'Optique, Universit\'e Laval, Qu\'ebec, QC G1V0A6, Canada}
\affiliation{$^5$ II. Institut f\"ur Theoretische Physik, Universit\"at Hamburg, Luruper Chaussee 149, D-22761 Hamburg}
\affiliation{Zentrum f\"ur Mathematische Physik, Universit\"at Hamburg, Bundesstrasse 55, D-20146 Hamburg }

\emailAdd{sebastian.harris@desy.de}
\emailAdd{akaviraj@iitk.ac.in}
\emailAdd{jeremy.mann@kcl.ac.uk}
\emailAdd{lorenzo.quintavalle.1@ulaval.ca}
\emailAdd{volker.schomerus@desy.de}

\date{January 2024}

\abstract{
We advance the multipoint lightcone bootstrap and compute anomalous dimensions of triple-twist operators at large spin. In contrast to the well-studied double-twist operators, triple-twist primaries are highly degenerate so that their anomalous dimension is encoded in a matrix. At large spin, the degeneracy becomes infinite and the matrix becomes an integral operator. We compute this integral operator by studying a particular non-planar crossing equation for six-point functions of scalar operators in a lightcone limit. The bootstrap analysis is based on new formulas for six-point lightcone blocks in the comb-channel. For a consistency check of our results, we compare them to perturbative computations in the epsilon expansion of $\phi^3$ and $\phi^4$ theory. In both cases, we find perfect agreement between perturbative results and bootstrap predictions. As a byproduct of our studies, we complement previous results on triple-twist anomalous dimensions in scalar $\phi^3$ and $\phi^4$ theory at first and second order in epsilon, respectively.}

\begin{document}
\addtolength{\baselineskip}{2mm}
\maketitle
\section{Introduction}
\label{sec:intro}

 \subsection{Motivation}
The modern conformal bootstrap offers a powerful new window into the dynamics of strongly coupled quantum field theories. 
Over the last decade, both numerical and analytical techniques have been advanced significantly, mostly in the context of 
four-point correlation functions. The latter are usually studied for a small set of scalar (or low-spin) field insertions.
Despite this restriction, the outcome has been truly impressive. This is particularly true 
for the flagship applications to the 3d critical Ising model, where the lowest scaling dimensions have been determined with
record precision \cite{El-Showk:2012cjh,El-Showk:2014dwa,Kos:2014bka}. In addition, with the input from numerical studies, 
the analytic bootstrap has provided accurate predictions for the CFT data of higher spin operators at low twist
\cite{Simmons-Duffin:2016wlq}. While a rigorous method has not yet been explored, the interplay between analytical
and numerical methods holds the potential to greatly enhance the efficiency and scope of the 
bootstrap~\cite{Simmons-Duffin:2016wlq,Caron-Huot:2020adz,Liu:2020tpf,Su:2022xnj}. However, in spite of impressive results in e.g.\ the $O(n)$ \cite{Kos:2013tga,Kos:2015mba,Chester:2019ifh,Liu:2020tpf} and GNY~\cite{Erramilli:2022kgp} models, the precision results for Ising remain somewhat 
singular. Even there, it seems that the existing methods are not able to push the frontier much further. At the same time, due to mixing effects and a lack of understanding of multi-twist operators, much of the CFT data at higher twist has 
eluded both numerical and analytical approaches to solving the bootstrap constraints. It is widely believed that the reason
for these limitations lies in the restricted set of observables that have been considered in the past.  In this context,
the inclusion of multipoint correlations with more than four fields could open the way for a new era of the bootstrap. 
 
All bootstrap studies, be they numerical or analytical, are based on a good knowledge of conformal 
blocks. In the case of four-point functions, it was this 
mathematical control that paved the way for the success of the modern bootstrap. Multipoint blocks with more 
than four external fields are not as well known, but several different approaches 
have been developed over the last few years. In $d=1$ dimensions, multipoint blocks are known for any number of legs and any 
topology~\cite{Rosenhaus_2019,Fortin:2020zxw,Alkalaev:2023axo,Fortin:2023xqq}. For higher dimensions, 
the results are less exhaustive, but there has been steady progress recently, based on the study of integral 
formulas, weight-shifting technologies, differential equations, and integrability, see e.g.\ \cite{Gon_alves_2019,Fortin:2020bfq,Hoback:2020syd,Poland_2021,Buric:2020dyz,Buric:2021ywo,Buric:2021ttm,
Buric:2021kgy,Fortin:2022grf}. These advances have enabled some first bootstrap studies including a numerical study of (truncated) five-point crossing equations \cite{Poland:2023bny,
Poland:2023vpn}, a semi-definite programming analysis of six-point crossing equations \cite{Antunes:2023dlk} 
and, more relevant to this work, several multipoint extensions of the lightcone 
bootstrap~\cite{Bercini:2020msp,Bercini:2021jti,Antunes:2021kmm,Kaviraj:2022wbw}.\footnote{Analytical studies of the 
crossing equation also include~\cite{Anous:2021caj}, where OPE data for heavy operators is obtained from 
the Euclidean OPE limit of higher-point correlators in the snowflake channel.} For the latter, the characterization of blocks through a system of differential equations -- constructed from limits of Gaudin models in 
\cite{Buric:2020dyz,Buric:2021ywo,Buric:2021ttm,Buric:2021kgy} -- seems particularly suitable. Indeed, while the expressions for some of the higher-order (Casimir and vertex) differential operators are very complex in general, drastic 
simplifications occur for certain limiting regimes in the space of multipoint cross-ratios.  Once sufficiently 
many distances are made lightlike, it is possible to write down explicit solutions of the differential equations. 
Much of this was discussed in \cite{Kaviraj:2022wbw}, where we illustrated the simplifications and constructed 
the resulting lightcone blocks in the context of five-point functions. The methods developed in that work were 
geared towards extensions to more than five fields, with some first results for six-point (comb-channel) 
lightcone blocks already present therein.

As outlined in \cite{Kaviraj:2022wbw}, an important motivation for the study of higher-point blocks and 
crossing equations near lightcone limits is the resolution of multi-twist operators and their dynamics. 
The simplest class of multi-twist operators, namely double-twist operators, has been extensively studied. Using the four-point lightcone bootstrap, the OPE structure of these operators came to light in the milestone 
works of \cite{Fitzpatrick:2012yx,Komargodski:2012ek}. They showed that if the identity operator in a 
Lorentzian CFT is separated from the rest of the spectrum by a twist gap, then crossing symmetry implies 
the existence of an infinite family of large-spin operators whose twists asymptote to the double-twist 
operators in a generalized free field (GFF) theory. In fact, there exist discrete families of such operators 
labeled by their twists that organize into Regge trajectories. Such twist accumulations at large spin have been rigorously proven in \cite{Pal:2022vqc}. In the wake of the original lightcone bootstrap papers, several follow-ups  \cite{Alday:2015ewa, Alday:2016njk, Kaviraj:2015cxa, Kaviraj:2015xsa, Hofman:2016awc, Li:2015itl, Li:2015rfa} 
eventually showed that the deviation from GFF values in the OPE data can be determined perturbatively in large spin in terms of the leading-twist operators in the spectrum. The convergence and analyticity of this expansion, as hinted by these previous works, was later established by the Lorentzian inversion formula \cite{Caron-Huot:2017vep} and the construction of light-ray operators \cite{Kravchuk:2018htv}. The results of lightcone bootstrap 
have also been studied in AdS/CFT \cite{Cornalba:2006xk, Cornalba:2006xm, Cornalba:2007zb, Fitzpatrick:2014vua, 
Fitzpatrick:2015qma}, where double-twist operators correspond to bound states of two massive objects that have a very small binding energy when rotating around each other at large angular momentum.

In contrast,  the CFT data of higher multi-twist operators has remained elusive due to its suppression in 
four-point functions, as well as the large degeneracy of its spectrum -- which grows with spin -- in 
perturbative approaches. As the targets of the bootstrap reach higher twists, these operators 
will play an increasingly important role. While we can only hope to produce asymptotic large-spin expansions 
of the CFT data from lightcone bootstrap, its impressive accuracy and universal convergence properties in the 
four-point case suggest that the generalization to higher points is worth testing.  This is further backed up by the recent insights of \cite{Homrich:2022mmd} and especially \cite{Henriksson:2023cnh}, resolving an apparent tension between multi-twist degeneracy and analyticity in spin. Independently of these considerations, the duality between multi-twist dynamics and the many-body problem in AdS entails an abundance of non-perturbative information that has yet to be studied. 
 
As a starting point, six-point functions and their comb-channel OPE decompositions provide privileged access to 
triple-twist operators. In perturbative examples, the latter are composites of three fundamental fields along with derivatives. It is well known that triple-twist 
fields are highly degenerate in free theory at large spin, i.e.\ there exist many such operators with identical scaling dimensions. The triple-twist anomalous dimension is therefore an operator (rather than a number) labeled by the spin, that describes how the degeneracy is lifted by interactions. In multi-scalar $\phi^p$ theory in $d=\frac{2p}{p-2}-2\ep$ dimensions with $O(n)$ symmetry, these anomalous dimension operators have been computed for certain families of triple-twist operators at arbitrary spin in the first leading orders of the small $\ep$ \cite{Kehrein:1992fn,Derkachov:1995zr,Kehrein:1995ia,Derkachov:1997uh} or large $n$ \cite{Derkachov:1997qv} expansion. In all of these examples, the anomalous dimension lifts the triple-twist degeneracy at first or second order in the perturbative expansion. In spite of these developments, the diagonalization problem at lower orders was not 
fully solved, and higher-loop corrections have not yet been explored. Instead, the methods developed in
\cite{Derkachov:1995zr,Derkachov:1997qv,Derkachov:1997uh} were extended to the large $N_c$ limit of QCD
\cite{Braun:1998id,Braun:1999te,Belitsky:1999ru}, where the integrability of the anomalous dimension operator 
at weak coupling allowed for a comprehensive description of the triple-twist spectrum at large spin. This 
approach contributed to the program of solving integrable conformal gauge theories in the planar limit using 
integrability \cite{Korchemsky:2010kj}, leading to particularly impressive results on non-perturbative, 
single-trace anomalous dimensions in $\mathcal{N}=4$ Super-Yang Mills theory, see e.g.\ \cite{Gromov:2023hzc} for the state-of-the-art. However, as these recent developments are limited to conformal gauge theory and are based on the integrability of their planar limit, 
it is not clear which properties can generalize to non-planar CFTs with a twist gap.

\subsection{Main Results}

\begin{figure}[ht]
    \centering
    \begin{tikzpicture}
        \node (E1) at (-1,-1) {$\phi(x_1)$};
        \node (E2) at (-1,1) {$\phi(x_6)$};
        \node (E3) at (1,1.2) {$\phi(x_5)$};
        \node (E4) at (2,1.2) {$\phi(x_2)$};
        \node (E5) at (4,1) {$\phi(x_3)$};
        \node (E6) at (4,-1) {$\phi(x_4)$};
        \node[shape=circle,fill=blue!20,draw=black, scale = 0.5] (V1) at (0,0) {};
        \node[shape=circle,fill=blue!20,draw=black, scale = 0.5] (V2) at (1,0) {$n_1$};
        \node[shape=circle,fill=blue!20,draw=black, scale = 0.5] (V3) at (2,0) {$n_2$};
        \node[shape=circle,fill=blue!20,draw=black, scale = 0.5] (V4) at (3,0) {};
    
        \draw (E1) -- (V1);
        \draw (E2) -- (V1);
        \draw (E3) -- (V2);
        \draw (E4) -- (V3);
        \draw (E5) -- (V4);
        \draw (E6) -- (V4);
       
        \draw[double, double distance = 1.8 pt]  (V1) -- node[above] {$\mathcal{O}_1$} ++ (V2);
        \draw[double, double distance = 3.33 pt]  (V2) -- (V3);
        \path[-, draw] (V2) edge node[above] {$\mathcal{O}_2$} (V3);
        \draw[double, double distance = 1.8 pt]  (V3) -- node[above] {$\mathcal{O}_3$} ++ (V4);

        \node[shape=circle,fill=blue!20,draw=black, scale = 0.5] (V1) at (0,0) {};
        \node[shape=circle,fill=blue!20,draw=black, scale = 0.5] (V2) at (1,0) {$n_1$};
        \node[shape=circle,fill=blue!20,draw=black, scale = 0.5] (V3) at (2,0) {$n_2$};
        \node[shape=circle,fill=blue!20,draw=black, scale = 0.5] (V4) at (3,0) {};
    
        \node (DC) at (1.5,-1) {Direct Channel};

        \def\x{9};
        \def\l{1};
        \node (A1) at (5,0) {};
        \node (A2) at (7,0) {};
        \draw[{Stealth[length=5mm]}-{Stealth[length=5mm]},thick] (A1) -- (A2);
        advances
        \node (CE1) at (\x-1,-1) {$\phi(x_1)$};
        \node (CE2) at (\x-1,1) {$\phi(x_2)$};
        \node (CE3) at (\x+1,1.2) {$\phi(x_3)$};
        \node (CE4) at (\x+2,1.2) {$\phi(x_4)$};
        \node (CE5) at (\x+4,1) {$\phi(x_5)$};
        \node (CE6) at (\x+4,-1) {$\phi(x_6)$};
        \node[shape=circle,fill=blue!20,draw=black, scale = 0.5] (CV1) at (\x,0) {};
        \node[shape=circle,fill=blue!20,draw=black, scale = 0.5] (CV2) at (\x+1,0) {$n_1$};
        \node[shape=circle,fill=blue!20,draw=black, scale = 0.5] (CV3) at (\x+2,0) {$n_2$};
        \node[shape=circle,fill=blue!20,draw=black, scale = 0.5] (CV4) at (\x+3,0) {};
    
        \draw (CE1) -- (CV1);
        \draw (CE2) -- (CV1);
        \draw (CE3) -- (CV2);
        \draw (CE4) -- (CV3);
        \draw (CE5) -- (CV4);
        \draw (CE6) -- (CV4);
        \draw[double, double distance = 1.8 pt]  (CV1) -- node[above] {$\mathcal{O}_1$} ++ (CV2);
        \draw[double, double distance = 3.33 pt]  (CV2) -- (CV3);
        \path[-, draw] (CV2) edge node[above] {$\mathcal{O}_2$} (CV3);
        \draw[double, double distance = 1.8 pt]  (CV3) -- node[above] {$\mathcal{O}_3$} ++ (CV4);

        \node[shape=circle,fill=blue!20,draw=black, scale = 0.5] (CV1) at (\x,0) {};
        \node[shape=circle,fill=blue!20,draw=black, scale = 0.5] (CV2) at (\x+1,0) {$n_1$};
        \node[shape=circle,fill=blue!20,draw=black, scale = 0.5] (CV3) at (\x+2,0) {$n_2$};
        \node[shape=circle,fill=blue!20,draw=black, scale = 0.5] (CV4) at (\x+3,0) {};
        
        \node (CC) at (\x + 1.5,-1) {Crossed Channel};
\end{tikzpicture}
\caption{OPE diagrams of the six-point crossing equation studied in this work.  We shall refer to 
the left-hand side as the \textit{direct channel} and to the right-hand side as the \textit{crossed channel}.} 
    \label{fig:combcrossing} \vspace*{-5mm} 
\end{figure}
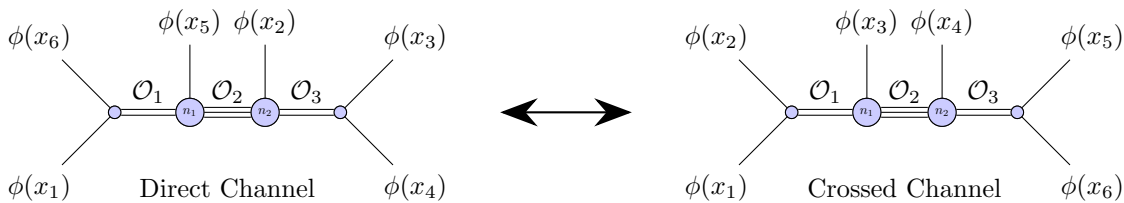

In this work, we shall advance a new bootstrap approach to triple-twist operators in general and their 
anomalous dimensions in particular. Following the usual lightcone bootstrap ideology, our analysis is 
based on the study of a certain crossing symmetry equation for correlation functions of six identical
scalar fields $\phi$ in an appropriate lightcone limit. We shall denote the scaling dimension of the 
external field by $\Dg_\phi = 2 h_\phi$. For six-point functions, the OPE decompositions belong to one of two possible topologies known as comb and 
snowflake. A lightcone bootstrap study extracting interesting large-spin asymptotics of 
conformal data from crossings of snowflake channels was performed in \cite{Antunes:2021kmm}. Here, 
we study a crossing equation in which both expansions use blocks of comb topology, see 
Figure~\ref{fig:combcrossing}. 
\medskip 

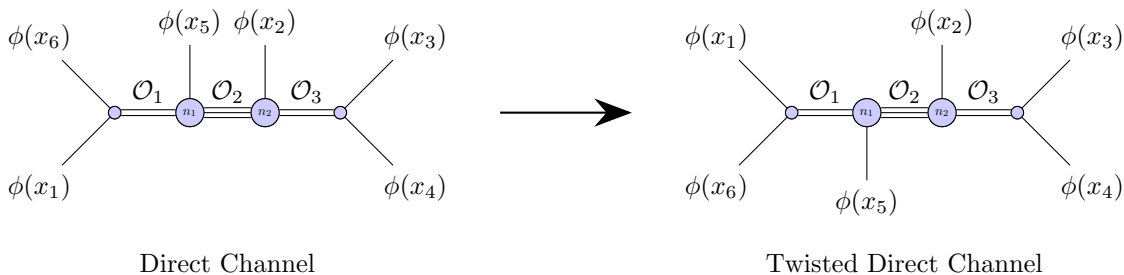
\begin{figure}[ht]
    \centering
    \begin{tikzpicture}
        \node (E1) at (-1,-1) {$\phi(x_1)$};
        \node (E2) at (-1,1) {$\phi(x_6)$};
        \node (E3) at (+1,1.2) {$\phi(x_5)$};
        \node (E4) at (+2,1.2) {$\phi(x_2)$};
        \node (E5) at (+4,1) {$\phi(x_3)$};
        \node (E6) at (+4,-1) {$\phi(x_4)$};
        \node[shape=circle,fill=blue!20,draw=black, scale = 0.5] (V1) at (0,0) {};
        \node[shape=circle,fill=blue!20,draw=black, scale = 0.5] (V2) at (1,0) {$n_1$};
        \node[shape=circle,fill=blue!20,draw=black, scale = 0.5] (V3) at (2,0) {$n_2$};
        \node[shape=circle,fill=blue!20,draw=black, scale = 0.5] (V4) at (3,0) {};
    
        \draw (E1) -- (V1);
        \draw (E2) -- (V1);
        \draw (E3) -- (V2);
        \draw (E4) -- (V3);
        \draw (E5) -- (V4);
        \draw (E6) -- (V4);
       
        \draw[double, double distance = 1.8 pt]  (V1) -- node[above] {$\mathcal{O}_1$} ++ (V2);
        \draw[double, double distance = 3.33 pt]  (V2) -- (V3);
        \path[-, draw] (V2) edge node[above] {$\mathcal{O}_2$} (V3);
        \draw[double, double distance = 1.8 pt]  (V3) -- node[above] {$\mathcal{O}_3$} ++ (V4);

        \node[shape=circle,fill=blue!20,draw=black, scale = 0.5] (V1) at (0,0) {};
        \node[shape=circle,fill=blue!20,draw=black, scale = 0.5] (V2) at (1,0) {$n_1$};
        \node[shape=circle,fill=blue!20,draw=black, scale = 0.5] (V3) at (2,0) {$n_2$};
        \node[shape=circle,fill=blue!20,draw=black, scale = 0.5] (V4) at (3,0) {};
    
        \node (DC) at (1.5,-2) {Direct Channel};

        \def\x{9};
        \def\l{1};
        \node (A1) at (5,0) {};
        \node (A2) at (7,0) {};
        \draw[
        -{Stealth[length=5mm]},thick] (A1) -- (A2);
        
        \node (E1) at (\x-1,1) {$\phi(x_1)$};
        \node (E2) at (\x-1,-1) {$\phi(x_6)$};
        \node (E3) at (\x+1,-1.2) {$\phi(x_5)$};
        \node (E4) at (\x+2,1.2) {$\phi(x_2)$};
        \node (E5) at (\x+4,1) {$\phi(x_3)$};
        \node (E6) at (\x+4,-1) {$\phi(x_4)$};
        \node[shape=circle,fill=blue!20,draw=black, scale = 0.5] (V1) at (\x+0,0) {};
        \node[shape=circle,fill=blue!20,draw=black, scale = 0.5] (V2) at (\x+1,0) {$n_1$};
        \node[shape=circle,fill=blue!20,draw=black, scale = 0.5] (V3) at (\x+2,0) {$n_2$};
        \node[shape=circle,fill=blue!20,draw=black, scale = 0.5] (V4) at (\x+3,0) {};
    
        \draw (E1) -- (V1);
        \draw (E2) -- (V1);
        \draw (E3) -- (V2);
        \draw (E4) -- (V3);
        \draw (E5) -- (V4);
        \draw (E6) -- (V4);
       
        \draw[double, double distance = 1.8 pt]  (V1) -- node[above] {$\mathcal{O}_1$} ++ (V2);
        \draw[double, double distance = 3.33 pt]  (V2) -- (V3);
        \path[-, draw] (V2) edge node[above] {$\mathcal{O}_2$} (V3);
        \draw[double, double distance = 1.8 pt]  (V3) -- node[above] {$\mathcal{O}_3$} ++ (V4);

        \node[shape=circle,fill=blue!20,draw=black, scale = 0.5] (V1) at (\x+0,0) {};
        \node[shape=circle,fill=blue!20,draw=black, scale = 0.5] (V2) at (\x+1,0) {$n_1$};
        \node[shape=circle,fill=blue!20,draw=black, scale = 0.5] (V3) at (\x+2,0) {$n_2$};
        \node[shape=circle,fill=blue!20,draw=black, scale = 0.5] (V4) at (\x+3,0) {};
    
        \node (DC) at (\x+1.5,-2) {Twisted Direct Channel};
\end{tikzpicture}
\caption{Cutting the direct-channel diagram at the middle leg, rotating the left half by $\pi$ and 
gluing the two halves back together restores the cyclic order of the external points.} 
    \label{fig:DCtwisting}
\end{figure}

By definition, OPE channels that share the same topology are related to each other by a permutation of the external operators. In this sense, the crossing equation visualized in Figure~\ref{fig:combcrossing} corresponds to the 
permutation $\varrho = (123)(123456)$. Application of $(123456)$ generates a planar duality. However, the other 
factor $(123)$ acts only on three legs, thereby destroying the cyclic order of the external points. We note 
that the cyclic order in the left diagram of Figure~\ref{fig:combcrossing} can be restored by cutting the 
diagram at the middle leg, rotating the left half of the diagram by $\pi$, and gluing it back to the right 
half, as shown in Figure \ref{fig:DCtwisting}.

Let us now discuss the relevant lightcone limits. The main goal is to isolate the leading-twist contributions in the 
two outer legs of the direct channel (see Fig.~\ref{fig:combcrossing} for a definition of the labels ``direct'' and ``crossed''). 
This is achieved by making the differences $x_{16}$ and $x_{34}$ lightlike, which amounts to setting $X_{16} =0, X_{34} =0$ in embedding-space coordinates. We only look at direct-channel contributions with at least one 
identity exchange, see Figure \ref{fig:six-pt-DC-leading}. For these leading contributions, no further limits 
are necessary in order to get leading twist in the middle leg of the direct channel. However, in order to remove 
higher-twist contributions in the crossed channel, we must introduce an additional triple-scaling limit where 
$X_{45},X_{56},X_{46}$ tend to zero at the same rate.\footnote{For Euclidean space, it was shown in \cite{Buric:2021kgy} that
this triple-scaling limit is equivalent to the OPE limit, which projects onto the lowest exchanged scaling 
dimensions, at the middle leg.} This adds up to three limits in our analysis of the crossing equation. A precise definition of the 
limiting regime in terms of cross-ratios will be given in Section \ref{sec:lightcone}, see in particular 
eqs.~(\ref{eq:three_epsilon}--\ref{eq:epsilon_limits2}).

Our analysis culminates in the matrix elements of the anomalous dimension operator $\gamma$ with respect to a particular basis of triple-twist operators. The latter is obtained by applying the construction of double-twist primaries twice, first on a pair of 
fields $\phi$ to obtain $[\phi\phi]_{0,\ell}$, then on $\phi$ and $[\phi\phi]_{0,\ell}$ to obtain $[\phi [\phi\phi]_{0,\ell}]_{0,J,\kappa}$. We refer to this basis of triple-twist operators as the \emph{double-twist basis}, labeled by the spin $\ell$ of the 
double-twist constituent. The expression for $\gamma_{k\ell}(J,\kappa)$ in eq.~\eqref{adim_master}, valid in the regime $1\ll k,\ell\ll J$, is the main result of this paper. While we restrict our focus to identical scalars, let us note that this result generalizes straightforwardly to the anomalous dimension of $[\phi_1\phi_2\phi_3]$, via the six-point function $\langle \phi_1\phi_2\phi_3\phi_3\phi_2\phi_1\rangle$ of three identical scalars. 

\subsection{Outline}

Before we conclude this introduction, let us briefly outline the content of the individual sections. 

Section \ref{sec:definitions} starts by introducing the relevant notation and describing the general 
setup, i.e.\ the crossing equation and the lightcone limits. There, we shall also evaluate the three 
direct-channel contributions that we want to analyze through the crossed-channel block expansion. Since all 
three terms involve at least one intermediate identity exchange, the relevant blocks in the direct 
channel contain at most four external fields and hence are well known. 

The study of the relevant crossed-channel blocks is the main subject of Section \ref{sec:blocks_cc}. 
There, we work out Casimir and vertex operators in the lightcone limits and apply them to direct-channel 
contributions in order to deduce the parameters of the crossed-channel blocks that contribute. 
The results are $h_1 = h_3 = 2 h_\phi, h_2 = 3 h_\phi$ where $h_a = h(\mathcal{O}_a)$ denotes the 
half-twist of the three intermediate operators. We conclude that the double-twist operators dominate 
the exchange in the outer legs while triple-twist exchange dominates at the middle leg of the crossed 
channel, as anticipated. The spin labels $J_a$ of the exchanged operators are all large, but they 
become large at different rates. While $J_1 \propto J_3 \propto X_{16}^{-1}$, the spin $J_2$ of the middle 
leg is sent to infinity as $J_2 \propto X_{16}^{-1}X_{34}^{-1}$. On the other hand, the mixed-symmetry tensor (MST) label 
$\kappa$ of the triple-twist field at the middle leg stays finite. Finally, the tensor 
structure labels $n_i$ restrict to their maximal allowed values $n_1 = J_1-\kappa$ and $n_2 = 
J_3-\kappa$. The relevant blocks are spelled out in eq.~\eqref{gtilde_to_besselK}.  

In Sections \ref{sec:param_3twist} and \ref{sec:GFFOPECoefficients}, we pause our analysis of the 
crossing equation and discuss some features of multi-twist operators in generalized free field 
(GFF) theories. First, Section \ref{sec:param_3twist} is devoted to the construction and parameterization 
of the triple-twist primaries in terms of certain multivariate polynomials. This enables us to introduce the 
double-twist basis by choosing a basis of polynomials, see eqs.~\eqref{dtPsi_sym_mst}, 
\eqref{dt_wf_mst} and \eqref{dt_wf} for explicit formulas. Section \ref{sec:GFFOPECoefficients}
is devoted to the calculation of triple-twist OPE coefficients for the operator product between a  
scalar and a double-twist primary in GFF. The general result is stated in eq.~\eqref{opec_PiWPi}
and then evaluated more explicitly in the large-spin limit that is relevant to the six-point lightcone bootstrap, see Subsection~\ref{ssec:Crossing-kernel-OPE-coeff}. As in the case of the four-point lightcone bootstrap, it will turn out that the 
dynamical data of scalar six-point functions in a generic CFT approach GFF values when the spins
become large. 

Section~\ref{sec:cft_data_3twist} returns to the analysis of the crossing equation in the 
lightcone limit. There, we shall derive how the three terms we consider in the direct channel are reproduced by summing crossed-channel lightcone blocks. At leading order, the term with two 
identity exchanges confirms the relation between large-spin triple-twist OPE coefficients 
and their values in GFF. Then, at next-to-leading order, the terms with a single identity exchange 
in the direct channel furnish our central new result: the large-spin, triple-twist anomalous dimension matrix $\gamma$, see eq.~\eqref{adim_master}. This matrix non-trivially mixes 
triple-twist operators in the double-twist basis. While the full diagonalization of $\gamma$ is beyond 
the scope of this work, we analyze its behavior near the asymptotic regime where triple-twist operators localize to elements of the double-twist basis. The first correction to the eigenvalues away from this regime is stated in eq.~\eqref{eq:adim_diagonal_limit}.
 
In Section~\ref{sec:checks_and_appl}, we compare our results with perturbative studies of 
$\phi^3$ theory in $6-2\epsilon$ and $\phi^4$ theory in $4-2\epsilon$ dimensions. For $\phi^3$ theory, where
the one-loop anomalous dimensions of triple-twist operators with vanishing $\kappa$ were computed by Derkachov and Manashov, we extend their results to non-vanishing $\kappa$. Following the same steps as in Derkachov and Manashov's work, we then perform a similar analysis for $\phi^4$ theory at two loops, deriving new analytic results for the anomalous dimension operator in the process. In both cases, we then verify that the large-spin limits of the resulting anomalous dimensions agree with our bootstrap results.

Section \ref{sec:conclusions} concludes the paper with a summary, followed by a list of open problems and future directions. Some technical background and further calculations are described in the appendices. 

\section{Non-planar Six-point Crossing Equation: Overview and Conventions}
\label{sec:definitions}

The six-point crossing equation we aim to study in this work compares two different sequences of 
operator product expansions that we denote $(16)5(2(34))$ and $(12)3(4(56))$, respectively. The two decompositions 
are depicted in Figure \ref{fig:combcrossing}. In order to spell out concrete formulas we need to first
introduce the relevant cross-ratios and the labels for the associated conformal blocks. The latter  
arise from the quantum numbers of the exchanged fields and the choice of tensor structures at the 
non-trivial vertices. Then we discuss the lightcone limit in which we would like to evaluate the 
crossing equation.
These are chosen to make leading-twist exchanges in the $(16)$ and 
$(34)$ OPEs dominate, thereby selecting $(16)5(2(34))$ as the direct channel (DC) of our lightcone bootstrap,
while projecting on triple-twist exchanges in the middle leg of $(12)3(4(56))$, which we then consider 
as our crossed channel (CC). The section concludes with concrete formulas for the three direct-channel 
contributions we would like to match by summing crossed-channel blocks.

\subsection{Conformally invariant cross-ratios}
\label{ssec:CrossRatios}
To describe the conformal blocks that appear in the DC and CC, we have to make a choice of conformally invariant 
cross-ratios. The most complex parts of our computations will involve the CC, thus our conventions are 
mostly tailored to this OPE channel. For a six-point correlator in sufficiently high dimension, i.e.\ $d >3$ 
there are nine independent cross-ratios. For most of our analysis, we choose these to be
\begin{equation}
	\begin{aligned}\dutchcal{u}_1&:=\frac{X_{12}X_{35}}{X_{13}X_{25}}\,, 
 & \dutchcal{u}_2&:=\frac{X_{13}X_{46}}{X_{14}X_{36}}\,, 
 & \dutchcal{u}_3&:=\frac{X_{24}X_{56}}{X_{25}X_{46}}\,, \\[2mm] 
	v_1&:=\frac{X_{14}X_{23}}{X_{13}X_{24}}\,, & v_2&:=\frac{X_{25}X_{34}}{X_{24}X_{35}}\,, 
 & v_3&:=\frac{X_{36}X_{45}}{X_{35}X_{46}}\,, \\[2mm] 
	\mathcal{U}_1&:=\frac{X_{15}X_{24}}{X_{14}X_{25}}\,, 
 & \mathcal{U}_2&:=\frac{X_{26}X_{35}}{X_{25}X_{36}}\,, 
 &  \mathcal{U}^{6}&:=\frac{X_{16}X_{24}X_{35}}{X_{14}X_{25}X_{36}}\, . \\[2mm]
\end{aligned}
\label{eq:curlycrossratios}
\end{equation}
There is a second set of cross-ratios that we shall use occasionally, namely the \emph{OPE cross-ratios} 
that were introduced in \cite{Buric:2021kgy}, These are useful to express the asymptotics of blocks in the 
Lorentzian OPE limit and they are particularly well suited to implement the additional Gram determinant relations 
that eliminate cross-ratios as we go to dimensions $d < 4$. The nine OPE cross-ratios are denoted by  $z_1$, 
$z_2$, $z_3$, $\bar{z}_1$, $\bar{z}_2$, $\bar{z}_3$, $w_1$, $w_2$, and $\Upsilon_0$. The relation with the cross 
ratios defined through eqs.~\eqref{eq:curlycrossratios} are somewhat complicated to spell out for generic 
kinematics. We shall only need this relation in the regime in which all three cross $\dutchcal{u}_j, \rightarrow
0$ go to zero. In terms of the OPE cross-ratios, this regime is mapped to the limit in which the three cross 
ratios $\bar{z}_j, j=1,2,3,$ vanish. In this limit, the relation between our two sets of cross-ratios reads 
\begin{equation}
\begin{gathered}
    \dutchcal{u}_2 v_1 v_3 = z_2 \bar{z}_2\,, \qquad v_2=1-z_2+\orm (\bar{z}_2)\,,\\[2mm]
    \dutchcal{u}_1 v_2 = z_1 \bar{z}_1\,, \quad v_1=1-z_1+\orm (\bar{z}_1)\,, \quad 
    \mathcal{U}_1 v_1 v_2=1-z_1-z_2+ w_1 z_1 z_2 +\orm (\bar{z}_i)\,,\\[2mm] 
    \dutchcal{u}_3 v_2 = z_3 \bar{z}_3\,, \quad v_3=1-z_3+\orm (\bar{z}_3)\,, \quad 
    \mathcal{U}_2 v_2 v_3=1-z_2-z_3+ w_2 z_2 z_3 +\orm (\bar{z}_j)\,,\\[2mm] 
    \mathcal{U}^{6} v_1 v_2 v_3=1-z_1-z_3+z_1 z_3-z_2\left[1-w_1 z_1-w_2 z_3-z_1 z_3 
    \left(\Upsilon _0-w_1 w_2\right)\right]+\orm (\bar{z}_j).
    \end{gathered}
    \label{eq:OPE_CR_as_us}
\end{equation}
Note that the $z_j$ and $\bar{z}_j$ are Dolan-Osborn type of variables~\cite{Dolan:2003hv} and $\Upsilon_0$ 
is a rescaled version of the cross-ratio $\Upsilon$ that was introduced in~\cite{Buric:2021kgy}, 
$$ \Upsilon_0 = \Upsilon \sqrt{w_1(1-w_1)w_2(1-w_2)}\ . $$ 
In terms of OPE cross-ratios, the crossed-channel OPE limit that we mentioned above corresponds to the regime 
in which 
\be
\text{OPE}^{(12)3(4(56))}: \bar{z}_1,\bar{z}_2,\bar{z}_3 \ll z_1,z_2,z_3,\Upsilon_0\ll 1\,.
\label{opelim_cr}
\ee
In this limit, the conformal blocks behave as
\begin{equation}
g_{\oo_1\oo_2\oo_3;n_1n_2}^{(12)3(4(56))} \stackrel{\mathrm{OPE}^{(12)3(4(56))}}{\sim} \prod_{i=1}^3 \bar z_i^{h_i} z_i^{\bar h_i} \Upsilon_0^\kappa (1-w_1)^{n_1}(1-w_2)^{n_2},
\label{psi_ope}
\end{equation}
where the $n_i$ labels parameterize the three-point tensor structures at the two innermost vertices, as we will discuss in Subsection~\ref{sec:TensorStructures}.

Since the relations \eqref{eq:OPE_CR_as_us} between our cross-ratios \eqref{eq:curlycrossratios} and the 
OPE cross-ratios are valid in the OPE limit, they can be used to determine the behavior of $\dutchcal{u}_j, 
v_j$ and $\mathcal{U}_1,\mathcal{U}_2, \mathcal{U}^6$ in the limiting regime \eqref{opelim_cr}. 

As we mentioned above the OPE cross-ratios are also well suited to describe the reduction of cross-ratios that 
occurs in the $d=3$ dimensions, due to the vanishing of the Gram determinant $\det(X_{ij})$\cite{Buric:2021kgy}. 
Indeed, in terms of OPE cross-ratios, the $d=3$ constraint simply reads
\begin{equation}\label{eq:Upsilon0constraint}
\Upsilon_0^2 = \frac{4z_2\bar z_2}{(z_2-\bar z_2)^2} w_1(1-w_1)w_2(1-w_2).
\end{equation}
This may be translated into a much more complicated-looking constraint on $\mathcal{U}_2$. The precise relation is 
not relevant to our scope. All we should keep in mind is that for $d=3$ the cross-ratio $\mathcal{U}_2$ may be expressed 
as some function $\mathcal{U}_2^*\!\left(\dutchcal{u}_i,v_i,\mathcal{U}_1,\mathcal{U}^6\right)$ of the other eight 
cross-ratios. 

Before we conclude this short discussion of cross-ratios we would like to add one short comment that will become 
relevant much later in Section \ref{sec:cft_data_3twist} when we compute the leading triple-twist anomalous dimensions. 
It will turn out that some of the direct-channel contributions can be most easily interpreted in terms of a crossed-channel expansion for some CC' that differs from CC by the permutation $\si:=(13)(46) \in S_6$. From the 
explicit construction \eqref{eq:curlycrossratios} it is not difficult to work the following action of $\si$ on 
our cross-ratios, 
\begin{equation}
\si: (\dutchcal{u}_1,v_1,\dutchcal{u}_2,v_2,\dutchcal{u}_3,v_3,\mathcal{U}_1,\mathcal{U}_2,\mathcal{U}^6)\longrightarrow \left(\mathcal{U}_1 v_1,  \frac{\dutchcal{u}_1}{\mathcal{U}_2},\dutchcal{u}_2, \frac{\mathcal{U}^6}{\mathcal{U}_1\mathcal{U}_2}, \mathcal{U}_2v_3,\frac{\dutchcal{u}_3}{\mathcal{U}_1},v_3,\mathcal{U}_2,\mathcal{U}_1, \mathcal{U}_1\mathcal{U}_2v_2 \right).
\end{equation}

\subsection{Tensor structures}\label{sec:TensorStructures}
In this subsection, we introduce the conventions and notation that we use for two- and three-point correlators of 
spinning fields, giving special attention to primary fields in mixed-symmetry tensor (MST) representations with two spin labels. Such fields can be exchanged in the middle leg of a six-point OPE channel of comb topology for 
$d \geq 4$. Since this is the leg in which we also expect to produce triple-twist fields, such representations are quite 
relevant for us. 

Throughout this work, we shall use the embedding space formalism, in which the insertion points $x$ are encoded
in lightlike vectors $X \in \mathbb{R}^{2,d}$. Fields in symmetric traceless tensor (STT) representations of the 
rotation group involve an additional lightlike vector $Z$ that is required to be orthogonal to $X$, i.e.\ $X \cdot Z =0$. 
For fields in MST representations with two spin labels, we finally need one more object $W \in \mathbb{C}^{d+2}$ 
that is required to have a vanishing norm and be perpendicular to both $X$ and $Z$, i.e.\ $W \cdot X = 0 = W 
\cdot Z$. For an MST operator $\mathcal{O}_i(X_i,Z_i,W_i)$ inserted at the embedding space position $X_i$ with 
polarization vectors $Z_i$ and $W_i$ labeled by some index $i$, we use the abbreviations
 \begin{align} \label{eq:Xij} 
  X_{ij} &:= X_i \cdot X_j, \\[2mm] H_{ij} &:=  (X_i \wedge Z_i) \cdot (X_j \otimes Z_j) = (X_i \cdot X_j)  
  (Z_i \cdot Z_j) -(X_i \cdot Z_j)  (Z_i \cdot X_j)  \\[2mm]  J_{i,jk} &:= (X_i \wedge Z_i) \cdot 
  (X_j \otimes X_k) \\[2mm]   
    U_{i,jk} &:=  (X_i \wedge Z_i \wedge W_i) \cdot (X_j \otimes Z_j \otimes X_k) 
    = (X_i \cdot X_j)(Z_i \cdot Z_j) (W_i \cdot X_k) + \dots \label{eq:Uijk} 
\end{align} 
With this notation, the normalized two-point function of two MST primaries reads
 \begin{align}\label{eq:twopointconv}
     \langle \mathcal{O}(X_1,Z_1,W_1)\mathcal{O}(X_2,Z_2,W_2) \rangle := X_{12}^{-(\Delta + J)} 
     H_{12}^{J-\kappa} [(X_1 \wedge Z_1 \wedge W_1) \cdot (X_2 \otimes Z_2 \otimes W_2)]^\kappa\ . 
 \end{align}
For STT primaries with $\kappa=0$ the dependence on the variables $W$ drops out. We can also use 
the objects defined in eqs.~(\ref{eq:Xij}-\ref{eq:Uijk}) to introduce the three-point tensor structures. 
For a three-point function of one scalar, one STT, and one MST primary, the three-point function
may be expanded as
 \begin{align} \label{eq:MST-STT-scalar-3pt}
  \langle \mathcal{O}_1(X_1,Z_1,W_1)\mathcal{O}_2(X_2,Z_2) \mathcal{O}_3(X_3) \rangle & = \\[2mm] 
  & \hspace*{-5cm} = \sum\limits_{n=0}^J C_{\mathcal{O}_1\mathcal{O}_2\mathcal{O}_3}^{(n)} 
  U_{1,23}^{\kappa_1} H_{12}^{n} J_{2,13}^{J_2-\kappa_1-n} J_{1,23}^{J_1-\kappa_1-n} 
  X_{12}^{-\bar h_{12;3}+\kappa_1} X_{13}^{-\bar h_{13;2}+\kappa_1+n-J_2} 
  X_{32}^{-\bar h_{3 2;1}+n-J_1}.
\nonumber
\end{align}
Here, $J_1$ and $J_2$ 
denote the spins of $\mathcal{O}_1$ and $\mathcal{O}_2$, $\kappa_1$ denotes the MST spin of the primary $\mathcal{O}_1$ i.e.~the length of the second row 
of the Young tableaux corresponding to the $SO(d)$ representation of $\mathcal{O}_1$, and $\bar{h}_{ij,k}$ is defined as
\begin{align}
\bar{h}_{ij,k} := \bar{h}_i + \bar{h}_j - \bar{h}_k && \text{with} &&\bar{h} := \frac{1}{2}(\Delta + J + \kappa).
\end{align}
The label $n$ that enumerates the possible three-point tensor structures, i.e.\ terms on the right-hand side of 
equation \eqref{eq:MST-STT-scalar-3pt}, runs over a finite set of non-negative integers with the upper bound 
$J$ given by the minimum of $J_1-\kappa_1$ and $J_2-\kappa_1$.

\subsection{Lightcone limits}
\label{sec:lightcone}

The lightcone limits we are going to study below are controlled by three different scales that dictate how fast various pairs of insertion points approach the lightcone. The first limits we shall 
take below are the ones that make leading-twist exchanges in the outer legs of the direct 
channel dominate. In terms of our embedding space variables, this means that we shall first take $X_{16}$ and
$X_{34}$ to zero. This is then followed by a limit that suppresses subleading-twist exchanges in the 
crossed channel. This is done by sending $X_{45}, X_{56}$ and $X_{46}$ to zero simultaneously. We 
shall control these limits by three dimensionless parameters $\epsilon_{16}, \epsilon_{34}$ and 
$\epsilon_{456}$ that multiply the various scalar products $X_{ij}$, 
\begin{equation}
    X^\epsilon_{16} \sim \epsilon_{16} X_{16}\,, \qquad X^\epsilon_{34}\sim \epsilon_{34} X_{34}\,, 
    \qquad X^\epsilon_{45},X^\epsilon_{46},X^\epsilon_{56}\sim 
    \epsilon_{456} X_{45}, \epsilon_{456} X_{56},\epsilon_{456} X_{46}\ . 
    \label{eq:three_epsilon}
\end{equation}
such that the lightcone limits may be performed by sending $\epsilon_{16},\epsilon_{34}$ and 
$\epsilon_{456}$ to zero. There are three different ways to do so that shall be relevant for us. 
These are distinguished by the relative scaling of the two direct-channel limits that involve 
$\epsilon_{16}$ and $\epsilon_{34}$. The first possibility is to send those two to zero at the 
same rate before sending $\epsilon_{456}$ to zero. This regime is denoted by 
\begin{equation} \label{eq:epsilon_limits}
\LCL^{(16,34)}: (\epsilon_{16}= \epsilon_{34}) \ll \epsilon_{456} \ll 1\ . 
\end{equation}
In addition, we shall also consider the regimes in which one of the two direct-channel limits 
is taken to zero much faster than the other. We denote these regimes by 
\begin{align}\label{eq:epsilon_limits1}
\LCL^{(16)}: \epsilon_{16} \ll \epsilon_{34} \ll \epsilon_{456} \ll 1, \\[2mm]
\LCL^{(34)}: \epsilon_{34}\ll \epsilon_{16} \ll \epsilon_{456}\ll 1.
\label{eq:epsilon_limits2}
\end{align}
By definition, the parameters $\epsilon$ determine how the scalar products $X_{ij}$ go to 
zero in the limiting regimes. Since all the differential operators and blocks are later written 
in terms of cross-ratios, it is most relevant for us to know how the latter behave in the 
lightcone limits. In terms of the cross-ratios we introduced in eqs.~\eqref{eq:curlycrossratios}, 
the three different regimes are characterized as  
\begin{equation}
\LCL^{(16)}: \mathcal{U}^6\ll v_2 \ll \dutchcal{u}_2 \ll 1\ , \quad 
\LCL^{(34)}: v_2 \ll \mathcal{U}^6 \ll \dutchcal{u}_2 \ll 1 \ . 
\end{equation}
with all the remaining cross-ratios $\dutchcal{u}_1,\dutchcal{u}_3,v_1,v_3,\mathcal{U}_1,
\mathcal{U}_2$ kept finite. Obviously, in order to reach the lightcone regime $\LCL^{(16,34)}$
we should take $\mathcal{U}^6$ and $v_2$ to zero at the same rate. Under the action of the 
permutation $\sigma=(13)(46)$ that we introduced and discussed briefly at the end of Subsection 
\ref{ssec:CrossRatios}, the two lightcone limits $\LCL^{(16)}$ and $\LCL^{(34)}$ are exchanged 
while the symmetric $\LCL^{(16,34)}$ is left invariant.

\subsection{Leading-twist expansion in the direct channel}\label{sec:leadingtwistexpansionDC}

With the conventions we have introduced in the previous subsections, the crossing equation of interest 
takes the form
\begin{equation} \label{eq:crossing}
\frac{\Om_\DC}{\Om_\CC} \sum_{\oo_i,n_i} P_{\oo_1\oo_2\oo_3} ^{(n_1,n_2)} 
g^\DC_{\oo_1\oo_2\oo_3;n_1n_2}(\dutchcal{u}_i,v_i,\mathcal{U}_i,\mathcal{U} ^6) = 
\sum_{\oo_i,n_i} P_{\oo_1\oo_2\oo_3} ^{(n_1,n_2)} g^\CC_{\oo_1\oo_2\oo_3;n_1n_2}
(\dutchcal{u}_i,v_i,\mathcal{U}_i,\mathcal{U} ^6) . 
\end{equation}
Here, $g^\DC$ and $g^\CC$ denote the six-point blocks in the direct and crossed channel, respectively, 
$P$ are the OPE coefficients that appear in the two expansions, and $\Omega_\CC$ and $\Omega_\DC$ are covariant 
prefactors. Throughout this paper, we adopt conventions in which the crossed-channel prefactor $\Omega_{CC}$ is 
given by 
\begin{equation}
\Omega_\CC = (X_{12}X_{34}X_{56})^{-\Dg_\phi} (v_1 \dutchcal{u}_2 v_3)^{-\frac{\Dg_\phi}{2}}, 
\label{eq:Omega_definition}
\end{equation}
Since we are dealing with six-point functions of identical scalars and both channels possess the same comb 
topology, the OPE coefficients are the same in both expansions.
The conformal blocks and the covariant prefactor, on the other hand, are not the same but we can obtain the 
direct-channel blocks and covariant prefactors from those of the crossed channel by the relevant (non-planar) 
permutation of the insertion points. With the help of eq.\ 
\eqref{eq:curlycrossratios} it is not difficult to determine the action of this permutation on our set of crossed-channel cross-ratios. The resulting relation between direct- and crossed-channel blocks reads
\begin{eqnarray}
  g^\DC_{\oo;n}(\dutchcal{u}_1,v_1,\dutchcal{u}_2,v_2,\dutchcal{u}_3,v_3,\mathcal{U}_1,\mathcal{U}_2,
\mathcal{U} ^6) & = & \\[2mm] 
& & \hspace*{-2cm} = g^\CC_{\oo;n} \left(\frac{\mathcal{U}^6}{\mathcal{U}_1}, 
\frac{\dutchcal{u}_1\dutchcal{u}_2\dutchcal{u}_3}{\mathcal{U}_1\mathcal{U}_2}, 
\frac{\mathcal{U}_1}{v_3\dutchcal{u}_1\dutchcal{u}_2}, \frac{1}{\mathcal{U}_2}, 
\mathcal{U}_2 v_2, v_1\dutchcal{u}_2v_3, \frac{\mathcal{U}_2}{\dutchcal{u}_1},\frac{1}{v_3},
\frac{\mathcal{U}_2}{v_3\dutchcal{u}_1\dutchcal{u}_2} \right).\nonumber 
\end{eqnarray}
Similarly, the covariant prefactor $\Omega$ for the direct channel can be obtained from that of the 
crossed channel as 
\begin{equation*}
\Omega_\DC = (X_{16}X_{25}X_{34})^{-\Dg_\phi} \left(\frac{\dutchcal{u}_1\dutchcal{u}_2\dutchcal{u}_3}{\mathcal{U}_1
\mathcal{U}_2} \cdot \frac{\mathcal{U}_1}{v_3\dutchcal{u}_1\dutchcal{u}_2}\cdot v_1\dutchcal{u}_2v_3
\right)^{-\frac{\Dg_\phi}{2}} = (X_{16}X_{25}X_{34})^{-\Dg_\phi} \left(\frac{v_1\dutchcal{u}_2
\dutchcal{u}_3}{\mathcal{U}_2}
\right)^{-\frac{\Dg_\phi}{2}}\ . 
\end{equation*}
In the crossing equation \eqref{eq:crossing} only the ratio of the two covariant prefactors appears. This 
is now easy to evaluate as a function of the nine cross-ratios.  
\smallskip 

\begin{figure}[htp]
    \centering
    \includegraphics[scale=0.8]{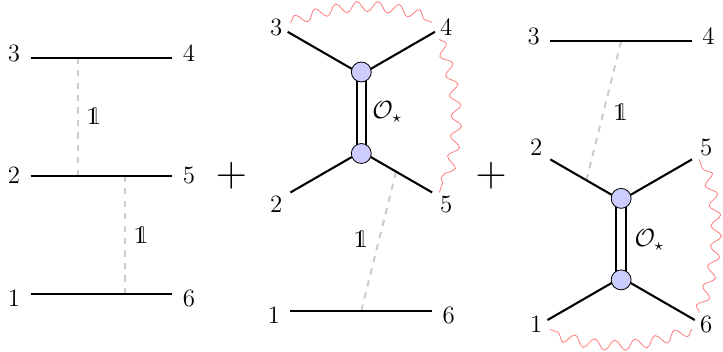}
    \caption{Leading contributions in the DC for the $\LCL^{(16,34)}$ limit, where the red wavy lines connect points that are null-separated under this limit. The exact same contributions can be produced from a $(16)2(5(34)$ or a $(16)(25)(34)$ direct channel.}
    \label{fig:six-pt-DC-leading}
\end{figure}

Let us now finally look at a few leading terms in the direct channel as we approach the lightcone regime 
we specified in the previous subsection. As we send $X_{16}$ and $X_{34}$ to zero, the direct channel is 
dominated by leading-twist exchanges in the operator products $\phi(X_1)\times\phi(X_6)$ and $\phi(X_4)
\times\phi(X_3)$. At leading order, both these OPEs are dominated by the identity exchange. We will 
analyze the crossing equation to next-to-leading order, i.e.\ also include contributions for which the 
identity operator is exchanged in only one of these OPEs while the second OPE involves the first leading operator with a twist above that of the identity. This implies that, in all 
direct-channel terms we consider, the corresponding conformal blocks reduce to four-point lightcone 
blocks, see Figure~\ref{fig:six-pt-DC-leading}.

Hence, at next-to-leading order where we take into account the leading non-trivial primary $\mathcal{O}_\ast$ 
that appears in the operator product $\phi(X_1)\times\phi(X_6)$ or $\phi(X_4)\times\phi(X_3)$, the direct-channel 
conformal block decomposition reads 
\begin{equation}
\sum_{\oo_i,n_i} P_{\oo_1\oo_2\oo_3} ^{(n_1,n_2)} g^\DC_{\oo_1\oo_2\oo_3;n_1n_2} \ \stackrel{\LCL} {\sim} \ F_{\mathds{1},\phi,\mathds{1}}+F_{\mathcal{O}_{\star},\phi,\mathds{1}}+F_{\mathds{1},\phi,\mathcal{O}_\star}+ \orm(X_{34}^{>h_\star},\!X_{16}^{>h_\star}),
\end{equation}
where the three terms on the right-hand side arise from the double identity exchange, the exchange of $\mathcal{O}_\ast$
in the $\phi(X_1)\times \phi(X_6)$ and in the $\phi(X_4) \times\phi(X_3)$ OPE, respectively. They are given by the 
following explicit formulas, 
\begin{align}
    F_{\mathds{1},\phi,\mathds{1}}&=\left(\frac{v_1 
    \dutchcal{u}_2\dutchcal{u}_3}{\mathcal{U}_2}\right)^{\frac{\Dg_\phi}{2}}\, ,\\[2mm] 
    F_{\mathcal{O}_{\star},\phi,\mathds{1}}&= C_{\phi\phi\oo_\star}^2 \left(\frac{v_1 \dutchcal{u}_2\dutchcal{u}_3}
    {\mathcal{U}_2}\right)^{\frac{\Dg_\phi}{2}} \left(\frac{\mathcal{U}^6}{\mathcal{U}_1\mathcal{U}_2}\right)^{h_\star} 
    \left(1-\frac{\dutchcal{u}_1\dutchcal{u}_2\dutchcal{u}_3}{\mathcal{U}_1\mathcal{U}_2}\right)^{J_\star}\hypg{\bar h_\star}
    {\bar h_\star}{2\bar h_\star}\left(1-\frac{\dutchcal{u}_1\dutchcal{u}_2\dutchcal{u}_3}{\mathcal{U}_1\mathcal{U}_2}\right)
    \, , \\[2mm] 
   F_{\mathds{1},\phi,\mathcal{O}_\star}&=C_{\phi\phi\oo_\star}^2\left(\frac{v_1 \dutchcal{u}_2\dutchcal{u}_3}{
   \mathcal{U}_2}\right)^{\frac{\Dg_\phi}{2}}  v_2^{h_\star} \left( 1-v_1\dutchcal{u}_2v_3\right)^{J_\star} 
   \hypg{\bar h_\star}{\bar h_\star}{2\bar h_\star}\left(1-v_1\dutchcal{u}_2v_3\right)\, ,
\end{align}
In writing these expressions we have not yet applied the lightcone limits that are controlled by $\epsilon_{456}$
and that are used to suppress higher-twist exchange in the crossed channel. Once we apply this limit as well, the 
hypergeometric functions can be approximated through their relevant asymptotic behavior. After multiplying with the 
covariant factors that appear on the left-hand side of the crossing equation and applying the full lightcone limit, 
we obtain the following terms in the direct-channel expansion
\begin{align}\label{eq:leading_DC}
&\frac{\Om_\DC}{\Om_\CC}\sum_{\oo_i,n_i} P_{\oo_1\oo_2\oo_3} ^{(n_1,n_2)} g^\DC_{\oo_1\oo_2\oo_3;n_1n_2} 
\ \stackrel{\LCL^{(16,34)}}{\sim} \ \omega^{h_\phi} 
\left[f_{\mathds{1},\phi,\mathds{1}}+f_{\mathcal{O}_{\star},\phi,\mathds{1}}
+f_{\mathds{1},\phi,\mathcal{O}_\star}+ \orm(X_{34}^{>h_\star},\!X_{16}^{>h_\star})\right],
\end{align}
where
\begin{align}
    \om &= v_1\dutchcal{u}_1^2 v_2^2\dutchcal{u}_2^3 v_3\dutchcal{u}_3^2 
    & 
    f_{\mathcal{O}_{\star},\phi,\mathds{1}}&=-\frac{C_{\phi\phi\oo_\star}^2}{\mathrm{B}_{\bar h_\star}}
    \left(\frac{\mathcal{U}^6}{\mathcal{U}_1\mathcal{U}_2}\right)^{h_\star}\frac{\log \dutchcal{u}_2+ \mathcal{O}
    ( \dutchcal{u}_2^0)}{(\mathcal{U}^6 v_2)^{2h_\phi}} \nonumber
    \\[2mm] 
    f_{\mathds{1},\phi,\mathds{1}}&=(\mathcal{U}^6 v_2)^{-2h_\phi} 
    &
    f_{\mathds{1},\phi,\mathcal{O}_\star}&=-\frac{C_{\phi\phi\oo_\star}^2}{\mathrm{B}_{\bar h_\star}}v_2^{h_\star}
    \frac{\log \dutchcal{u}_2 + \mathcal{O}( \dutchcal{u}_2^0)}{(\mathcal{U}^6 v_2)^{2h_\phi}}\,.
    \label{eq:list_DC_contrib}
\end{align}
These are the three terms of the direct-channel expansion that we would like to express in terms of the crossed-channel lightcone expansion. But before we can do so, we need some extensive preparation. In the next section, we 
shall first analyze the relevant crossed-channel lightcone blocks. These are much harder to determine than for 
the direct channel since they involve a non-trivial exchange in all three intermediate channels.

\section{Crossed-channel Blocks in the Lightcone Limit}
\label{sec:blocks_cc}

\begin{figure}[htp]
    \centering
    \includegraphics[scale=0.8]{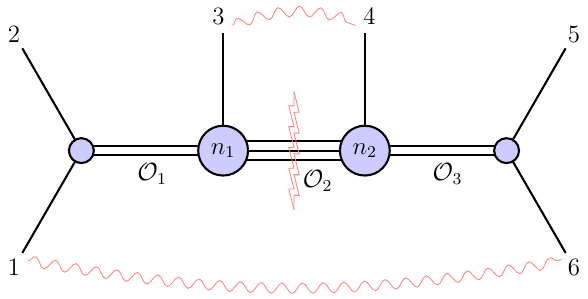}
    \caption{OPE diagram associated with the CC expansion. The $\epsilon_{16},\epsilon_{34}\to 0$ limits, represented as wavy lines, entail large-spin exchanges at the internal legs that they cover, while the $\epsilon_{456}\to 0$ limit (conformally equivalent to a $\epsilon_{123}\to 0$ limit for the first three fields) is represented as a line that cuts the middle leg, where contributions with lowest twist become more relevant.}
    \label{fig:six-pt-CC-limits}
\end{figure}

The goal of this section is to derive an explicit formula for the lightcone conformal blocks relevant for the crossed-channel expansion, see Figure~\ref{fig:six-pt-CC-limits}. Our main motivation for this is that expanding the leading-twist DC terms discussed in Section~\ref{sec:leadingtwistexpansionDC} into these blocks allows us to extract asymptotic conformal data from the crossing equation. 

In the first subsection (Sec.~\ref{sect:Casimirs-Vertex}), we spell out 
explicit formulas for the leading terms of the differential operators in the relevant 
limit. Then we use these formulas to determine the scaling of eigenvalues in 
the lightcone limit. In the resulting regime, we shall encounter a very 
remarkable surprise: it turns out that the spectrum of the two vertex 
operators can be calculated exactly in terms of the usual tensor structure 
labels. This will allow us to construct the lightcone limit of conformal 
blocks in the usual basis for tensor structures, see eq.~\eqref{eq:MST-STT-scalar-3pt}. Since 
the expressions are obtained by solving a system of differential equations, 
the overall normalization must be determined through continuation from the 
OPE limit as explained in \cite{Kaviraj:2022wbw}. This analysis is described 
in the final subsection. Let us stress, however, that the normalizing  
prefactors are not needed in the subsequent determination of anomalous 
dimensions. 

\subsection{Casimir and vertex operators}
\label{sect:Casimirs-Vertex}
In this subsection, we state our conventions regarding Casimir and vertex operators
and explicitly spell out the leading terms (in the sense of \cite{Kaviraj:2022wbw}) 
of several key differential operators in our lightcone limits. 

We shall denote the usual conformal generators acting on the coordinates of the 
$i^{\textrm{th}}$ insertion point by $\mathcal{T}_i$. In the embedding space formalism, the conformal generators take the following simple form 
\begin{align}
    (\mathcal{T}_i)^{A B} := X_i^A \partial^B_i -X_i^B \partial^A_i.
\end{align} 
In terms of these generators,
we can construct seven Casimir operators of the following form 
\begin{align}
    \mathcal{D}_{ij}^2 := \frac{1}{4} \text{tr}(\mathcal{T}_i + \mathcal{T}_j)^2 && 
    \mathcal{D}_{ijk}^2 
    := \frac{1}{4} \text{tr}(\mathcal{T}_i + \mathcal{T}_j+ \mathcal{T}_k)^2 \\[2mm] 
    \mathcal{D}_{ij}^4 := \frac{1}{2} \text{tr}(\mathcal{T}_i + \mathcal{T}_j)^4 - 
    2\left(\mathcal{D}_{ij}^2\right)^2&& \mathcal{D}_{ijk}^4 := \frac{1}{2} 
    \text{tr}(\mathcal{T}_i + 
    \mathcal{T}_j+ \mathcal{T}_k)^4 - 2\left(\mathcal{D}_{ijk}^2\right)^2
\end{align}
\begin{equation}
     \mathcal{D}_{ijk}^6 :=  
    \text{tr}(\mathcal{T}_i + 
    \mathcal{T}_j+ \mathcal{T}_k)^6-4 \left(\mathcal{D}_{ijk}^2\right)^3-3\left(\mathcal{D}_{ijk}^4\right)\!\left(\mathcal{D}_{ijk}^2\right)-2d^2\left(\mathcal{D}_{ijk}^2\right)^2
\end{equation}
where the indices $ij$ are either $ij=12$ or $ij=56$ and $ijk = 123$. These operators can be used to label the representations of the fields exchanged in the internal legs of the diagram. In addition, 
we can construct two vertex differential operators that label the tensor structures at the two innermost vertices of the comb. We shall work with the following pair\footnote{compared to the previous works~\cite{Buric:2021ywo, Buric:2021ttm, Kaviraj:2022wbw}, we are here considering a definition for these vertex operators shifted by quadratic and quartic Casimirs. This form makes the vertex operators behave well under $\mathbb{Z}_2$ transformations that exchange their spinning legs, and display a simpler form under lightcone and large-spin limits.}  
\begin{align}
	    \mathcal{V}_1 := \frac{1}{2}
	    \text{tr}\left[\left(\mathcal{T}_1+\mathcal{T}_2\right)^3
	    \left(\mathcal{T}_3\right)\right] +\frac14 \tr \left(\mathcal{T}_{1}+\mathcal{T}_2\right)^4+\left(\frac12 \tr \mathcal{T}_3^2-\mathcal{D}_{123}^2\right)\mathcal{D}_{12}^2+\frac{d(d-1)}{2}\left[ \frac14 \tr \mathcal{T}_3^2 - \mathcal{D}_{12}^2 \right]\,,\\[2mm]
        \mathcal{V}_2 := \frac{1}{2}
	    \text{tr}\left[\left(\mathcal{T}_6+\mathcal{T}_5\right)^3
	    \left(\mathcal{T}_4\right)\right] +\frac14 \tr \left(\mathcal{T}_{6}+\mathcal{T}_5\right)^4+\left(\frac12 \tr \mathcal{T}_4^2-\mathcal{D}_{123}^2\right)\mathcal{D}_{56}^2+\frac{d(d-1)}{2}\left[ \frac14 \tr \mathcal{T}_4^2 - \mathcal{D}_{56}^2 \right]\,.
\end{align} 
After inserting these expressions into our formulas for the nine differential operators and 
conjugating with the leg factor $\Omega_{\mathrm{CC}}$ that we defined in eq.~\eqref{eq:Omega_definition},
we obtain a set of reduced differential operators that act on the nine cross-ratios instead of the six 
embedding space vectors. In a slight abuse 
of notation, we will continue to denote these dressed operators by $\mathcal{D}$ and $\mathcal{V}$. 
Furthermore, we will often conjugate by some additional factor $\om$ in order to simplify the expressions 
we are dealing with, i.e.\ we introduce a new set of blocks $\tilde g$ by splitting off a simple function 
of the cross-ratios as follows
\begin{equation}
    g_{\oo_1\oo_2\oo_3;n_1n_2}^\CC(\dutchcal{u}_a,v_a,\mathcal{U}_i,\mathcal{U}^6) = 
    \om^{h_\phi}  \tilde{g}_{\oo_1\oo_2\oo_3;n_1n_2}(\dutchcal{u}_a,v_a,\mathcal{U}_i,\mathcal{U}^6)\ , 
    \quad    \quad  \om := v_1\dutchcal{u}_1^2 v_2^2\dutchcal{u}_2^3 v_3\dutchcal{u}_3^2\,. 
    \label{eq:rescaled_blocks_omega}
\end{equation}
When acting on the functions $\tilde g$, the leading terms of the quadratic Casimir operators read 
\begin{multline}
    \widetilde{\mathcal{D}}_{12}^2=\epsilon_{16}^{-1} \mathcal{U}_2 \partial_{\mathcal{U}^6} \!\Bigl[\left(1-\mathcal{U}_1\right) \left(\vartheta _{\dutchcal{u}_3}+\vartheta _{\mathcal{U}_1}+\vartheta _{\mathcal{U}^6}-\vartheta _{v_2}\right)-\mathcal{U}_1 \left(\vartheta _{\dutchcal{u}_1}+\vartheta _{\mathcal{U}_2}+2h_\phi \right)\\[2mm] 
    -\vartheta _{1-v_1-\frac{\dutchcal{u}_1}{\mathcal{U}_2}}+\orm\!\left(\epsilon_{456}\right)\Bigr]
    +\orm\!\left(\epsilon_{16}^{0}\right),
    \label{eq:Cas12epsilon_b}
\end{multline}
\begin{multline}
    \widetilde{\mathcal{D}}_{56}^2= \epsilon_{16}^{-1}\mathcal{U}_1 \partial _{\mathcal{U}_6} \Bigl[\left(1-\mathcal{U}_2\right) \left(\vartheta _{\dutchcal{u}_1}+\vartheta _{\mathcal{U}_2}+\vartheta _{\mathcal{U}_6}-\vartheta _{v_2}\right)-\mathcal{U}_2 \left(\vartheta _{\dutchcal{u}_3}+\vartheta _{\mathcal{U}_1}+2h_\phi\right)\\[2mm] 
    -\vartheta _{1-v_3-\frac{\dutchcal{u}_3}{\mathcal{U}_1}}+\orm\!\left(\epsilon_{456}\right)\Bigr]
    +\orm\!\left(\epsilon_{16}^{0}\right),
    \label{eq:Cas56epsilon_b}
\end{multline}
\begin{align}
   \text{and} \hspace{1 cm} \widetilde{\mathcal{D}}_{123}^2=\epsilon_{16}^{-1}\epsilon_{34}^{-1}\left[\partial_{v_2}\partial_{\mathcal{U}^6}+\orm\!\left(\epsilon_{456} \right)\right]+\orm\!\left(\epsilon_{16}^{-1},\epsilon_{34}^0 \right)+\orm\!\left(\epsilon_{16}^{0},\epsilon_{34}^{-1} \right),
    \label{eq:Cas123epsilon_b}
\end{align}
where we used the shorthand notation for Euler operators $\vt_x:=x\partial_x$.
The fourth-order Casimir operator $\widetilde{\mathcal{D}}_{123}^4$ can be conveniently expressed in terms of $\widetilde{\mathcal{D}}_{123}^2$ as
\begin{equation}
\begin{aligned}
    \widetilde{\mathcal{D}}_{123}^4=& \Bigl[2 (1-\mathcal{U}_1-\mathcal{U}_2) 
    \left(3 d-12 h_{\phi }-4 \vartheta _{\dutchcal{u}_2}-6\right) \left(2 h_{\phi }+
    \vartheta _{\dutchcal{u}_1}+\vartheta _{\dutchcal{u}_3}+\vartheta _{\mathcal{U}_1}+\vartheta _{\mathcal{U}_2}+
    \vartheta _{\mathcal{U}_6}-\vartheta _{v_2}\right)\\[2mm] 
    &\quad +8 \vartheta _{\dutchcal{u}_2} \left(-d+6 h_{\phi }+\vartheta _{\dutchcal{u}_2}+1\right)+
    (d-1) \left(3 d-24 h_{\phi }-4\right)+72 h_{\phi }^2+\orm\!\left(\epsilon_{456}\right)\Bigr]\widetilde{\mathcal{D}}_{123}^2\\[2mm] 
    &\hspace{8cm}+\orm\!\left(\epsilon_{16}^{-1},\epsilon_{34}^0 \right)+\orm\!\left(\epsilon_{16}^{0},\epsilon_{34}^{-1} \right).
    \end{aligned}
    \label{eq:quarticepsilon_b}
\end{equation}
The remaining quartic Casimirs $\widetilde{\mathcal{D}}_{12}^4$ and $\widetilde{\mathcal{D}}_{56}^4$, the sixth-order Casimir 
$\widetilde{\mathcal{D}}_{123}^6 $, as well as the two vertex operators $\widetilde{\mathcal{V}}_1$ and $\widetilde{\mathcal{V}}_2$ 
have a more complicated form that we will not display explicitly. 

\subsection{Large-spin behavior of the eigenvalues}

As our notations for the conformal blocks $\tilde g$ indicate, the eigenfunctions of our differential operators 
carry nine quantum numbers. These are the weights $h_i$ and spins $J_i$ of the two outer intermediate operators, the 
weight $h_2$ and spins $(J_2,\kappa)$ of the intermediate operator in the center, as well as the tensor structure labels $n_1,n_2$ of the two non-trivial vertices. As we shall show 
in the next subsection, the expansion of the direct-channel terms through crossed-channel blocks requires that all 
intermediate spins $J_i$ become large such that the spins scale as 
\begin{equation}  \label{eq:largespinregime} 
  1 \ll J_1,J_3 \ll J_2 \  
\end{equation} 
while keeping $\kappa$ finite. For later use, we want to spell out some of the well-known eigenvalues of the various 
Casimir operators in this regime. Let us begin with the second- and fourth-order Casimir operators 
$\widetilde{\mathcal{D}}_{123}^{p}, p=2,4$ of the middle leg. In the large-spin regime their eigenvalue 
equations read 
\begin{equation}
    \widetilde{\mathcal{D}}_{123}^2 \tilde{g}_{\oo_1\oo_2\oo_3;n_1n_2}= \left(J_2^2+\orm (J_2)\right)
    \tilde{g}_{\oo_1\oo_2\oo_3;n_1n_2} 
\end{equation}
and
\begin{equation}
    \widetilde{\mathcal{D}}_{123}^4 \tilde{g}_{\oo_1\oo_2\oo_3;n_1n_2}= \left(\dutchcal{c}(h_2,\kappa)J_2^2+\orm(J_2)\right)\tilde{g}_{\oo_1\oo_2\oo_3;n_1n_2} 
    \label{eq:Quartic_123_eigenvalue}
\end{equation}
with
\begin{equation}
    \dutchcal{c}(h,\kappa)= 8 h(h+\kappa-d+1)-6(d-2)\kappa +3d^2-7d+4\,.
\end{equation}
For the remaining Casimirs $\widetilde{\mathcal{D}}_{12}^p$ and $\widetilde{\mathcal{D}}_{56}^p$ that are associated with 
the outer legs, the Casimir eigenvalue equations take the same form, with the obvious replacements $J_2\rightarrow J_1, J_3$ 
and $\dutchcal{c}(h_2,\kappa)\rightarrow \dutchcal{c}(h_1,0),\dutchcal{c}(h_3,0)$ respectively. 
\smallskip 

While the blocks $g_{\oo_1,\oo_2,\oo_3;n_1,n_2}$ are eigenfunctions of the Casimir operators that measure the quantum numbers of the three intermediate fields $\oo_a$, they need not be eigenfunctions of the two vertex operators in general. 
Put differently, the eigenvalue equations of the vertex operators $\mathcal{V}$ select a different basis of blocks that 
is not aligned with the basis labeled by the conventional tensor structure labels $n_1,n_2$. For a generic choice of spin labels, the eigenvalues and eigenfunctions of the six-point vertex operators are not known. But in the regime in which the 
spin $J_2$ is much larger than $J_{1,3}$ there is a very surprising twist to this general description: when applied 
to the usual basis of blocks that is labeled by the tensor structures $n_1 \leq J_1-\kappa$ and $n_2\leq J_3-\kappa$, 
the two vertex operators turn out to be upper triangular! Since all matrix elements, and in particular the ones on the 
diagonal, can be computed explicitly, this implies that the spectrum of the vertex operators is explicitly known 
in the regime of large $J_2$. In addition, there is one block, namely the one with the largest possible tensor 
structure labels $n_1 = J_1 - \kappa$ and $n_2 = J_3-\kappa$, that is an eigenfunction of the vertex differential 
operators. 

For later use, we want to spell out new closed-form expressions for the eigenvalues and eigenvectors of the vertex differential operators 
in the regime of large $J_2$, see Appendix \ref{app:VertexOperator} for details. We parameterize the spectra of 
$\mathcal{V}_1$ and $\mathcal{V}_2$ as $\mathfrak{t}(h_1,J_1,h_2,J_2,\kappa,h_\phi,d;\nu_1)$ and $\mathfrak{t}(h_3,J_3,h_2,J_2,
\kappa,h_\phi,d;\nu_2)$, respectively, where $\nu_{1,2}$ are integers that go from $0$ to $J_{1,3} - \kappa$, respectively. As 
discussed in the appendix, their large $J_2$ expansion reads 
\begin{align}\label{eq:largeJ2Vertex}
    \mathfrak{t}(h_i,J_i,h_2,J_2,\kappa,h_\phi,d;\nu)=\frac{J_2^2}{2}\Bigl[[\nu]_{\bar{h}_i}\!\left([\nu]_{\bar{h}_i}-d\right)-\kappa([\nu]_{\bar{h}_i}-2)\Bigr] + \orm(J_2)
\end{align}
for the eigenvalues $\mathfrak{t}$ with $i=1,3$. In order to render the expression a bit more compact, we have 
introduced the following shorthand, 
\begin{align}
    [\nu]_{\bar{h}_i} = 2 (\bar h_i - \nu).
\end{align}
The vertex operator eigenfunction corresponding to the eigenvalue labeled by $\nu_1,\nu_2$ is, to leading order in $J_2$, 
a simple linear combination of standard basis tensor structures with label $n_1,n_2$. Both sets of integers range from $0$
to $J_i-\kappa$ (see eq.~\eqref{eq:vertexwavefunction}). Furthermore, it turns out that the eigenfunction of the vertex 
differential operators associated with the largest value of the tensor structure labels, i.e.\ $\nu_1 = J_1 - \kappa$ 
and $\nu_2 = J_3 - \kappa$ coincides with the blocks with tensor structure labels $n_1 = \nu_1, n_2 = \nu_2$, respectively. 
This remarkable observation will be crucial for the reasoning below when we construct the lightcone limit of crossed 
channel blocks.

\subsection{Casimir singularity of the direct channel}
\label{ssec:Casimir_singularity}

In order to identify the asymptotics of quantum numbers exchanged in the CC OPE, we shall follow the approach of 
\cite{Alday:2015ewa,Alday:2016njk,Simmons-Duffin:2016wlq}, see also \cite{Kaviraj:2022wbw} for the extension to multipoint lightcone bootstrap, 
and act with the CC differential operators on the leading DC contributions. In terms of the rescaled blocks 
$\tilde{g}_{\oo_1\oo_2\oo_3;n_1n_2}$ we introduced in eq.~\eqref{eq:rescaled_blocks_omega} and using 
the leading DC contributions identified in eq.~\eqref{eq:leading_DC}, we can rewrite the crossing equation 
as
\begin{equation}
f_{\mathds{1},\phi,\mathds{1}}+f_{\mathcal{O}_{\star},\phi,\mathds{1}}+f_{\mathds{1},\phi,\mathcal{O}_\star}+ 
\orm(X_{34}^{>h_\star},\!X_{16}^{>h_\star})=
\sum_{\oo,n}\! P_{\oo_1\oo_2\oo_3}^{(n_1n_2)} \, \tilde{g}_{\oo_1\oo_2\oo_3;n_1n_2}.
\end{equation}
Acting on the terms on the left-hand side with the Casimir and vertex operators we described in the first subsection, 
we find that they all satisfy
\begin{gather}
    \frac{\widetilde{\mathcal{D}}_{12}^4 f_{a,\phi,b}}{\widetilde{\mathcal{D}}_{12}^2 f_{a,\phi,b}}\simeq
    \frac{\widetilde{\mathcal{D}}_{56}^4 f_{a,\phi,b}}{\widetilde{\mathcal{D}}_{56}^2 f_{a,\phi,b}}\simeq
    \dutchcal{c}(2h_\phi,0) \,, \label{eq:action_DC_quartics_sides}
    \\[2mm]
    \frac{\widetilde{\mathcal{D}}_{123}^4 f_{a,\phi,b}}{\widetilde{\mathcal{D}}_{123}^2 f_{a,\phi,b}}\simeq
    \dutchcal{c}(3h_\phi,(2h_\phi-\abs{h_a+h_b})(\mathcal{U}_1+\mathcal{U}_2-1))\,,
    \\[2mm]
\frac{\left(\widetilde{\mathcal{D}}_{123}^4-6\widetilde{\mathcal{V}}_s\right) 
f_{a,\phi,b}}{\widetilde{\mathcal{D}}_{123}^2f_{a,\phi,b}} \simeq-4(-1+d-6h_\phi+3dh_\phi-6h_\phi^2), 
\qquad \mathrm{for} \quad s=1,2\,,\label{eq:action_DC_Vertex-Quart}
\end{gather}
where $(a,b)\in \{(\mathds{1},\mathds{1}),(\mathds{1},\mathcal{O}_\star),(\mathcal{O}_\star,\mathds{1})\}$ and  ``$\simeq$'' stands for equality at leading order in the $\epsilon$ scaling parameters. Thus, 
if we assume the spins $J_i$, $i=1,2,3$ of the exchanged operators to be large, the leading DC contributions are 
reproduced by CC conformal blocks with twists $h_i$, MST spin $\kappa$, and tensor structures $\mathfrak{t}_i$ 
fixed to be
\begin{gather}
    h_1=h_3=2h_\phi\,, \qquad h_2=3h_\phi\,, \qquad \kappa = \orm(1)
    \label{eq:hiCC}\\[2mm]
    \mathfrak{t}_1=\mathfrak{t}_2=\mathfrak{t}_\star =J_2^2
    \left[\left(h_1+\kappa\right)\left(2h_1-d\right)+2\kappa \right].
\label{ts_constr_from_dc}
\end{gather}
The large-spin assumption is confirmed by the behavior of the quadratic Casimirs, which give
\begin{equation}
    \widetilde{\mathcal{D}}_{123}^2 f_{a,\phi,b}\simeq\frac{2h_\phi 
    \left(2 h_{\phi}-\abs{h_a+h_b}\right)}{v_2\mathcal{U}^6}f_{a,\phi,b} \,,
\end{equation}
again for $(a,b)\in \{(\mathds{1},\mathds{1}),(\mathds{1},\mathcal{O}_\star),(\mathcal{O}_\star,\mathds{1})\}$, 
as well as
\begin{gather}
    \widetilde{\mathcal{D}}_{12}^2 f_{\mathds{1},\phi,\mathds{1}}\simeq \widetilde{\mathcal{D}}_{56}^2 
    f_{\mathds{1},\phi,\mathds{1}}\simeq\frac{4 h_{\phi}^2\mathcal{U}_1\mathcal{U}_2}
    {\mathcal{U}^6}f_{\mathds{1},\phi,\mathds{1}}\,,\label{eq:equal_spins_two_identity}
    \\[4mm] 
    \widetilde{\mathcal{D}}_{12}^2 f_{\mathcal{O}_{\star},\phi,\mathds{1}}\simeq 
    \widetilde{\mathcal{D}}_{56}^2 f_{\mathcal{O}_{\star},\phi,\mathds{1}}\simeq 
    \frac{\left(2 h_{\phi}-h_\star\right)^2\mathcal{U}_1\mathcal{U}_2}
    {\mathcal{U}^6}f_{\mathcal{O}_{\star},\phi,\mathds{1}}\,,\\[4mm]
    \widetilde{\mathcal{D}}_{12}^2 f_{\mathds{1},\phi,\mathcal{O}_{\star}}\simeq 
    \left[\widetilde{\mathcal{D}}_{56}^2 +\frac{2h_\phi h_\star\left(\mU_1-\mU_2\right)}{\mU^6}\right] 
    f_{\mathds{1},\phi,\mathcal{O}_{\star}}\simeq \frac{2h_\phi \left(2 h_{\phi}-h_\star\right)\mathcal{U}_1
    \mathcal{U}_2+2h_\phi h_\star \mathcal{U}_2}{\mathcal{U}^6}f_{\mathds{1},\phi,\mathcal{O}_{\star}}\,.
\end{gather}
In fact, given the scalings $\mathcal{U}^6\propto \epsilon_{16}$ and $v_2\propto \epsilon_{34}$, the action of these quadratic Casimirs implies that the most relevant contributions to the CC correspond to exchanged 
operators whose spins scale as
\begin{equation} \label{eq:Jscaling} 
    J_1^2=\orm \!\left(\epsilon_{16}^{-1}\right)\,, \qquad J_2^2=\orm \! \left(\epsilon_{16}^{-1}\epsilon_{34}^{-1}\right)\,, \qquad J_3^2=\orm \! \left(\epsilon_{16}^{-1}\right),
\end{equation}
subject to the extra constraint $J_1=J_3$ for blocks that reproduce $f_{\mathds{1},\phi,\mathds{1}}$ and 
$f_{\mathcal{O}_{\star},\phi,\mathds{1}}$. This is indeed the regime we anticipated in the previous subsection, 
see eq.\ \eqref{eq:largespinregime}, when we discussed the spectrum of the various differential operators. 

In particular, since we are indeed driven into a regime in which $J_2$ is much larger than all other quantum 
numbers, we can use the large $J_2$ spectrum of the vertex differential operators as given in eq.~\eqref{eq:largeJ2Vertex}. 
Comparing with eq.~\eqref{ts_constr_from_dc}, we deduce
\begin{align}
    [\nu_1]_{\bar{h}_1} = 2(2h_\phi+J_1+\kappa-\nu_1) \stackrel{!}{=}  4 h_\phi +2 \kappa && \Longrightarrow &&  
    \nu_1 = J_1 - \kappa  
\end{align}
and likewise $\nu_2 = J_3 -\kappa$. By the observation made at the end of the previous subsection, this implies 
that the CC lightcone blocks are fixed to have maximal tensor structure labels in the $n$-basis of three-point tensor structures 
defined previously, i.e.\ we have
\begin{equation}
 (n_1,n_2)=(J_1-\kappa,J_3-\kappa).
\end{equation}
We have thus seen that, for the specific quantum numbers dictated by the crossing equation, the CC lightcone blocks in the usual $n$-basis of tensor structures satisfy the eigenvalue 
equations for the vertex differential operators. Therefore, we will be able to determine the lightcone 
limit of the relevant blocks from differential equations alone, at least up to an overall normalization.

\subsection{Explicit form of six-point triple-twist blocks}
In the previous subsection, we understood which quantum numbers must appear in the CC conformal blocks in order 
to reproduce the DC contributions to the crossing equation~\eqref{eq:list_DC_contrib}. We now aim to leverage 
the differential eigenvalue equations for these conformal blocks to determine their functional form. While 
these equations are hard to solve in full generality, in the lightcone limits we consider and for the twists 
relevant to the crossing equation $h_1=h_3=2h_\phi$ and $h_2=3 h_\phi$, the problem greatly simplifies. When defining $\tilde{g}_{\oo_1\oo_2\oo_3;n_1n_2}$ in eq.~\eqref{eq:rescaled_blocks_omega}, we extracted out 
the lightcone OPE asymptotics of the blocks, i.e.\ a factor $\left(\du_1^2 \du_2^3 \du_3^2\right)^{h_\phi}$ for 
$\du_{\{i\}}\to 0$, Hence, we now look for solutions of the differential operators presented in 
eqs.~\ref{sect:Casimirs-Vertex} whose leading term in the regime $\du_{\{i\}}\to 0$ does not depend on 
$\du_{i}$. In particular, since the limits we are interested in already include $\du_2\to 0$ as a 
consequence of $X_{45},X_{56},X_{46}\ll 1$, we conclude that only the leading term in $\du_2$ of the 
blocks will contribute, and thus the function $\tilde{g}_{\oo_1\oo_2\oo_3;n_1n_2}$ has to be 
independent of~$\du_2$. This is perfectly compatible with the fact that all the relevant differential 
operators for the blocks commute with the Euler operator $\vt_{\du_2}$, and thus are diagonalized by 
any power law in $\du_2$, including the case of a constant. 

Before we start our diagonalization procedure we shall perform the following simple change of 
variables 
\begin{equation}
    \begin{array}{c}
      v_1 \quad \longrightarrow \quad r_1=1-v_1-\frac{\du_1}{\mU_2}\,, \\[2mm]
      v_3 \quad \longrightarrow \quad  r_3=1-v_3-\frac{\du_3}{\mU_1}\,,
    \end{array}
\end{equation}
Once performed, it is easy to see that the differential operators spelled out in eqs.\ 
\eqref{eq:Cas12epsilon_b}, \eqref{eq:Cas56epsilon_b}, \eqref{eq:Cas123epsilon_b}, 
and~\eqref{eq:quarticepsilon_b} commute with the four Euler operators $\vt_{\du_1}$, 
$\vt_{\du_3}$, $\vt_{r_1}$, $\vt_{r_3}$. We can therefore work temporarily in an eigenbasis 
of the latter, 
\begin{equation} \label{eq:Eulerbasis} 
    \begin{gathered}
    \left(\begin{array}{c}
       \vt_{u_1}    \\
        \vt_{u_3}\\
        \vt_{r_1}\\
        \vt_{r_3}
    \end{array}\right)G_{(J_1,J_2,\kappa,J_3);m_1m_2}^{p_1,p_3}=\left(\begin{array}{c}
       p_1    \\
        p_3\\
        m_1\\
        m_3
    \end{array}\right) G_{(J_1,J_2,\kappa,J_3);m_1m_2}^{p_1,p_3}\,.
    \end{gathered}
\end{equation}
Let us stress that this basis is \emph{not} compatible with the diagonalization of the two quartic Casimirs
$\widetilde{\mathcal{D}}_{12}^4$, $\widetilde{\mathcal{D}}_{56}^4$ of the outer intermediate
legs and of the two vertex operators. 

Let us now aim to diagonalize the quadratic Casimirs and the quartic Casimir $\widetilde{\mathcal{D}}^4_{123}$
in the basis \eqref{eq:Eulerbasis}. Diagonalizing the Euler operators by simple power laws, and enforcing $\widetilde{\mathcal{D}}_{123}^2 \rightarrow J_2^2$ in the expression~\eqref{eq:quarticepsilon_b} 
of $\widetilde{\mathcal{D}}_{123}^4$, we conclude that the eigenvalue equation~\eqref{eq:Quartic_123_eigenvalue} 
for the quartic Casimir simply reduces to a first-order differential equation that constrains the auxiliary 
functions $G$ to take the form
\begin{equation}
    G_{(J_1,J_2,\kappa,J_3);m_1m_3}^{p_1,p_3}\!=v_2^{2h_\phi+\kappa+p_1+p_3}(\mathcal{U}_1+\mathcal{U}_2-1)^{\kappa}
    \! \left(1-v_1-\frac{\du_1}{\mU_2}\right)^{\hspace{-1pt}m_1}\hspace*{-4pt}\left(1-v_3-\frac{\du_3}{\mU_1}\right)^{\hspace{-1pt} m_3}
    \hspace*{-2pt} \widetilde{G}(v_2 \mU_1,v_2 \mU_2,v_2 \mU^6)\,.\label{Gp1p2m1m2}
\end{equation}
We can now plug this Ansatz into three eigenvalue equations for the quadratic Casimir. By taking an 
appropriate linear combination of the form 
\begin{equation}
    \mathcal{D}_{123}^2+\frac{\mathcal{D}_{12}^2}{v_2\mathcal{U}_2}+\frac{\mathcal{D}_{56}^2}{v_2\mathcal{U}_1}
\end{equation}
one obtains a relatively simple differential equation that constrains the dependence of the solutions on 
the variable $v_2 \mathcal{U}^6$, 
\begin{equation} \label{eq:BesselClifford} 
   \left[ \partial_{v_2 \mathcal{U}^6}\left(v_2 \mathcal{U}^6\partial_{v_2 \mathcal{U}^6} +
   (-2h_\phi -m_1-m_3-\kappa)\right)-\left(\frac{J_1^2}{v_2\mathcal{U}_2}+J_2^2+
   \frac{J_3^2}{v_2\mathcal{U}_1}\right)
   \right]\widetilde{G}(v_2 \mU_1,v_2 \mU_2,v_2 \mU^6)=0\,. 
\end{equation}
We can recognize this as a Bessel-Clifford differential equation~\cite[Appendix B.1]{Kaviraj:2022wbw}, up 
to a change of variables. The solutions can be further constrained by imposing the eigenvalue equations 
for any two of the quadratic Casimir operators. When applied to solutions of the Bessel-Clifford equation 
\eqref{eq:BesselClifford} one obtains two first-order constraints which are easily solved by
\begin{multline}
    \widetilde{G}(v_2 \mU_1,v_2 \mU_2,v_2 \mU^6)=\left(\frac{\mU^6}{v_2\mU_1\mU_2}\right)^{2h_\phi +\kappa} 
    \left(\frac{\mU^6}{\mU_2}\right)^{m_1}\left(\frac{\mU^6}{\mU_1}\right)^{m_3} \left(v_2 
    \mathcal{U}_2\right)^{-p_1} \left(v_2 \mathcal{U}_1\right)^{-p_3} 
    \\[2mm]  
    \mathcal{K}_{2h_\phi+m_1+m_3+\kappa} \left(\left[\frac{J_1^2}{\mathcal{U}_2}+J_2^2v_2+
    \frac{J_3^2}{\mathcal{U}_1}\right]\mathcal{U}^6 \right).
\end{multline}
With this step, we completely determined the auxiliary functions $G$ which we introduced in eq.\ 
\eqref{eq:Eulerbasis}. By construction, these functions provide a basis of the function space we 
are interested in. While generic lightcone conformal blocks correspond to linear combinations of 
these auxiliary functions, one may check that the constraints~\eqref{eq:action_DC_quartics_sides} and~\eqref{eq:action_DC_Vertex-Quart} are satisfied by conformal blocks that are built up of a single 
function $G_{(J_1,J_2,\kappa,J_3);0,0}^{0,0}$, with the parameters $m_1=p_1=m_3=p_3=0$. In 
conclusion we have shown that the relevant lightcone blocks are given by 
\begin{equation}
    \tilde{g}_{(J_1,J_2,\kappa,J_3);J_1J_3} \sim\mathcal{N}_{J_1(J_2,\kappa)J_3}  
    (\mathcal{U}_1+\mathcal{U}_2-1)^\kappa \left(\frac{\mathcal{U}^6}{\mU_1 \mU_2}\right)^{2h_\phi+\kappa} \mathcal{K}_{2h_\phi+\kappa} \left(\left[\frac{J_1^2}{\mathcal{U}_2}+J_2^2 v_2+\frac{J_3^2}{\mathcal{U}_1}\right]\mathcal{U}^6 \right)
    \label{gtilde_to_besselK}
\end{equation}
up to an overall normalization $\mathcal{N}_{J_1(J_2,\kappa)J_3}$. Here we have suppressed the dependence 
on the quantum numbers $h_i$ since these are determined by the conformal weight $h_\phi$ of the external 
scalar through eq.\ \eqref{eq:hiCC}. Note also that we have dropped the shift by $-\kappa$ from the 
tensor structure indices since we consider lightcone blocks in the regime where $n_1 = J_1-\kappa \sim 
J_1$ and $n_2 \sim J_3$, see eq.\ \eqref{eq:Jscaling}. Let us stress once again that the conformal blocks have been obtained from the differential equations alone, without any additional input. The only quantity that 
requires input from the OPE limit of conformal blocks is the normalization. In the next subsection, we 
will show that this normalization is given by
\begin{align}
\mathcal{N}_{J_1(J_2,\kappa)J_3} &= \lim_{J_1,J_3\rightarrow \infty}\frac{\Gamma(2\Dg_\phi+2J_1)}
{\Gamma(\Dg_\phi+J_1)}\frac{\Gamma(2\Dg_\phi+2J_3)}{\Gamma(\Dg_\phi+J_3)}\lim_{J_2\rightarrow\infty} 
\frac{2\,\Gamma(3\Dg_\phi+2J_2+\kappa)}{\Gamma(2\Dg_\phi+J_1+J_2+\kappa)\Gamma(2\Dg_\phi+J_2+J_3+\kappa)} 
\nonumber \\[2mm] 
&= \pi^{-\frac{1}{2}}2^{2(J_1+J_2+J_3)+7\Dg_\phi+\kappa-1}\, 
e^{-(J_1+J_3)}J_1^{J_1+\Dg_\phi}J_2^{\frac{1}{2}-(J_1+J_3+\kappa+\Dg_\phi)} J_3^{J_3+\Dg_\phi}.
\label{norm_6ptblock}
\end{align}
The formulas \eqref{gtilde_to_besselK} and \eqref{norm_6ptblock} for the triple-twist CC lightcone 
blocks are the main result of this section and they are key to the lightcone bootstrap analysis that 
we will describe in the remaining sections of this work.  

While the focus of this paper is identical scalars, the same procedure can be used to derive the conformal blocks for the channel $((\phi_1\phi_2)\phi_3)(\phi_3(\phi_2\phi_1))$ that involves the triple-twist primaries $[\phi_1\phi_2\phi_3]_{0,J_2,\kappa}$ at the middle leg and the double-twist primaries $[\phi_1\phi_2]_{0,J_i}$, $i=1,3$ at the two sides. Keeping track of all dependencies of the differential equations on the three conformal dimensions $\Delta_1$, $\Delta_2$ and $\Delta_3$, one obtains the following generalization of~\eqref{gtilde_to_besselK} for non-identical scalars:
\begin{equation}
    \tilde{g}_{(J_1,J_2,\kappa,J_3);J_1J_3}^{(\Delta_1,\Delta_2,\Delta_3)} \sim\mathcal{N}_{J_1(J_2,\kappa)J_3}^{(\Delta_1,\Delta_2,\Delta_3)}  
    \frac{(\mathcal{U}_1+\mathcal{U}_2-1)^\kappa}{(\mathcal{U}^6)^{\Delta_1-\Delta_3}} \left(\frac{\mathcal{U}^6}{\mU_1 \mU_2}\right)^{\Delta_2+\kappa} \mathcal{K}_{\Delta_2+\Delta_3-\Delta_1+\kappa} \left(\left[\frac{J_1^2}{\mathcal{U}_2}+J_2^2 v_2+\frac{J_3^2}{\mathcal{U}_1}\right]\mathcal{U}^6 \right).
\end{equation}

\subsection{Normalization of six-point blocks}
\label{ssec:normalization_6ptblocks}

The goal of this section is to derive the formula \eqref{norm_6ptblock} for the normalization of 
CC lightcone blocks for $(h_1,h_2,h_3)=(2,3,2)\times h_\phi$ and $(n_1,n_2)=(J_1-\kappa,J_3-\kappa)$.  
Readers solely interested in the derivation of the triple-twist anomalous dimensions can skip this 
subsection. Indeed, the normalization~\eqref{norm_6ptblock} for six-point blocks at GFF twists 
follows from the results of Section~\ref{sec:param_3twist}, where we determine GFF triple-twist 
OPE coefficients exactly, and in Section~\ref{sec:cft_data_3twist}, where we determine the 
large-spin asymptotics of the GFF OPE coefficients in terms of the normalization from the 
six-point crossing equation. Nonetheless, the method used to derive the normalization in this 
section also applies to six-point blocks with twists and tensor structures not necessarily 
equal to the GFF values.

Our derivation of the normalization~\eqref{norm_6ptblock} follows the same strategy that was 
detailed in our previous work for the case of five-point lightcone blocks.  The derivation 
starts from the following observation: a conformal block corresponding to the basis element 
$(n_1,n_2)$ of monomial tensor structures is the unique solution to the comb-channel 
Casimir equations with OPE limit boundary condition~\eqref{psi_ope},
where the OPE limit $\mathrm{OPE}^{(12)3(4(56))}$ following the pattern of coincident points 
$x_1\rightarrow x_2; x_4\leftarrow x_5\leftarrow x_6$, is defined by eq.~\eqref{opelim_cr} 
in terms of OPE cross-ratios.  Starting from the OPE limit boundary condition~\eqref{psi_ope} 
for the tensor structures $(n_1,n_2)=(J_1-\kappa,J_3-\kappa)$,  we then use the Casimir equations to 
interpolate this solution with the lightcone limit $\LCL^{(16),(34)}$.   
\bigskip

The lack of analytic results on higher-point conformal blocks makes the interpolation procedure 
difficult. However, the problem is greatly simplified within two limiting regimes.
\begin{enumerate}
\item The maximal subset of lightcone limits in $\LCL^{(16),(34)}$ that still contains the OPE limit, i.e.
\begin{equation}
 \ep_{12},\ep_{456},\ep_{56}\rightarrow 0 \iff \bar z_1,\bar z_2,\bar z_3\rightarrow 0.
 \label{ltlim}
\end{equation}
This limit restricts the sum over descendants in the conformal block to those with minimal twists $(2h_1,2h_2,2h_3)$. 
However, applying these limits before taking $\ep_{16},\ep_{34}$ to zero is opposite to the order~\eqref{eq:epsilon_limits} prescribed by 
$\LCL^{(16),(34)}$, which is the order necessary for leading-twist exchange in the direct channel. Therefore, the interpolation is only valid if the limits commute at the level of single blocks. This assumption is supported by the fact that all comb-channel Casimir operators $\mathcal{D}$ satisfy this property at leading order, that is to say
\begin{equation}
\left( \lim_{\ep_{12,456,56}\rightarrow 0} \lim_{\ep_{16,34}\rightarrow 0}- \lim_{\ep_{16,34}\rightarrow 0}\lim_{\ep_{12,456,56} \rightarrow 0}\right) \vec{\ep}^{\,\,\vec{m}} \mathcal{D}^{\vec{\ep}} = 0,
\end{equation}
where $\vec{\ep}^{\,\,-\vec{m}} $ is the leading scaling of the differential operator.  
\item Three-dimensional kinematics,  where the embedding space vectors $X_i$ (respectively the spacetime vectors $x_i$) are restricted to a $\Rs^{2,3}$ subspace (respectively a $\Rs^{1,2}$ subspace). In the leading-twist limit~\eqref{ltlim}, this amounts to taking $\Upsilon_0\rightarrow 0$ in OPE cross-ratios or equivalently $\mathcal{U}_2\rightarrow 1-\mathcal{U}_1$ in the cross-ratios of eq.~\eqref{eq:curlycrossratios}. 
\end{enumerate} 
At leading order in both of these limits,  the normalized conformal blocks then take the form
\begin{equation}
g_{\oo_1\oo_2\oo_3;n_1n_2}\stackrel{\bar z,\Upsilon_0\rightarrow 0}{\sim} \prod_{i=1}^3 \bar z_i^{h_i} z_i^{\bar h_i} \Upsilon_0^\kappa (1-w_1)^{n_1}(1-w_2)^{n_2} \tilde{F}_{(h_i,J_i,\kappa;n_1,n_2)}(z,w),
\label{psi_to_Ftilde}
\end{equation}
where $\tilde{F}$ is a power series in five variables.  The coefficients of this power series are straightforward to derive from the three second-order Casimir equations spelled out in eq.~\eqref{D1_lc}--\eqref{D2_lc}, with boundary condition $\tilde{F}(0,w)=1$.  We now solve the interpolation problem in the two remaining limits of $\LCL^{(16),(34)}$ using the limiting form of the second-order Casimir equations. 
\begin{enumerate}
\item In the $\ep_{34}\rightarrow 0$ limit with $J_2^2=\orm(\ep_{34}^{-1})$,  we find
\begin{align}
\tilde{F}_{(h_i,J_i,\kappa;n_1,n_2)}&\left(z_1,1-\frac{v_2}{J_2^2},z_3,\frac{w_1}{J_2^2},\frac{w_2}{J_2^2}\right) \stackrel{J_2 \rightarrow \infty}{\sim} \mathcal{N}^{\mathrm{4pt}}_{(h_i,J_i,\kappa;n_1,n_2)} \nonumber \\
&G_1(z_1,-z_1(1-w_1)\ds_{v_2})G_3(z_3,-z_3(1-w_2)\ds_{v_2})\mathcal{K}_{\ag_2}(v_2 J_2^2),
\end{align}
where $\ag_2:= h_{1\phi}+h_{3\phi}+n_1+n_2+\kappa$ and $G_a(z_a,y_a)$ for $a=1,3$ are power series in two variables given by eq.~\eqref{Ga_of_zaya}. The normalization $\mathcal{N}^{\mathrm{4pt}}$ is obtained by comparing the limits $z_1,z_3\rightarrow 0$, $w_1,w_2\rightarrow 1$ before and after $\ep_{34}\rightarrow 0$. On the one hand, this limit reduces the system to a spinning four-point lightcone block, with $\tilde{F}(0,z_2,0,1,1)$ given by a ${}_2 F_1$ hypergeometric function in $z_2$. On the other hand, the large-spin limit at $z_1,z_3=0$ simplifies to a single modified Bessel-Clifford function multiplied by $\mathcal{N}^{\mathrm{4pt}}$ because $G_a(0,0)=1$. Comparing both expressions, we deduce that
\begin{equation}
\frac{2 \Gamma(2\bar h_2-\kappa)}{\Gamma(\bar h_2-h_{1\phi}-n_1-\kappa)\Gamma(\bar h_2-h_{3\phi}-n_2-\kappa)} \stackrel{\bar h_2\rightarrow \infty}{\sim} \mathcal{N}^{\mathrm{4pt}}_{(h_i,\bar h_i-h_i,\kappa;n_1,n_2)}.
\label{norm_34}
\end{equation}
\item In the $\ep_{16}\rightarrow 0$ limit,  we can relate the cross-ratios $w_1,w_2$ to $\mathcal{U}^6=\orm(\ep_{16})$ via the relation
\begin{equation}
w_1=1-\frac{(1-z_1)v_2}{z_1(1-v_2)}(1-\mU_1-\mU^6)+\orm(\bar z),\quad w_2 =1-\frac{v_2(1-z_3)}{(1-v_2)z_3}\mU_1+\orm(\bar z).
\end{equation}
We then take the limit $\ep_{16}\rightarrow 0$ with $J_1^2,\ep_{34}J_2^2,J_3^2=\orm(\ep_{16}^{-1})$ and $(n_1,n_3)=(J_1-\kappa-\dg n_1,J_3-\kappa-\dg n_3)$, where $\dg n_i$ are positive integers in general, while the blocks in eq.~\eqref{gtilde_to_besselK} correspond to $\dg n_{1,3}=0$. In this limit, we employ an alternative power series representation~\eqref{app:tildeGa} of $G_a(z_a,y_a)$, as well as the integral representation of the Bessel-Clifford function in \cite[eq.~(B.4)]{Kaviraj:2022wbw}, to extract the same leading asymptotics of blocks as $G^{h_1-2h_\phi,h_3-2h_\phi}_{\dg n_1 \dg n_2}$ in eq.~\eqref{Gp1p2m1m2}, times additional factors of $\Gamma(2\bar h_a)/\Gamma(\bar h_a)$, $a=1,3$, and $J_2^{-2\ag_2}$.  Combining these factors with the normalization $\mathcal{N}^{\mathrm{4pt}}$ in eq.~\eqref{norm_34} and setting $(h_1,h_2,h_3;\dg n_1,\dg n_3)=(2h_\phi,3h_\phi,2h_\phi;0,0)$, we obtain the formula for $\mathcal{N}_{J_1(J_2,\kappa)J_3}$ in eq.~\eqref{norm_6ptblock}. 
\end{enumerate}

\section{Triple-twist Operators in Generalized Free Field Theory}
\label{sec:param_3twist}

As in the four-point lightcone bootstrap, the leading terms of the six-point lightcone bootstrap 
turn out to coincide with generalized free field theory (GFF). For this reason, it is useful to understand 
the GFF triple-twist operators and their OPE coefficients first, before analyzing the lightcone limit of 
the crossing equations in more general CFTs. In the 
current section, focus on the subspace of lowest dimension operators at a fixed spin. The dimension of this subspace, that is to say the number of linearly independent operators with the same quantum numbers, grows linearly with the spin. We introduce a particular basis of triple-twist operators 
and then determine the exact form of four-point functions with three scalar and one triple-twist 
insertion. In reference to \cite{Derkachov:1995zr,Derkachov:2010zza}, we call this four-point function the ``Derkachov-Manashov wave function" of triple-twist operators.

\subsection{Minimal-twist descendants in CFT}
The first step in our endeavor to study triple-twist operators is related to studying the spaces 
$\hh_{\oo}^\tau$ of minimal-twist descendants of a primary operator $\mathcal{O}$. The latter serve as the building blocks of multi-twist 
operators. For MST primaries, the minimal-twist descendants fall into irreducible 
representations of the Lie algebra $\mathfrak{su}(1,2)$. This will allow us later to fully characterize 
multi-twist operators as lowest weights in the tensor product of lowest-weight representations of 
$\mathfrak{su}(1,2)$. To simplify the exposition, we will focus on the case where all operators 
are STTs, i.e.\ $\kappa=0$, which reduces the relevant representations to those of 
$\mathfrak{su}(1,1)\subset \mathfrak{su}(1,2)$.

\subsubsection{Some background from representation theory}

As we shall show below, the space of so-called minimal twist descendants of an STT primary, see next 
subsection for a precise definition, carries an irreducible representation of the Lie algebra 
$\mathfrak{su}(1,1)$. In order not to clutter the presentation of the CFT constructions too much, 
we use this section to first collect some representation theoretic facts. 

To begin, let us introduce the vector space $\mathcal{H}_{\bar \tau}$ that is spanned by polynomials 
$\psi$ in a single variable $\alpha$. On this space, we introduce an $\mathfrak{su}(1,1)$ action through 
the following standard differential operators  
\begin{eqnarray} \label{eq:Sp}
  {S}_+ {\psi}(\alpha)&:=& - \partial_\alpha \psi(\alpha), \\[2mm] 
  {S}_0\, \psi(\alpha)&:=& (\alpha\ds_\alpha+\frac12 \bar \tau){\psi}(\alpha), 
  \label{eq:Sn} \\[2mm] 
  {S}_-\psi(\alpha) &:=& (\alpha^2 \ds_\alpha+\bar\tau \alpha) \psi(\alpha)\ . 
   \label{eq:Sm} 
\end{eqnarray}
As is well known, these differential operators obey the conditions 
\begin{equation}
S_\pm^{\dagger} = - S_{\mp}\ ,\quad S_0^\dagger = S_0
\end{equation}
with respect to the following  $\mathfrak{su}(1,1)$-invariant scalar product 
\begin{eqnarray} \label{eq:dualSP}
\langle \psi_1,\psi_2\rangle_{\bar\tau} & = & \frac{\bar\tau-1}{\pi} \int_{\mathbb{D}} \dd^2\ag\, 
(1-\bar\ag\ag)^{\bar\tau-2} \bar\psi_1(\bar\ag)\psi_2(\ag),\quad \dd^2\ag:= \dd(\mathrm{Re}\,\ag)
\dd(\mathrm{Im}\,\ag)\ .
\end{eqnarray}
Here, $\mathbb{D}:=\{ \ag\in \Cs\,\vert\, \bar\ag\ag\leq 1\}$ is the unit disk, and the 
normalization of the measure follows from $\langle 1,1\rangle_{\bar\tau}=1$.
The subscript for scalar products and the associated norms denotes an explicit dependence on the 
weight $\bar\tau$ that characterizes the highest-weight representation of $\mathfrak{su}(1,1)$. We 
will suppress this dependence whenever it is clear from context\footnote{We have already done 
this for the generators $S_{+},S_{0},S_{-}$, which also exhibit an explicit dependence on 
$\bar\tau=2\bar h$.}. For later use, we introduce the orthogonal monomial basis $\psi_M(\alpha) 
= \ag^M$ of the space $\mathcal{H}_{\bar \tau}$ and note that the norm of these elements is given by 
\begin{equation}
\norm{\psi_M}_{\bar\tau}^2  = \frac{M!}{(\bar\tau)_M}.
\label{monom_norm}
\end{equation}
Let us now discuss a dual representation of the Lie algebra $\mathfrak{su}(1,1)$ on the space of one-variable polynomials. We refer to this representation as $\check{\mathcal{H}}
_{\bar \tau}$ and to its elements as $\check \psi$ to distinguish them from the polynomials $\psi$ in the representation $\mathcal{H}
_{\bar \tau}$. The dual 
action of $\mathfrak{su}(1,1)$ is given by the following set of differential operators $\{\check{S}_a\}_{a=0,\pm}$
\begin{eqnarray} \label{eq:cSp}
  \check{S}_+\check{\psi}(x)&:=& x\check{\psi}(x), \\[2mm] 
  \check{S}_0\, \check{\psi}(x)&:=& (x\ds_x+\frac12 \bar \tau)\check{\psi}(x), 
  \label{eq:cSn} \\[2mm] 
  \check{S}_-\check{\psi}(x) &:=& -(x \ds_x^2+\bar\tau \ds_x)\check{\psi}(x).
   \label{eq:cSm} 
\end{eqnarray}
We can think of the elements $\check{\psi}\in \check{\mathcal{H}}_{\bar \tau}$ in the dual vector space as linear functionals on $\mathcal{H}_{\bar \tau}$ using the following 
$\mathfrak{su}(1,1)$-invariant dual pairing
\begin{equation}  \label{eq:pairingsu11}
(\check{\psi}_1, \psi_2 ) = \check\psi_1(\partial_\alpha) \psi_2 (\alpha) |_{\alpha =0} \ .   
\end{equation} 
Note that the duality relation is essentially given by a Fourier transform which sends $x$ to 
a derivative $\partial_\alpha$ with respect to the dual variable. More precisely, there exists an 
an isomorphism $\mathcal{F}_{\bar \tau}$ that maps a polynomial $\check{\psi}\in \check{\mathcal{H}}_{\bar\tau}$ 
to a polynomial $\mathcal{F}_{\bar \tau}[\check{\psi}] \in \mathcal{H}_{\bar \tau}$ defined by
\begin{equation}  \label{eq:Ftrafo}
\mathcal{F}_{\bar \tau}  [\check{\psi}](\alpha) :=  
\check{\psi}(\ds_{\bar\ag}) (1-\bar\ag\ag)^{-\bar\tau}\vert_{\bar\ag=0}. 
\end{equation} 
It is easy to see that $\mathcal{F}_{\bar\tau}$  transforms monomials to monomials. Concretely,
\begin{align}
\check{\psi}_M(x) := \frac{x^M}{(\bar\tau)_M}  
&&
\text{is mapped to}
&& \mathcal{F}_{\bar \tau}[\check{\psi}_M](\alpha)  = \alpha^M=:\psi_M(\alpha)\ .  \label{eq:psi_to_checkpsi}
\end{align}
The transform $\mathcal{F}$ is designed to relate the scalar product \eqref{eq:dualSP} with pairing 
\eqref{eq:pairingsu11} in the sense that 
\begin{equation} 
\langle \mathcal{F}_{\bar \tau}[\check{\psi}_1],\psi_2\rangle_{\bar \tau} = (\check{\psi}_1,\psi_2)  \ .   
\end{equation} 
Hence, pulling back the invariant scalar product on  $\mathcal{H}_{\bar \tau}$ to an invariant product on $\check{\mathcal{H}}_{\bar \tau}$ via $\mathcal{F}_{\bar{\tau}}$ gives
\begin{equation}
 \langle \check{\psi}_1,\check{\psi}_2\rangle_{\bar \tau} := \langle \mathcal{F}_{\bar\tau}[\check{\psi}_1],\mathcal{F}_{\bar\tau}[\check{\psi}_2]\rangle_{\bar \tau} = (\check{\psi}_1, \mathcal{F}[\check{\psi}_2]) =   
 \check{\psi}_1(\ds_{\bar\ag})\check{\psi}_2(\ds_{\ag}) (1-\bar\ag\ag)^{-\bar\tau} \vert_{\bar\ag,\ag=0}\ . 
 \label{SP_sl2_alpha}
\end{equation}
This concludes our brief discussion of $\mathfrak{su}(1,1)$ and the two highest-weight 
representations $\check{\mathcal{H}}_{\bar \tau}$ and $\mathcal{H}_{\bar \tau}$.

\subsubsection{Minimal-twist descendants and their wave functions}
\def\cT{\mathcal{T}}

Consider some primary field $\oo$ with conformal weight and MST spin given by $(\Delta,J,\kappa)$. The twist $\tau$ of 
$\oo$ is the quantity $\tau \equiv \Dg - J - \kappa$. We shall denote the associated space of descendants by $\hh_\oo$. 
Note that this space carries an irreducible representation of the conformal algebra. Upon restriction to the subalgebra 
generated by dilations and rotations, the space $\hh_\oo$ decomposes into an infinity of irreducible subrepresentations
which we enumerate by some integer $i = 0,1,\dots$. Within each subrepresentation, the descendant fields possess 
well-defined conformal weight and spin. Hence we can also associate a twist $\tau_i$ to each of these subrepresentations. 
It is not difficult to see that the twists of descendant fields are bounded from below by the twist $\tau$ of the primary, 
i.e.\ $\tau = \tau_0 \leq \tau_i$. The subspace of states for which the twist $\tau_i$ assumes the minimal possible 
value $\tau_i=\tau$ is denoted by $\hh_\oo^\tau$. We shall refer to elements $\hh_\oo^\tau$ as minimal-twist states 
and to the associated fields as \textit{minimal-twist fields}. By definition, all primary fields are minimal-twist in 
this sense. Note that this concept of minimal-twist fields is purely kinematical and it should not be confused with the 
dynamical concept of lowest-twist primary. We also stress that the notion of minimal-twist fields makes no reference 
to general free fields, as opposed to the notions of double- and triple-twist fields we are going to study below. 
Our goal in this subsection is to analyze the space $\hh_\oo^\tau$ of minimal-twist states, at least for primary fields 
$\oo$ with STT spin, i.e.\ for which $\kappa=0$. We shall address the more general case in the fourth subsection below. 

In order to do so, we shall describe our STT field $\oo = \oo(X,Z)$ as a homogeneous function in embedding space.
Here, $X$ and $Z$ denote two orthogonal, lightlike vectors in embedding space $\mathbb{R}^{2,d}$, i.e.\ $X^2 = 0 = 
Z^2$ and $X\cdot Z=0$. For a field $\oo$ of twist $\tau$, we demand the degree of homogeneity under simultaneous rescalings 
of $X$ and $Z$ to be given by given by $-\tau$, 
\begin{equation} 
\oo(\varpi X,\varpi Z) = \varpi^{-\tau} \oo(X,Z)\  \quad \textrm{ for } \quad \varpi \in \mathbb{R}^*\ .  
\end{equation} 
Such a field is primary of conformal weight $\Delta$ and spin $J$ if and only if it satisfies the following condition 
\begin{equation}
\oo(\la^{-1} X+\ag Z, \la Z) = \la^{\bar\tau} \oo(X,Z),
\label{stt_prim}
\end{equation}
where $\bar\tau:=\Dg+J$. This property has an interpretation based on the subgroup of conformal transformations that 
preserve the null space $\mathrm{Span}(P,Z)\subset \Rs^{2,d}$, namely
\begin{equation} \label{eq:su11}
\mathfrak{su}(1,1)=\langle \cT_- = X\ds_Z, \cT_0 = \frac{1}{2}(Z\ds_Z-X\ds_X),\cT_+ = Z\ds_X\rangle. 
\end{equation}
The infinitesimal version of the eq.~\eqref{stt_prim} that characterized the primary field is given by the 
two conditions $\cT_- \oo=0$ and $\cT_0 \oo = \bar h\oo$, where the eigenvalue $\bar h$ is given by $\bar h := 
\bar \tau/2= \tau+J$.  Since the operator $\cT_-$ lowers $\bar h$ by one unit, the first conditions demand $\oo$ 
to possess lowest weight. The corresponding lowest-weight representation is then spanned by the operators $\cT_+^M 
\oo$ with $M=0,1,\dots,\infty$. To make contact with representation theory, we follow the approach of 
\cite{Derkachov:1995zr} and express states of the minimal-twist Hilbert space as
\begin{equation} \label{eq:checkpsipar}
\oo_{\check{\psi}}(X,Z) := \check{\psi}(\ds_{\bar\ag}) \oo(X-\bar\ag Z,Z)\vert_{\bar\ag=0}, 
\end{equation} 
where $\check{\psi}(x)$ is a power series in $x$, often called the ``coefficient function". With this definition, the action of the generators $\cT_a, 
a=0,\pm,$ defined in eq.~\eqref{eq:su11} can be written as
\begin{equation} 
\cT_a \oo_{\check{\psi}}(X,Z) = \oo_{\check{S}_a\check{\psi}}(X,Z)
\end{equation} 
where the operators $\check{S}_a, a=0,\pm,$ were defined in eq.\ \eqref{eq:cSp}. So, in more mathematical 
terms we say that the prescription \eqref{eq:checkpsipar} defines as isomorphism between the space
$\check{\mathcal{H}}_{\bar\tau}$ and the space of minimal-twist fields. This isomorphism intertwines 
the actions of $\mathfrak{su}(1,1)$ on the two spaces. 
\smallskip 

Under the isomorphism between the highest-weight representation $\check{\mathcal{H}}_{\bar \tau}$ 
and the space $\hh_\oo^\tau$ of minimal-twist states, the scalar product is sent to the two-point function 
of the associated operators. More precisely one finds that 
\begin{align}
\langle \check{\psi}_1,\check{\psi}_2\rangle_\tau =\check{\psi}_1(\ds_{\bar\ag})\check{\psi}_2(\ds_{\ag}) 
\langle \oo(X_1^\star-\bar\ag Z_1^\star,Z_1^\star)\oo(X_2^\star+\ag Z_2^\star,Z_2^\star)\rangle
\end{align}
where the two-point function on the right-hand side is to be computed in the gauge in which the embedding 
space coordinates of the two fields satisfy the additional conditions $X_1^\star\cdot X_2^\star=Z_1^\star\cdot Z_2^\star
= 1$ as well as $X_i^\star\cdot Z_j^\star=0$. 
\smallskip

The transform $\mathcal{F}$ we defined in eq.~\eqref{eq:psi_to_checkpsi} has a simple interpretation in terms of 
correlation functions in the conformal field theory\footnote{We thank Petr Kravchuk for introducing this parameterization to us.}. In fact, one can check that 
\begin{equation}
 \psi(\ag)= \mathcal{F}_{\bar \tau}(\check{\psi}) = 
 \frac{\langle \oo(X_1,Z_1) \oo_{\check{\psi}}(X_2,Z_2)\rangle}
 {\langle \oo(X_1,Z_1)\oo(X_2,Z_2)\rangle} \quad  \textrm{ with } 
 \quad \ag:= \frac{X_1\cdot Z_2}{X_1\cdot X_2}.
\label{desc_to_hatpsi}
\end{equation}
Thereby, we have now assembled all the background that is necessary to work with minimal-twist descendants 
and in particular to construct multi-twist primaries.

\subsection{Triple-twist operators and their wave functions}
\label{ssec:triple_twist_wavefunction}

Let us now turn our attention to the triple-twist primaries that can be constructed from the external scalar 
$\phi$. We shall first apply the content of the previous subsection to construct double-twist primaries, which will then be useful to construct a ``double-twist basis'' for triple-twist primaries. In the absence of transverse spin, where the symmetry algebra reduces to $\su(1,1)$, this basis has already appeared in several areas of the CFT literature, see e.g.\ \cite{Derkachov:1995zr,Braun:1999te,Belitsky:1999ru,Derkachov:2010zza}.

\subsubsection{Double-twist primaries}

Whereas the notion of minimal-twist fields and the associated structure we discussed in the previous 
subsubsection was entirely general, we shall focus on generalized free fields (GFF) from now on. Our 
first goal is to construct double-twist primaries. 

\paragraph{Parameterization.} Let $\oo_1$ and $\oo_2$ to be two primaries of spin $\ell_1$ and $\ell_2$, 
respectively. Their twists will be denoted by $\tau_a = \Delta_a - \ell_a$. Double-twist operators $\oo$ 
of weight $\Delta$ and spin $J$ as well as their descendants can always be expressed as linear combinations 
of products of derivatives of $\oo_1$ and $\oo_2$, 
\begin{align}
(Z\ds_X)^M[\oo_1\oo_2]_{0,J}(X,Z) &= \check{\psi}_{L,M}(\ds_{\bar \ag_1},
\ds_{\bar \ag_2}):\mathrel{\oo_1(X-\bar \ag_1 Z,Z)\oo_2(X-\bar \ag_2 Z,Z)}:
\vert_{\bar \ag_a=0}, \label{phiphi_to_psi} \\[2mm]
&= [\oo_{1}\oo_{2}]_{\check{\psi}_{L,M}}(X,Z) . 
\end{align}
Here, $L$ is given by $L:=J-\ell_1-\ell_2$ and the so-called ``coefficient function" $\check{\psi}_{L,M}$ is a homogeneous polynomial of degree 
$L+M$ in its two arguments. The derivative $\ds_{\bar \ag_a}$ acts as the operator 
$Z_a\ds_{X_a}$ on the field $\oo_a$ and it increases both the spin and the conformal dimension by one.  
Similarly, the action of $(Z\ds_X)^M$ on the primary $[\oo_1\oo_2]_{0,J}$ does not change the twist, 
so the corresponding descendants remain at leading twist. Finally, the normal ordered product 
$:\mathrel{(-)}:$ that ensures finiteness of eq.~\eqref{phiphi_to_psi} is equivalent to the 
identity-subtracted GFF OPE:
\begin{equation}
:\mathrel{\oo_1(X_1)\oo_2(X_2)}: \,:= \oo_1(X_1)\oo_2(X_2)-\langle \oo_1(X_1)\oo_2(X_2) \rangle \mathbf{1}.
\label{subtract_id}
\end{equation}
In the leading-twist sector, the twist of the operator $\oo$ is given by $\tau = \Dg-J=\tau_1+\tau_2$.  
Since the action of the generators \eqref{eq:su11} on the operators defined in eq.~\eqref{phiphi_to_psi} 
reduces to the action of the generators $\check{S}_{1a}(x_1,\ds_{x_1})+\check{S}_{2a}(x_2,\ds_{x_2})$ on 
the polynomials $\check{\psi}(x_1,x_2)$, we deduce that the set of $\check{\psi}_{L,M}$ forms a basis of 
states in the tensor product of two $\mathfrak{su}(1,1)$ lowest-weight representations, each given by the 
minimal-twist Hilbert spaces of $\oo_{1,2}$ and their descendants. Moreover, the condition that $[\oo_1
\oo_2]_{0,J}$ be a primary translates to $\check{\psi}_{L,0}$ being lowest-weight vectors of weight 
$\bar h=h_1+h_2+J$. 

In close analogy to eq.\ \eqref{desc_to_hatpsi}, we can express the dual polynomials $\psi_{L,0}$ in 
terms of the double-twist three-point function as 
\begin{equation}
\psi_{L,0}\left(\frac{X_1\cdot Z_a}{X_{1a}}, \frac{X_2\cdot Z_a}{X_{2a}}\right)
= 
\frac{\langle \oo_1(X_1,Z_1)\oo_2(X_2,Z_2) [\oo_1\oo_2]_{0,J}(X_a,Z_a)\rangle}
{\langle \oo_1(X_1,Z_1)\oo_1(X_a,Z_a)\rangle \langle \oo_2(X_2,Z_2)\oo_2(X_a,Z_a)\rangle } \ , 
\label{psihat_def}
\end{equation}
where $\psi_{L,0}$ is again a homogeneous polynomial of degree $L$ in its two arguments.  This generic 
form is obtained by inserting eq.~\eqref{phiphi_to_psi} into the three-point function above, setting 
$(X,Z)=(X_a,Z_a)$.  The action of the generators \eqref{eq:su11} on the eq.~\eqref{psihat_def} reduces 
to the action of the generators $S_{1a}(\ag_1,\ds_{\ag_1})+S_{2a}(\ag_2,\ds_{\ag_2})$ on $\psi(\ag_1,
\ag_2)$. Now, the lowest-weight condition reads 
\begin{equation}\label{eq:STTcondition}
\psi_{L,0}(\la \ag_1+\ag_0,\la \ag_2+\ag_0)=\la^L\psi_{L,0}(\ag_1,\ag_2) 
\end{equation}
which can be easily solved through 
\begin{equation} \label{eq:psiL}
\psi_{L,0}(\ag_1,\ag_2) = C_L^{(\bar\tau_1,\bar\tau_2)} (\ag_1-\ag_2)^L = C_L^{(\bar\tau_1,\bar\tau_2)} \ag_{12}^L 
\end{equation} 
The corresponding lowest-weight vector $\check{\psi}_{L,0}$ is obtained by applying the inverse 
of relation~\eqref{eq:psi_to_checkpsi} to each monomial in $\ag_1,\ag_2$, 
\begin{equation}
\check{\psi}_{L,0}^{(\bar\tau_1,\bar\tau_2)}(\ds_{\ag_1},\ds_{\ag_2}) =\frac{C_L^{(\bar\tau_1,\bar\tau_2)}}{1+\dg_{\oo_1,\oo_2}} 
\sum_{k=0}^L \frac{(-L)_k}{k!} \frac{\ds_{\ag_1} ^k}{(\bar\tau_1)_k} 
\frac{\ds_{\ag_2}^{L-k}}{(\bar\tau_2)_{L-k}}.
\label{dt_checkpsi}
\end{equation}
Here we have re-introduced the explicit dependence of $\check{\psi}_{L,0}$ on the lowest weights 
$\bar\tau_i$ via the superscript. In case the two fields $\oo_1$ and $\oo_2$ are identical we can 
implement permutation symmetry by demanding additionally that the polynomials $\psi_{L,0},
\check{\psi}_{L,0}$ are symmetric with respect to an exchange of its two arguments. It is easy 
to check from the above expressions that this permutation of arguments acts as multiplication 
by $(-1)^L$ on the latter. This means that only double-twist operators with even $L$ define 
a primary field in generalized free field theory. 

\paragraph{Normalization.} The constant prefactor $C_L^{(\bar\tau_1,\bar\tau_2)}$ that appeared in 
our formulas \eqref{eq:psiL} and \eqref{dt_checkpsi} may be determined from the normalization of 
the double-twist two-point function, i.e.
\begin{equation}
\langle [\oo_1\oo_2]_{0,J}(X_1,Z_1)[\oo_1\oo_2]_{0,J}(X_2,Z_2)\rangle = 
\frac{1+\dg_{\oo_1,\oo_2}(-1)^L}{1+\dg_{\oo_1,\oo_2}}\frac{H_{12}^J}{X_{12}^{\tau_1+\tau_2+2 J} }.\label{2pt_2tw_norm}
\end{equation}
The second term, proportional to $(-1)^L$, comes from the extra Wick contraction between $\oo_1$ and 
$\oo_2$ in the two-point function. Starting with the case $\oo_1\neq \oo_2$, we can insert 
eq.~\eqref{phiphi_to_psi} to compute the desired two-point function from eq.~\eqref{psihat_def}. 
We thereby deduce that the normalization condition \eqref{2pt_2tw_norm} of the two-point function is equivalent to 
\begin{equation*}
(\check{\psi}_{L,0}^{(\bar\tau_1,\bar\tau_2)},\psi_{L,0}^{(\bar\tau_1,\bar\tau_2)})=\norm{\psi_{L,0}}_{\bar\tau_1,\bar\tau_2}^2 = 1
\end{equation*}
In deriving this condition, it is convenient to use variables $X_i^*,Z_i^*$ that obey the additional 
gauge conditions $X_1^* X_2^* = Z_1^* Z_2^* = 1$ and $X_a^*Z_b^*=0$, as before. We conclude that the 
coefficients $C$ in eq.~\eqref{dt_checkpsi} read
\begin{equation}  C_L^{(\bar\tau_1,\bar\tau_2)}=\frac{1+\dg_{\oo_1,\oo_2}(-1)^L}
{\sqrt{1+\dg_{\oo_1,\oo_2}}}\norm{\ag_{12}^L}_{\bar\tau_1,\bar\tau_2}^{-1}
\label{C_pref}
\end{equation} 
with the relevant norm of $\ag^L_{12}$ given by 
\begin{equation} 
\norm{\ag_{12}^L}_{\bar\tau_1,\bar\tau_2}^2 = \sum_{k=0}^L \frac{(-L)_k^2}{k!^2} 
\frac{k!}{(\bar\tau_1)_k} \frac{(L-k)!}{(\bar\tau_2)_{L-k}} = 
\frac{L! (\bar\tau_1+\bar\tau_2+L-1)_L}{(\bar\tau_1)_L(\bar\tau_2)_L}.
\label{dt_norm2}
\end{equation}
This lowest-weight norm is computed using the orthogonality of the monomial basis 
$\ag_1^{k_1}\ag_2^{k_2}$, along with eq.~\eqref{monom_norm} for the norm of monomials. To extend this 
result to identical fields $\oo_1=\oo_2$, we note that the three-point function of $\oo_1,\oo_2,$ and 
$[\oo_1\oo_2]$ in definition~\eqref{psihat_def} of $\psi_{L,0}$ will also contain one extra Wick 
contraction when $\oo_1=\oo_2$. Thus, in accordance with the normalization convention~\eqref{2pt_2tw_norm} 
for the two-point function of double-twist fields, the resulting value of the prefactor $C$ coincides with 
the value we gave in eq.\ \eqref{C_pref} for even $L$ and it vanishes otherwise.

Now that we have completely determined the functions $\psi_{L,0}$ and thereby the GFF three-point function 
that appears on the right-hand side of eq.\ \eqref{psihat_def}, we can compare with the general form of the 
spinning three-point function expressed in an extended monomial basis:
\begin{align*}
& \langle \oo_1(X_1,Z_1)\oo_2(X_2,Z_2)\oo_3(X_3,Z_3)\rangle = \Om_{J_1J_2J_3} \\[2mm]
& \times \sum_{n_{12},n_{23},n_{31}} C_{\oo_1\oo_2\oo_3}^{(n_{12},n_{23},n_{31})} \left(\frac{H_{12}X_{31}X_{23}}{J_{1,23}J_{2,31}}\right)^{n_{12}}\left(\frac{H_{23}X_{12}X_{31}}{J_{2,31}J_{3,12}}\right)^{n_{23}}\left(\frac{H_{31}X_{23}X_{12}}{J_{3,12}J_{1,23}}\right)^{n_{31}},
\end{align*}
where
\begin{equation}
    \Omega_{J_1,J_2,J_3}=
    \frac{J_{1,23}^{J_1}J_{2,31}^{J_2}J_{3,12}^{J_3}}
    {X_{12}^{\frac12\left(\bar \tau_1+\bar \tau_2- \tau_3\right)}
    X_{23}^{\frac12 \left(\bar \tau_2+\bar \tau_3-\tau_1\right)}
    X_{31}^{\frac12 \left(\bar \tau_3+\bar \tau_1-\tau_2\right)}}\ . 
\end{equation}
The comparison shows that the double-twist three-point function in eq.~\eqref{psihat_def} has only a single 
non-vanishing three-point (as opposed to two-point) tensor structure, namely 
$$ \frac{J_{3,12}}{X_{13}X_{23}}= \ag_{12}\ . $$
Consequently, we find that the OPE coefficient is non-vanishing only when the three tensor structure 
labels $n_{ij}$ take the values $(n_{12},n_{23},n_{31})=(0,\ell_2,\ell_1)$. For this special choice, the 
OPE coefficient is given by\footnote{Throughout this work, we denote the OPE coefficients of GFF with lower 
case letters in order to distinguish them from the OPE coefficients in generic CFTs which are denoted by 
upper case letters.} 
\begin{equation}
\cgff_{\oo_1\oo_2[\oo_1\oo_2]_{0,J}}^{(0,\ell_2,\ell_1)}=
\frac{1+\dg_{\oo_1,\oo_2}(-1)^{J-\ell_1-\ell_2}}{\sqrt{1+\dg_{\oo_1,\oo_2}}}
\norm{\ag_{12}^{J-\ell_1-\ell_2}}_{\bar\tau_1,\bar\tau_2}^{-1}.
\label{gff_dt_opec}
\end{equation}
In writing this result, we have expressed $L=J-\ell_1-\ell_2$ in terms of the quantum numbers that appear on the left-hand 
side. The result is valid for identical and non-identical fields and it uses the 
norm $\norm{\ag^L_{12}}$ we computed in eq.~\eqref{dt_norm2}. 

\subsubsection{Double-twist basis of triple-twist primaries}

We now come to the main topic of this section, namely the construction of triple-twist operators in GFF. 
As we explained before, triple-twist operators in GFF exhibit a spin-dependent degeneracy that leaves freedom in the choice of a basis of operators. Here we shall make some particular choice that we shall refer to as the double-twist basis. This will turn out to be convenient in our discussion of the lightcone bootstrap 
below. We shall treat the cases of non-identical and identical scalars separately. Recall that this subsection remains restricted to STT spins. 

\paragraph{Non-identical scalars.} Consider three non-identical scalars $\phi_1,\phi_2,\phi_3$. For 
simplicity, we shall assume that they possess the same conformal weight $\Delta_\phi$. Their triple-twist 
primaries can be constructed from the two-fold iteration of eq.~\eqref{phiphi_to_psi}: first for $(\oo_1,
\oo_2)=(\oo_\ell,\phi_3)$, where $\oo_\ell:=[\phi_2\phi_1]_{0,\ell}$, and then for $(\oo_1,\oo_2)=(\phi_2,
\phi_1)$. The result of this iteration is
\begin{equation}
[\phi_3\oo_\ell]_{0,J}(X,Z) = \check{\psi}_{J-\ell,0}(\ds_{\bar\ag_1}+\ds_{\bar\ag_2},\ds_{\bar\ag_3}) 
\check{\psi}_{\ell,0}(\ds_{\bar\ag_1},\ds_{\bar\ag_2}) \phi_3(X-\bar\ag_3 Z)
\phi_2(X-\bar\ag_2Z)\phi_1(X-\bar\ag_1 Z) \vert_{\bar\ag_i=0}.
\label{dt_iterated}
\end{equation}
On the other hand, the most general parameterization of a triple-twist operator at minimal twist is
\begin{equation}
[\phi_3\phi_2\phi_1]_{0,J}(X,Z) = \check{\Psi}_{J}(\ds_{\bar\ag_1},\ds_{\bar\ag_2},\ds_{\bar\ag_3}) \phi_3(X-\bar\ag_3 Z)\phi_2(X-\bar\ag_2 Z)\phi_1(X-\bar\ag_1 Z) \vert_{\bar\ag_i=0},
\label{phiphiphi_to_Psi}
\end{equation}
where the ``coefficient function" $\check{\Psi}$ is a lowest-weight vector of weight $3\Dg_\phi/2+J$ in the triple tensor product 
of lowest-weight representations $\mathcal{H}_{\bar\tau}$ spanned by minimal-twist descendants of $\phi$.  
Comparing the expressions in the previous two equations, we obtain the lowest-weight vector 
\begin{equation}\label{eq:defPsicheck}
\check{\Psi}_{\ell,J}^{(12)}(\ds_{\bar\ag_i})= \check{\psi}_{J-\ell,0}
(\ds_{\bar\ag_{1}}+\ds_{\bar\ag_{2}},\ds_{\bar\ag_{3}}) 
\check{\psi}_{\ell,0}(\ds_{\bar\ag_1},\ds_{\bar\ag_2}).
\end{equation}
Here we placed a superscript $(12)$ at the triple-twist polynomials $\check{\Psi}$ to stress they 
these are associated with a particular choice of basis that comes with the special iterative 
construction \eqref{dt_iterated} of triple-twist operators, in which we form double-twist operators 
$\oo_\ell$ of $\oo_1=\phi_1$ and $\oo_2=\phi_2$ first. We shall refer to the polynomials 
$\check{\Psi}_{\ell,J}^{(12)}$ as a \textit{double-twist basis} of triple-twist operators. Note that 
the basis vectors of the double-twist basis are enumerated by the spin $\ell$ of the intermediate 
double-twist operator.  

As in our discussion of minimal-twist and double-twist operators, we can pass from the polynomials 
$\check{\Psi}_J$ to a dual set of polynomials $\Psi_{J}$ with the help of the triple-twist four-point 
function. The general prescription is 
\begin{align}\label{eq:defPsiJ}
\Psi_{J} \left(\frac{X_1\cdot Z_b}{X_{1b}}, \frac{X_2\cdot Z_b}{X_{2b}}, \frac{X_3\cdot Z_b}
{X_{3b}}\right):=\frac{\langle\prod\limits_{i=1}^3\phi_i(X_i) [\phi_3\phi_2\phi_1]_{0,J}(X_b,Z_b)  \rangle}
{\langle\prod\limits_{i=1}^3\phi_i(X_i)\phi_i(X_b)\rangle}.
\end{align}
Homogeneity and gauge-invariance of the correlation function then imply that the dual state $\Psi_J$ is 
a translation-invariant and homogeneous polynomial of degree $J$, i.e.
\begin{equation}
\Psi_J(\la\ag_i+\ag_0) = \la^J \Psi_J(\ag_i) \Rightarrow \Psi_J(\ag_1,\ag_2,\ag_3) = \ag_{31}^J 
\,\Psi_J\left(0,\frac{\ag_{12}}{\ag_{13}},1\right).
\label{Psi_red}
\end{equation}
In Appendix~\ref{app:rep_theory}, we explicitly compute the exact form of these polynomials for our 
double-twist basis, by inserting eq.~\eqref{eq:defPsicheck} into eq.~\eqref{eq:defPsiJ}. The 
final result of this calculation is
\begin{equation}
\Psi_{\ell,J}^{(12)}(\ag_1,\ag_2,\ag_3) = \norm{g_{\ell,J}^{(12)}}^{-1} g_{\ell,J}^{(12)}(\ag_1,\ag_2,\ag_3), 
\label{dtPsi_sym}
\end{equation}
with functions $g$ of the form 
\begin{equation}
 g_{\ell,J}^{(12)}(\ag_1,\ag_2,\ag_3) = \alpha_{21}^{\ell}\alpha_{31}^{J-\ell} 
 \hypg{\ell-J}{\Dg_\phi+\ell}{2\Dg_\phi+2\ell}\left(\frac{\ag_{12}}{\ag_{13}}\right).
  \label{dt_wf}
\end{equation}
In complete analogy with our discussion of double-twist operators, the normalization of the polynomial 
$\Psi_{\ell,J}^{(12)}$ which we displayed in eq.~\eqref{dtPsi_sym} can be read off from the normalization
of the two-point functions of triple-twist operators. The relevant norm of $g_{\ell,J}^{(12)}$ can be 
given explicitly, 
\begin{equation}
\norm{g_{\ell,J}^{(12)}}^2=\norm{\ag_{12}^\ell}_{\Dg_\phi,\Dg_\phi}^2
\norm{\ag_{a3}^{J-\ell}}_{2\Dg_\phi+2\ell,\Dg_\phi}^2. 
\label{dt_wf_norm2}
\end{equation}
The norm that appears on the right-hand side can be found in eq.~\eqref{dt_norm2}. Before we conclude 
our discussion of non-identical scalars we note that any permutation $\si \in S_3$ of the three scalar fields
gives rise to a double-twist basis by the same construction. The double-twist basis that is associated with the 
triple-twist operators $[\phi_{\si(3)}[\phi_{\si(2)}\phi_{\si(1)}]_\ell]$ is described by the polynomials 
$$\Psi_{\ell,J}^{(\si(1)\si(2))}(\ag_i):= \Psi_{\ell,J}^{(12)}(\ag_{\si(i)})\ . $$

\paragraph{Identical scalars.} In the case of three identical scalars $\phi_{1,2,3}=\phi$, the permutation 
of scalar fields yields an equivalent basis. This is reflected by the extra Wick contractions in the four-point 
function~\eqref{eq:defPsiJ} that act as a symmetrizer on the wave functions $\Psi_J$, such that
\begin{equation}
\Psi_{\ell,J}(\ag_i) = \frac{1}{\sqrt{6}}\sum_{\si\in S_3} \Psi_{\ell,J}^{(\si(1)\si(2))}(\ag_1,\ag_2,\ag_3).
\end{equation}
Similarly, the extra Wick contractions in the two-point function of $[\phi\,[\phi\phi]_{0,\ell}]_{0,J}$ lead 
to the expression
\begin{align}
\langle [\phi\,[\phi\phi]_{0,k}]_{0,J}(X_1^\star,Z_1^\star)[\phi\,[\phi\phi]_{0,\ell}]_{0,J}(X_2^\star,Z_2^\star)\rangle 
= \frac{1}{6}\sum_{\si\in S_3} \langle \Psi^{(12)}_{k,J},\Psi^{(\si(1)\si(2))}_{\ell,J}\rangle.
\end{align}
Finally, imposing permutation symmetry of the triple-twist field and a normalization consistent with the 
two-point function, the dual polynomial in the double-twist basis take the form 
\begin{equation}
\check{\Psi}_{\ell,J}(\ds_{\ag_i}) = \frac{1}{6\sqrt{6}}\sum_{\si\in S_3} \check{\psi}_{J-\ell,0}
(\ds_{\bar\ag_{\si(1)}}+\ds_{\bar\ag_{\si(2)}},\ds_{\bar\ag_{\si(3)}}) \check{\psi}_{\ell,0}
(\ds_{\bar\ag_{\si(1)}},\ds_{\bar\ag_{\si(2)}})\,.
\end{equation}
In summary, considering three identical scalars imposes a complete symmetrization on wave functions. In accordance with this change, the two-point function of triple-twist operators also contains a projector onto the permutation-symmetric subspace, 
very much analogous to our discussion of double-twist operators for identical scalars, see in particular
eq.~\eqref{2pt_2tw_norm}. This projection reduces the dimension of the space of triple-twist primaries as 
follows:
\begin{equation}
\mathrm{dim}\,[\phi_3\phi_2\phi_1]_{0,J}=\begin{cases}
    & J,\quad \phi_1\neq \phi_2\neq \phi_3, \\[2mm]
    & J/2, \quad \phi_1=\phi_2, \\[2mm]
    & \left\lfloor \frac{J+2}{2}  \right\rfloor-\left\lfloor \frac{J+2}{3}\right\rfloor,\quad \phi_1=\phi_2=\phi_3\,.
\end{cases}
\end{equation}
Note that for three identical scalars and large values of the spin $J$, the triple-twist degeneracy  $\mathrm{dim}\,[\phi_3\phi_2\phi_1]_{0,J}$ behaves as $J/6 + \orm(1)$.

\subsection{Generalization to arbitrary MST spin}
\label{ssec:MST_spin_generalization}

The generalization of previous results to multi-twist operators with non-zero MST spin is tantamount to an 
extension of $\mathfrak{su}(1,1)$ to $\mathfrak{su}(1,2)$. This extension is straightforward, but leads to 
lengthier formulas due to the addition of two more raising operators that do not commute amongst themselves. 
To avoid these technicalities, we will restrict ourselves to a brief summary of results and refer to 
Appendix~\ref{app:sl3} for the full derivations.  

\paragraph{Minimal-twist descendants.} In embedding space, MST primaries $\oo(X,Z,W)$ are described in terms
of three pairwise-orthogonal null vectors $X,Z,W$. While the first two vectors $X,Z \in \Rs^{2,d}$ are assumed 
to be real, components of the third vector $W\in \Cs^{d+2}$ can be complex. At fixed twist $\tau$, the condition 
to be a primary of MST spin $\kappa$ is now
\begin{equation}
\oo(\la^{-1} X+\ag Z+\bg W, \la\zeta^{-1} Z-\cg W, \zeta W) = \la^{\bar\tau-\kappa} \zeta^{\kappa-J} \oo(X,Z,W). 
\end{equation}
This is a lowest-weight condition for a representation of $\mathfrak{su}(1,2)$. The latter contains five new 
generators, in addition to the three generators of the subalgebra $\mathfrak{su}(1,1)$ we displayed in eq.\ 
\eqref{eq:su11}. The additional generators take the form 
\begin{equation} 
X\ds_W,Z\ds_W,W\ds_W-Z\ds_Z,W\ds_Z,W\ds_X\ . 
\end{equation} 
Since the Lie-algebra $\mathfrak{su}(1,2)$ includes three raising operators, its lowest-weight representations 
can be realized on a space of polynomials in three variables which we shall denote as $\alpha,\beta,\gamma$. In 
complete analogy with eq.~\eqref{eq:checkpsipar} we can parameterize minimal-twists descendants in terms of polynomials 
$\check{\psi}$, 
\begin{equation}
\oo_{\check{\psi}}(X,Z,W) = \check{\psi}(\ds_{\bar\ag},\ds_{\bar\bg},\ds_{\bar\cg}) 
\oo(X-\bar\ag Z-\bar\bg W,Z+\bar\cg W,W) \vert_{\bar\ag,\bar\bg,\bar\cg=0}.
\end{equation}
Once again, we can pass to the dual wave function $\psi$ by computing the two-point function of a primary with 
the descendent field $\oo_{\check{\psi}}$, see eq.~\eqref{eq:checkpsipar}, 
\begin{equation}
\psi\left(\frac{X_1\cdot Z_2}{X_{12}},\frac{X_1\cdot W_2}{X_{12}}, 
\frac{X_1\otimes Z_1 \cdot X_2\wedge W_2}{H_{12}}\right)
= \frac{\langle \oo(X_1,Z_1,W_1) \oo_{\check{\psi}}(X_2,Z_2,W_2) \rangle}
{\langle \oo(X_1,Z_1,W_1) \oo(X_2,Z_2,W_2)\rangle }\ . 
\end{equation} 
The duality relation between $\check{\psi}$ and $\psi$ is given more explicitly by the following simple 
prescription that generalizes eq.\ \eqref{eq:pairingsu11}, 
\begin{equation}  
\psi(\ag,\bg,\cg) = \check{\psi}(\ds_{\bar\ag},\ds_{\bar\bg},\ds_{\bar\cg})
(1-\bar\ag\ag-\bar\bg\bg)^{\kappa-\bar\tau}(1+\bar\cg\cg-(\bar\cg\bar\ag-\bar\bg)(\cg\ag-\bg))^{J-\kappa} 
\vert_{\bar\ag,\bar\bg,\bar\cg=0}.
\label{psi_to_psihat_mst}
\end{equation}
There exists an obvious pairing $(\cdot, \cdot)$ between a wave function $\check{\psi}$ and its dual $\psi$
of the form 
\begin{equation} 
(\check{\psi}_1,\psi_2) = \check{\psi}_1(\partial_\ag,\partial_\bg,\partial_\cg) \psi_2(\ag,\bg,\cg)|_{\ag,\bg,\cg=0}\ .  
\end{equation} 
As in our previous discussion of STT representations, this pairing determines an $\mathfrak{su}(1,2)$ invariant 
scalar product on lowest-weight modules and their duals through 
\begin{equation} 
\langle \check{\psi}_1,\check{\psi}_2\rangle_{\bar\tau-\kappa,J-\kappa} =(\check{\psi}_1,\psi_2)
=\langle \psi_1,\psi_2\rangle_{\bar\tau-\kappa,J-\kappa}
\end{equation} 
The norms of monomials that are relevant for the applications of this paper are computed in 
Appendix~\ref{app:sl3}, see in particular eqs.~\eqref{eq:normab} and \eqref{norm_YtoX}.

\paragraph{Spinning double-twist primaries.} For applications to triple-twist composites of 
scalar fields, we consider double-twist operators of the form $[\phi\oo_\ell]_{0,J,\kappa}$,  
where $\oo_\ell$ is an STT primary of STT spin $\ell$. The double-twist operator can be 
expressed in terms of a tensor product of representations of  $\su(1,2)$ as follows:
\begin{align}
& [\phi\oo_\ell]_{0,J,\kappa}(X,Z,W)  = \label{defPsicheck_mst} \\[2mm] & \quad
= \check{\psi}_{J-\ell,\kappa,0}(\ds_{\bar\ag_a},\ds_{\bar\bg_a},\ds_{\bar\cg_a},
\ds_{\bar \ag_3},\ds_{\bar\bg_3}) :\mathrel{\phi(X-\bar\ag_3 Z-\bar\bg_3 W) 
\oo_\ell(X-\bar\ag_a Z-\bar\bg_a W,Z+\bar\cg_a W)}: \vert_{\bar\ag_i,\bar\bg_i,\bar\cg_i=0}
\nonumber \end{align}
The dual wave functions  $\psi_{J-\ell,\kappa,0}(\ag_a,\bg_a,\cg_a,\ag_3,\bg_3)$ can be 
computed from the double-twist three-point functions as, see also eq.~\eqref{psihat_def}:
\begin{align*}
\psi_{J-\ell,\kappa,0}\left(\textstyle{\frac{X_a\cdot Z_b}{X_{ab}},\frac{X_a\cdot W_b}{X_{ab}},
\frac{X_a\otimes Z_a \cdot X_b\wedge W_b}{H_{ab}},\frac{X_3\cdot Z_b}{X_{3b}}, 
\frac{X_3\cdot W_b}{X_{3b}}}\right) = \frac{\langle\oo(X_a,Z_a)\phi(X_3)[\phi\oo_\ell]_{0,J,\kappa}
(X_b,Z_b,W_b)\rangle}{\langle\oo_\ell(X_a,Z_a) \oo_\ell(X_b,Z_b) \rangle
\langle\phi(X_1)\phi(X_b)\rangle } .
\end{align*}
The lowest-weight condition (which follows from homogeneity and gauge-invariance of the 
three-point function) is now
\begin{equation}\label{eq:conditionMST}
\psi_{L,\kappa,0}(\la \ag_1+\ag_0,\zeta \bg_1+\bg_0+\cg_0\ag_1,\la \ag_2+\ag_0,
\zeta \bg_2+\bg_0+\cg_0\ag_2,\la^{-1}\zeta\cg_2+\cg_0) = \la^{L}\zeta^{\kappa} 
\psi_{L,\kappa,0} (\ag_1,\bg_1,\ag_2,\bg_2,\cg_2).
\end{equation}
This condition fixes solutions up to a multiplicative constant, see eq.~\eqref{eq:psiL} for 
the related result in the case of STT representations, 
\begin{equation}
\psi_{L,\kappa,0}(\ag_a,\bg_a,\cg_a,\ag_3,\bg_3) =\psi_{L,\kappa,0}(0,0,0,1,1) \, 
\ag_{a3}^{J-\ell} (\bg_{a3}-\cg_a\ag_{a3})^\kappa.
\label{defDTellkappa}
\end{equation}
The multiplicative constant is determined by requiring that $\psi_{L,\kappa,0}$ be of unit 
norm, just as in our previous discussion. Thereby we find  
$$ \psi_{L,\kappa,0}(0,0,0,1,1) =\norm{ \ag_{12}^{J-\ell} (\bg_{12}-\cg_2\ag_{12})^\kappa}^{-1}\ . 
$$ 
The square of the norm on the right-hand side can be expressed as a finite sum of the quantities
we defined in eq.~\eqref{dt_norm2}. The explicit expression is given by eq.~\eqref{dtnorm2_mst_finsum} 
for $(L,\bar\tau_a,\bar\tau_3)=(J-\ell,2\Dg_\phi+2\ell,\Dg_\phi)$. In the case $\oo_\ell=[\phi\phi]_{0,\ell}$ 
and in the large-spin limit $J>\ell\gg 1$, it is easy to check that the terms labeled by $q>0$ in 
eq.~\eqref{dtnorm2_mst_finsum} are subleading and of relative order $\ell^{-q}$, such that
\begin{equation}
\psi_{J-\ell,\kappa,0}(0,0,0,1,1) = \kappa!^{-1} \norm{\ag_{a3}^{J-
\ell}}_{2\Dg_\phi+2\ell,\Dg_\phi+\kappa}^{-1}\left(1+\orm(\ell^{-1})\right).
\label{dtnorm2_mst_ls}
\end{equation}
Hence, in this limiting regime we can compute the prefactor of the wave function \eqref{defDTellkappa}
by computing the norm \eqref{dt_norm2} for $\alpha_{a3}^{J-\ell}$ instead of $\alpha_{12}^L$. 

\paragraph{Double-twist basis of triple-twist operators.} Let us now finally construct the double-twist 
basis for triple-twist operators with arbitrary MST spin, the most general we can obtain from three 
external scalars. Following our reasoning in the discussion of STT spins, we can define MST triple-twist 
wave functions in terms of the corresponding triple-twist four-point function as
\begin{equation}
\Psi_{J,\kappa} 
\left(\frac{X_i\cdot Z_b}{X_{ib}},\frac{X_i\cdot W_b}{X_{ib}}\right) = 
\frac{\langle\prod_{i=1}^3 \phi(X_i) [\phi\phi\phi]_{0,J,\kappa}(X_b,Z_b,W_b) \rangle}
{\prod_{i=1}^3 \langle \phi(X_i)\phi(X_b)\rangle} .
\label{eq:defPsiJ_mst}
\end{equation}
As it stands, the formula holds for any triple-twist operator. If we want the wave function $\Psi_{J,\kappa}
(\ag_i,\bg_i)$ to describe a primary, we have to impose that it is a highest-weight vector in the tensor product 
of three $\kappa=0$ representations of $\su(1,2)$. This condition is equivalent to the homogeneity and 
gauge-invariance of the four-point function in eq.~\eqref{eq:defPsiJ_mst}, just as in our previous 
discussion, see eq.~\eqref{Psi_red}, 
\begin{equation}
\Psi_{J,\kappa}(\la \ag_i+\ag_0;\zeta\bg_i+\bg_0+\cg_0\ag_i)=\la^J\zeta^\kappa\Psi_{J,\kappa}(\ag_i;\bg_i).
\label{hw_PsiJkappa}
\end{equation}
The dependence in $\bg_i$ is easy to solve and takes the form 
\begin{align}
\Psi_{J,\kappa}(\ag_1,\bg_1,\ag_2,\bg_2,\ag_3,\bg_3) &= \ag_{21}^{J-\kappa} \left(\ag_{21}
\bg_{31}-\ag_{31}\bg_{21}\right)^\kappa\Psi_{J,\kappa}\left(0,0,1,0,\frac{\ag_{31}}{\ag_{21}},1\right).
\label{Psi_red_mst} 
\end{align}
Using the properties \eqref{hw_PsiJkappa} of the wave function $\Psi$, we can absorb the factor $\ag_{21}^J$ 
into $\Psi$ to obtain 
\begin{align} \label{eq:Psiababstar}
\Psi_{J,\kappa}(\ag_i;\bg_i) &= \omega^\kappa \, \Psi_{J,\kappa}(\ag_i; \bg_i^\star) \qquad \text{with} \ \  \  
\om(\ag_i;\bg_i) := \frac{\left(\ag_{13}\bg_{12}-\ag_{12}\bg_{13}\right)^\kappa}{\ag_{12}^\kappa}
\end{align}
and $(\bg_i^\star) = (\bg_1^\star,\bg_2^\star, \bg_3^\star) := (0,0,1)$. 
For the double-twist basis $[\phi\oo_\ell]_{0,J,\kappa}$ with $\oo_\ell:=[\phi\phi]_{0,\ell}$ of triple-twist 
operators, the wave functions $\Psi_{\ell,J,\kappa}(\ag_i;\bg_i)$ are obtained by inserting eqs.~\eqref{defPsicheck_mst}
and \eqref{Psi_red_mst} into eq.~\eqref{eq:defPsiJ_mst}. The calculation is performed in Appendix~\ref{app:sl3}, with 
the result
\begin{equation}
\Psi_{\ell,J,\kappa}(\ag_i;\bg_i)=\frac{1}{\sqrt{6}}\norm{g_{\ell,J,\kappa}^{(12)}}^{-1} \sum_{\si\in S_3} g_{\ell,J,\kappa}^{(12)}(\ag_{\si(i)};\bg_{\si(i)}),
\label{dtPsi_sym_mst}
\end{equation}
where 
\begin{equation}
g_{\ell,J,\kappa}^{(12)}(\ag_i;\bg_i) = \frac{\left(\ag_{12}\bg_{13}-\ag_{13}\bg_{12}\right)^\kappa}{\ag_{12}^\kappa} 
g_{\ell,J}^{(12)}(\ag_i).
\label{dt_wf_mst}
\end{equation}
This formula expresses the wave function for the double-twist basis of MST triple-twist primaries for three identical 
scalars through the function $g_{\ell,J}^{(12)}$ we constructed in order to obtain STT triple-twist operators, see  
eq.~\eqref{dt_wf}. The norm that appears as a prefactor in eq.~\eqref{dtPsi_sym_mst} is now given by
\begin{equation}
\norm{g_{\ell,J,\kappa}^{(12)}}^2 = \norm{\ag_{12}^\ell}_{\Dg_\phi,\Dg_\phi}^2
\norm{\ag_{a3}^{J-\ell}(\bg_{a3}-\cg_a\ag_{a3})}^2_{2\Dg_\phi+2\ell,\ell,\Dg_\phi}. 
\label{eq:norm_g12_kappa}
\end{equation}
Explicit formulas for the first and second factor on the right-hand side can be found in eqs.~\eqref{dt_norm2} 
and~\eqref{dtnorm2_mst_finsum}, respectively. Note that the non-normalized wave function at $\kappa>0$ reduces to 
its STT counterpart, i.e.
\begin{equation} \label{eq:gred}
g_{\ell,J,\kappa}^{(12)}(\ag_1,0,\ag_2,0,\ag_3,1) \equiv g_{\ell,J}^{(12)}(\ag_i;\bg_i^\star)
=g_{\ell,J}^{(12)}(\ag_i)
\end{equation} 
However, the norm of the MST wave function does not reduce to the norm of the STT 
wave function in general. Nonetheless, at leading order in the large-spin limit $J>\ell\gg 1$, the 
identity~\eqref{dtnorm2_mst_ls} implies the simple shift relation 
\begin{equation}
\norm{g_{\ell,J,\kappa}^{(12)}}^2=\norm{g_{\ell-\kappa,J-\kappa}^{(12)}}^2\Bigl
\vert_{\Dg_\phi\rightarrow\Dg_\phi+\kappa}\left(1+\orm(\ell^{-1})\right).
\label{norm2_shiftrel}
\end{equation}
Since we are only going to analyze a few leading terms of the crossing symmetry equation in the lightcone limit, 
this behavior of the normalization will be sufficient for what comes below.

\section{Triple-twist OPE Coefficients in Generalized Free Field Theory}\label{sec:GFFOPECoefficients}
In the previous subsection, we computed the four-point functions of the GFF triple-twist operators 
$[\phi [\phi\phi]_{0,\ell}]_{0,J,\kappa}$ in the double-twist basis, see  eq.~\eqref{dtPsi_sym} for 
$\kappa=0$ and eq.~\eqref{dtPsi_sym_mst} for $\kappa>0$. In this section, our goal is to expand the 
result in four-point blocks and to determine the associated OPE coefficients of GFF. The latter 
are given in eq.\ \eqref{opec_PiWPi}. In the first subsection, we shall review some relevant background 
on triple-twist conformal blocks in the lightcone limit. We shall see that these are equivalent to 
the functions $g^{(12)}$ used before when we constructed the triple-twist wave functions. This 
observation will then allow us, in the second subsection, to compute the OPE coefficients directly 
from eq.~\eqref{dtPsi_sym_mst}. In the final subsection, we evaluate the triple-twist OPE 
coefficients explicitly in the large-spin limit.

\subsection{Conformal blocks at GFF twists}\label{sec:blocksGFFtwist}

The four-point functions with three scalar fields $\phi$ and one 
triple-twist insertion of the form $[\phi\phi\phi]_{0,J,\kappa}$ can be directly expanded in lightcone 
blocks. Indeed, since in GFF all scalar fields must be Wick-contracted with one of the constituents 
of the triple-twist operator, the correlation function does not exhibit any dependence in the 
distances $X_{12},X_{23},X_{13}$. The absence of these singularities implies that, among all the 
operators that appear in the OPE of any two of the scalar fields, only the leading double-twist 
families contribute to the triple-twist four-point function. It is exactly these terms that are 
captured by the lightcone blocks.

Four-point blocks for correlation functions of three scalar fields and one field in an MST 
representation depend on three cross-ratios which we shall denote by $z,\bar z$ and $w$. The 
cross-ratios $z, \bar z$ are constructed from the insertion points $X_i$ of the four fields 
as usual. The third cross-ratio $w$ involves the embedding space coordinate $Z$ that  
is associated with the spin of the fourth field. Since the lightcone limit is obtained by sending 
$\bar z$ to zero, the desired lightcone blocks have a non-trivial dependence on $z$ and 
$w$ only. Before taking the lightcone limit, the blocks depend on three quantum numbers: the weight $\Delta_1$ and spin $J_1$ of the intermediate field in the OPE of the scalar 
$\phi$ with itself, and the tensor structure label $n$ at the non-trivial vertex. In the 
lightcone limit, $\Delta_1= 2\Delta_\phi+J_1$ and it therefore suffices to label lightcone 
blocks by two quantum numbers $J_1$ and $n$. Before we state the result, we 
recall that the triple-twist fields $[\phi\phi\phi]_{0,J,\kappa}$ we insert at the fourth 
point are labeled by $J,\kappa$. We will only need the lightcone blocks for the tensor 
structure $n=J_1-\kappa$. These blocks are given by 
\begin{equation} 
g_{J_1,n=J_1-\kappa}(z,w) = \left(\frac{z}{1-z}\right)^{\Dg_\phi} (1-w)^{-\kappa}
\left(\frac{z(1-w)}{1-z}\right)^{J_1} \hypg{\Dg_\phi+J_1}{J_1-J}{2\Dg_\phi+2J_1}
\left(\frac{z(w-1)}{1-z}\right)\ . \label{eq:4ptlightconeblock}
\end{equation} 
This formula for the four-point lightcone blocks is derived in Appendix~\ref{app:form_blocks_GFF}. 
Our analysis there starts from the conformal block decomposition of the scalar six-point 
function in OPE cross-ratios, 
\begin{equation}\label{eq:sixpointblockdecomp}
\langle \phi(X_1)\dots\phi(X_6)\rangle = \sum_{\oo_1,\oo_2,\oo_3;n_1,n_2} P_{\oo_1\oo_2\oo_3}^{(n_1n_2)}
\frac{g_{\oo_1\oo_2\oo_3;n_1n_2}(\bar z_1,z_1,\bar z_2,z_2,\Upsilon_0,\bar z_3,z_3,w_1,w_2)}
{(X_{12}X_{34}X_{56})^{\Dg_\phi} (\bar z_2 z_2)^{\Dg_\phi/2}}.
\end{equation}
The blocks of the scalar six-point functions that appear on the right-hand side depend on nine 
cross-ratios. First, by taking the OPE limit in the fields that are inserted at $X_5$ and $X_6$, we 
remove the two cross-ratios $z_3$ and $\bar z_3$. Next, the second OPE limit 
with the scalar field inserted at $x_4$ further removes the three cross-ratios $z_2,\bar z_2,\Upsilon_0$ and factorizes the monomial tensor structure $(1-w_2)^{n_2}$. It is this second OPE that produces the triple-twist 
field we are after. Finally, we take the lightcone limit $\bar z_1\rightarrow 0$, whose only effect is a factor of $\bar z_1^{\Dg_\phi}$ to account for double-twist exchange in the first leg. Hence, after this sequence of limits, the blocks depend only on two cross-ratios $z_1$ and $w_1$, while the dependence on all other cross-ratios is reduced to power laws. The original six-point 
blocks are labeled by nine quantum numbers. We fix these to the values they take in 
our triple-twist four-point function:
\begin{align} \label{eq:GFFtwists}
(h_1,\bar h_1) & = (\Dg_\phi,\Dg_\phi+J_1),\quad \quad (h_2,\bar h_2)=\left(\frac{3}{2}\Dg_\phi,
\frac{3}{2}\Dg_\phi+J_2+\kappa \right) \\[2mm] 
(h_3,\bar h_3) & = (\Dg_\phi,\Dg_\phi+J_3),\quad \quad (n_1,n_2)=(J_1-\kappa,J_3-\kappa).
\end{align}
For this choice of quantum numbers of intermediate fields and tensor structures, one finds in the above limit
\begin{align}
g_{\oo_1\oo_2\oo_3;n_1n_2}\  \stackrel{\bar z_{i},z_{2,3},\Upsilon_0\rightarrow 0}{\sim} &\  (\bar z_2 z_2)^{\frac{3}{2}\Dg_\phi} z_2^{J_2+\kappa} 
\Upsilon_0^\kappa (1-w_2)^{J_3-\kappa} (\bar z_3 z_3)^{h_3} z_3^{J_3} \label{6pt_to_4pt_block_gff} 
 \bar z_1^{\Dg_\phi} g_{J_1,n_1=J_1-\kappa}(z_1,w_1), 
\end{align}
where $g_{J_1,J_1-\kappa}$ on the right-hand side is the function we introduced in 
eq.~\eqref{eq:4ptlightconeblock}. We have thereby shown that eq.~\eqref{eq:4ptlightconeblock}
gives the lightcone limit of four-point blocks of three scalars and one MST primary. 

\smallskip 

Remarkably, our result \eqref{eq:4ptlightconeblock} for the triple-twist lightcone block resembles 
the expressions for the triple-twist wave functions 
$g_{J_1,J_2,\kappa}^{(12)}(\ag_i,\bg_i)$ derived in the previous subsection, see eqs.\ 
\eqref{dt_wf_mst} and \eqref{dt_wf}. Indeed, the two formulas agree provided one makes the identification 
\begin{equation}
z_2 \equiv \ag_1-\ag_3,\quad \frac{z_1(w_1-1)}{1-z_1} \equiv \frac{\ag_1-\ag_2}{\ag_1-\ag_3}. 
\label{eq:z1_z2_w_as_alphas}
\end{equation}
The result~\eqref{6pt_to_4pt_block_gff} and the identification~\eqref{eq:z1_z2_w_as_alphas} are 
derived in Appendix~\ref{app:crossratios_prefactor_from_OPElimit}. Let us stress, in particular, that the normalization of 
the $g^{(12)}$-functions matches the normalization of blocks. Indeed, in the $(12)$ OPE limit where 
$z_1$ goes to zero, the lightcone blocks behave as
\begin{equation}
g_{J_1;J_1-\kappa}(z_1,w_1) \stackrel{z_1\rightarrow 0}{\sim} 
z_1^{\Dg_\phi} z_1^{J_1} (1-w_1)^{J_1-\kappa} + \dots.   
\end{equation}
On the other hand, according to the identification \eqref{eq:z1_z2_w_as_alphas}, the OPE limit corresponds to $\ag_1 \rightarrow \ag_2$. In this limit, the function $g^{(12)}$ behaves as 
\begin{equation} 
 g_{J_1,J_2,\kappa}^{(12)}
(\ag_1,0,\ag_2,0,\ag_3,1)\stackrel{\ag_1\rightarrow \ag_2}{\sim} \ag_{12}^{J_1}+ \dots.  
\label{opelim_alphas}
\end{equation}
Inserting the identification rules \eqref{eq:z1_z2_w_as_alphas} then confirms that the 
relation between blocks and $g^{(12)}$ is compatible with the canonical normalization~\eqref{psi_ope} of conformal 
blocks. 

In conclusion, we have shown that the conformal block decomposition for the GFF four-point 
function with one insertion of the triple-twist field $[\phi [\phi\phi]_{0,J_3}]_{0,J_2,\kappa}$ 
takes the form 
\begin{equation}
\Psi_{J_3,J_2,\kappa}(\ag_i,\bg_i) =  \sum_{J_1=0}^{J_2} \pgff_{J_1}^{J_3}(J_2,\kappa)\, 
g_{J_1,J_2,\kappa}^{(12)}(\ag_i,\bg_i). 
\label{3t_cb_dec}
\end{equation}
where
\begin{equation}
\pgff_k^\ell(J,\kappa) :=\pgff_{[\phi\phi]_{0,k}}^{(\ell-
\kappa)}\left(\ell,J,\kappa\right) :=  
\cgff_{\phi\phi [\phi\phi]_{0,k}} \cgff_{[\phi\phi]_{0,k} \phi [\phi[\phi\phi]_{0,\ell}]_{0,J,\kappa}}^{(\ell-\kappa)}
\label{opec_matrix}
\end{equation}
is the product of GFF OPE coefficients in the triple-twist four-point function, and we have 
expressed the lightcone conformal blocks through the function $g$ using the identification
\eqref{eq:z1_z2_w_as_alphas}. 

\subsection{OPE coefficients in terms of Racah coefficients}
\label{ssec:OPE-coeff-Racah}

In the previous subsection, we obtained the conformal block decomposition~\eqref{3t_cb_dec} of the 
triple-twist wave functions $\Psi_{J_1,J_2,\kappa}$. The blocks were expressed in terms of the same 
polynomial $g_{J_1,J_2,\kappa}^{(12)}(\ag_i,\bg_i)$ that we used to construct the wave-functions in
Subsections~\ref{ssec:triple_twist_wavefunction} and~\ref{ssec:MST_spin_generalization}, which establishes a close similarity between 
eq.~\eqref{dt_wf_mst} and eq.~\eqref{3t_cb_dec}. However, our two formulas for the triple-twist wave function are not quite the same. Indeed, equation~\eqref{dtPsi_sym_mst} expresses the wave function as a sum of $g$-polynomials with permuted 
arguments:
\begin{equation}
\Psi_{J_3,J_2,\kappa} = \frac{1}{\sqrt{6}}\norm{g_{J_3,J_2,\kappa}^{(12)}}^{-1}
\sum_{\si \in S_3}g^{(\si(1)\si(2))}_{J_3,J_2,\kappa} \quad \textrm{ with } \quad 
 g^{(\si(1)\si(2))}(\ag_i,\bg_i) := g^{(12)}(\ag_{\si(i)},\bg_{\si(i)}).
\label{def_Wmatrices}
\end{equation}
Moreover, the derivation in Section~\ref{sec:param_3twist} provided a concrete formula for the normalizing prefactor in this equation (namely eq.~\eqref{eq:norm_g12_kappa}).  In contrast to this, the expansion
\eqref{3t_cb_dec} of the wave function into blocks does not involve a summation over permutations and contains OPE 
coefficients that we have not yet computed. 

We now close this gap by bringing eq.~\eqref{def_Wmatrices}
into the form of the expansion~\eqref{3t_cb_dec}. This is achieved via an expansion of the polynomials $g_{J_3,J_2,
\kappa}^{(\si(1)\si(2))}$ into a linear combination of polynomials $g_{J_1,J_2,\kappa}^{(12)}$, where $J_1$ 
denotes the spin of the double-twist field exchanged in the s-channel of the four-point function.  

The coefficients that appear in this expansion can be expressed in terms of the orthogonal matrices $W^\sigma$ defined by the change of basis
\begin{equation}
\Psi^{(\si(1)\si(2))}_{\ell,J,\kappa} = \sum_{k=0}^{J} W_{k\ell}^{\si} 
\Psi^{(12)}_{k,J,\kappa}  \quad \textrm{ for } \quad \si \in S_3.
\label{Wracah_def}
\end{equation}
Here, $\Psi^{(ij)}_{\ell,J,\kappa}$ denotes the normalized $g$-function which we introduced in eq.~\eqref{dtPsi_sym_mst}. The specific, $i,j$-independent normalization is given in eq.~\eqref{eq:norm_g12_kappa}. Since  $W^\sigma$ describes a transformation between orthonormal bases, we can express its components as scalar products
\begin{equation}
W_{k\ell}^{\si}(J,\kappa) = \langle \Psi^{(12)}_{k,J,\kappa},
\Psi^{(\si(1)\si(2))}_{\ell,J,\kappa}\rangle \quad \textrm{ for } \quad  \si \in S_3.
\label{Wracah_defalt}
\end{equation}
The matrices $W^\sigma$ furnish an orthogonal representation of the permutation group $S_3$ on $\Rs^{J}$. Hence,
\begin{align}
W^{\si \si'}_{km}= \sum_{\ell=0}^{J_2} W^\si_{k\ell} W^{\si'}_{\ell m}
&&
\text{and}
&&
(W^{\si})^{-1}_{k\ell} = W_{\ell k}^\si.
\label{grouprels_S3}
\end{align}
As $S_3$ is generated by the transpositions $(12)$ and $(13)$, the group of matrices $\{W^\sigma\}_{\sigma \in S_3}$ is generated by $W^{(12)}$ and $W^{(13)}$. In particular,
\begin{align}
W^{(23)} = W^{(12)} W^{(13)} W^{(12)}, && W^{(123)}= W^{(12)} W^{(13)} && \text{and} && W^{(132)} = W^{(13)} W^{(12)}. 
\end{align}
Moreover, note that the matrix  $W^{(12)}$ is diagonal: 
\begin{equation}
\sum_{k=0}^{J} W^{(12)}_{k\ell} \Psi_{k,J,\kappa}^{(12)} = \Psi_{\ell,J,\kappa}^{(21)} = (-1)^{\ell} 
\Psi_{\ell,J,\kappa}^{(12)}\Longrightarrow W^{(12)}_{k\ell} = (-1)^{\ell} \dg_{k\ell}.
\end{equation}
Thus, we can compute all matrix elements $W^\sigma_{k \ell}$ for any choice of $\sigma$ from the 
matrix elements of $W^{(13)}$. We shall often omit the superscript $(13)$ when referring to 
this special generating $W$-matrix, i.e.  
\begin{align}
    W_{k\ell} := W_{k\ell}^{(13)}.
\end{align} 
The matrix elements $W_{k\ell}(J_2,\kappa)$ are 
Racah coefficients for 
$\mathfrak{su}(1,1)$ when $\kappa=0$ and for $\mathfrak{su}(1,2)$ in case $\kappa>0$. The Racah 
coefficients of $\mathfrak{su}(1,1)$ are known exactly and can be expressed in terms of $_4 F_3(1)$ 
hypergeometric functions, see e.g.~\cite[eq.~(36)]{Belitsky:1999ru}. In the large-spin limit $J_2>
J_1,J_3\gg 1$, the formula $g^{(12)}_{J_1,J_2,\kappa}(\ag_1,0,\ag_2,0,\ag_3,1)=g^{(12)}_{J_1,J_2}
(\ag_1,\ag_2,\ag_3)$ combined with the shift relation~\eqref{norm2_shiftrel} implies 
\begin{equation} 
W_{J_3J_1}(J_2,\kappa) \sim W_{J_3J_1}(J_2,0)\vert_{\Dg_\phi\rightarrow\Dg_\phi+\kappa}\ .
\end{equation} 
In the next section, we will derive a formula for the large-spin asymptotics of the Racah coefficients
from the functional form of six-point comb-channel lightcone blocks in that regime. 
\smallskip 

Coming back to the expansion \eqref{def_Wmatrices} of the wave functions, we can now insert eq.\ 
\eqref{Wracah_def} to obtain
\begin{equation} 
 \Psi_{J_3,J_2,\kappa} = \frac{1}{\sqrt{6}}\sum_{J_1=0}^{J_2}\left\{\norm{g_{J_1}^{(12)}}^{-1} 
 \sum_{\si\in S_3} W^\si_{J_1J_3}\right\} g_{J_1,J_2,\kappa}^{(12)}.
 \label{PsiJ3_to_g12J1}
\end{equation}
Note that here the summation over the permutation group only applies to the matrices $W^\sigma$. 
The sum is therefore proportional to the $S_3$ projection operator 
\begin{equation}
\mathbb{S}_{k\ell} := \frac{1}{3!} \sum_{\si \in S_3} W^\si_{k\ell}(J,\kappa) = \frac13 
\left(\Pi_{k\ell}+2(\Pi W \Pi)_{k\ell}\right), \quad  \textrm{ with } \ \ 
\Pi_{k\ell} = \frac{1+(-1)^\ell}{2} \dg_{k\ell}.
\label{projectors_J1J3}
\end{equation}
In rewriting the sum over permutations, we have used the relations of eq.~\eqref{grouprels_S3} to 
express all six terms through $W^{(12)}$ and $W^{(13)} = W$. Note that $\Pi$ is simply the projector 
for the $S_2$ subgroup generated by $W^{(12)}$. Now, the only $g^{(ij)}$-function that appears in
eq.~\eqref{PsiJ3_to_g12J1} is $g^{(12)}$, just as in the case of the conformal block 
expansion~\eqref{3t_cb_dec}. By comparing the two expansions, we therefore deduce that the GFF 
OPE coefficients of eq.~\eqref{opec_matrix} take the form 
\begin{equation}
\pgff_{J_1}^{J_3}(J_2,\kappa) =\sqrt{6}\,\norm{g_{J_1,J_2,\kappa}^{(12)}}^{-1} \mathbb{S}_{J_1J_3}(J_2,\kappa).
\label{opec_PiWPi}
\end{equation}
This concludes the derivation of the main result of this subsection. 
\smallskip 

\subsection{Large-spin crossing kernel and OPE coefficients}
\label{ssec:Crossing-kernel-OPE-coeff}

In order to prepare for our analysis of the crossing equation, we finally want to determine the triple-twist 
OPE coefficients~\eqref{opec_PiWPi} of the six-point function in the limit of large spins. This requires, in 
particular, to determine the large-spin limit of the Racah coefficients $W^{(13)}_{k\ell}(J,\kappa)$. We shall 
first do so by combining some existing results in the literature. For convenience, we shall then sketch a second 
derivation that is based on the study of six-point blocks with GFF scaling dimensions. The results on the 
large-spin limit of the triple-twist OPE coefficients are collected at the end.

\subsubsection{Hankel transform from large-spin Racah coefficients}
\label{ssec:large-spin-racah}

The goal here is to evaluate the Racah matrix elements $W_{k\ell}^{(13)}(J,\kappa)\equiv W_{k\ell}(J,\kappa)$ 
which were defined by eq.~\eqref{Wracah_def}, in the following large-spin limit
\begin{equation}
J,k,\ell\rightarrow\infty, \quad \frac{k\ell}{J} =\mathrm{finite}.  
\label{lim_Jsimell2}
\end{equation}
To describe the limiting form of the Racah matrix elements, we first use the fact that to leading order 
in the limit~\eqref{lim_Jsimell2}, the dependence of $W$ on the MST spin variable $\kappa$ can 
simply be absorbed in a shift of the conformal weight, 
\begin{equation} 
W_{k\ell}^{(\Dg_\phi)}
(J,\kappa) = W_{k\ell}^{(\Dg_\phi+\kappa)}(J,0)
\end{equation} 
This property of the Racah matrix coefficients is explained below eq.~\eqref{opec_PiWPi}. This reduces the 
study of the large-spin limit for the Racah coefficients of $\mathfrak{su}(1,2)$ to the well-studied case 
of $\mathfrak{su}(1,1)$. After defining $\nu:=\Dg-1:=\Dg_\phi+\kappa-1$, we can now use the result in 
\cite[eq.~(5.1.9)]{Derkachov:1997uh} (see also \cite[eq.~(39)]{Belitsky:1999ru}) on the asymptotics of 
Racah coefficients in the intermediate regime 
\begin{equation}
\ell=\tau J, \quad J\rightarrow \infty, \quad \tau \in [0,1), \quad k \in \Zs_+.
\end{equation}
In \cite[App.~B]{Derkachov:1997uh}, this regime is studied by direct computation of the scalar product in 
eq.~\eqref{Wracah_def}. Equivalently, in \cite[Sec.~3.5]{Belitsky:1999ru} (also in \cite[App.~A]{Braun:1999te} 
for $\Dg=2$), the asymptotics are derived from a scaling limit of the second-order recurrence relation satisfied 
by the Racah coefficients. This recurrence relation follows from the representation of the $(13)$ Casimir operator 
of $\mathfrak{su}(1,1)$ as a tridiagonal matrix in the orthonormal eigenbasis $\Psi_{k,J}^{(12)}$ of the 
corresponding $(12)$ Casimir operator, see e.g.\ \cite[eq.~(38)]{Belitsky:1999ru}. The result is a second-order 
differential equation in $\tau=k/J$ with polynomial solution 
\begin{equation}
W_{k,\tau J}^{(13)} \sim (-1)^{J}\sqrt{\frac{2\tau}{J}} \left(\tau^2(1-\tau^2)\right)^{\frac{\nu}{2}} 
\left(\frac{\Gamma_{k+2\nu+1}(2k+\nu+1)}{k!\,\Gamma_{\nu+k+1}^2} \right)^{\frac{1}{2}} P_k^{(\nu,\nu)}(1-2\tau^2),
\label{W_to_jacobi}
\end{equation}
where $P_k^{(\ag,\bg)}$ are the Jacobi polynomials. The multiplicative prefactor ensures that the functions 
$W_{k(\tau J)}^{(13)}$, when seen as polynomials in $\tau$, are unit-normalized with respect to the Jacobi 
polynomials' scalar product, which is the limiting form of the $\mathfrak{su}(1,1)$-invariant scalar product 
in the orthonormal basis $\psi_{\tau J}^{(23)}$. We can now retrieve the large-spin limit~\eqref{lim_Jsimell2} 
we are interested in by setting $\tau = \orm(k^{-1})$ and taking $k\rightarrow \infty$. In this limit, the 
leading asymptotics of the Jacobi polynomials are given by a Mehler-Heine type 
formula \cite[Thm.~8.1.1]{szeg1939orthogonal}:
\begin{equation}
P_k^{(\ag,\bg)}\left(1-\frac{z^2}{2 k^2}\right) \stackrel{k\rightarrow \infty}{\sim} \left(\frac{2k}
{z}\right)^{\ag}J_\ag(z),
\end{equation}
where $J_\nu(z)$ is the Bessel function of the first kind. Setting $\ag=\nu=\Dg-1$ and $z=2 k\tau = 2k\ell/J$, 
we obtain 
 \begin{equation}\label{Wkappa}
W_{k\ell}^{(\Dg)}(J,\kappa) = (-1)^J\sqrt{\frac{2}{J}} \sqrt{\frac{2k\ell}{J}} J_{\Dg+\kappa-1}
\left( 2\frac{k\ell}{J} \right),
\end{equation}
Note that $W_{k\ell}$ in eq.~\eqref{Wkappa} is the kernel of an integral transform named after Hankel 
that has been well explored in the theory of special functions, see e.g.\ \cite[\S~10.22(v)]{dlmf}:
\begin{equation}
\mathcal{H}_\nu[f](y) := \int_0^\infty \dd x \, \sqrt{xy} J_\nu(xy) f(x), 
\end{equation}
It is well known that the Hankel transform $\mathcal{H}_\nu$ squares to the identity, i.e.\ it satisfies 
$\mathcal{H}_\nu \circ \mathcal{H}_\nu =1$. This ensures that the limiting form of the Racah coefficients in eq.~\eqref{Wkappa} continues to define an orthogonal representation of the $S_2$ generated by the 
permutation $(13)$.

\subsubsection{Six-point crossing kernel}\label{sec:sixptcrossingkernel}

To define the action of the integral transform, we first rewrite the product of blocks and GFF OPE coefficients 
at leading order as
\begin{equation}
P_{\oo_1\oo_2\oo_3}^{(J_1,J_3)}g_{\oo_1\oo_2\oo_3;J_1J_3}(\dutchcal{u}_i,v_i,\mU_i,\mU^6) = 
\om(\dutchcal{u}_i,v_i)^{h_\phi} \frac{J_2^{2\Dg_\phi-1}}{\kappa!} \frac{32}{\Gamma_{\Dg_\phi}^3} 
\dg(J_1-J_3)  f_{J_1(J_2,\kappa)J_3}(\mathcal{U}_0,v_2;\mU_1,\mU_2),
\label{Ppsi_to_f}
\end{equation}
where the prefactor is given by $\om(\dutchcal{u}_i,v_i)=\dutchcal{u}_1^2\dutchcal{u}_2^3\dutchcal{u}_3^2 v_1 v_2^2 v_3$. 
In addition, we have introduced a new name for the cross-ratio $\mU_0:=\mU^6/(\mU_1\mU_2)$ that appears here and will 
do so frequently in the next section. The conformal blocks in GFF normalization, i.e.\ with the GFF OPE coefficients 
are finally given by  
\begin{equation}
f_{J_1(J_2,\kappa)J_3}(\mathcal{U}_0,v_2;\mU_1,\mU_2):=(J_1J_3)^{\frac{\Dg_\phi+\kappa-1}{2}}
\mU_0^{\Dg_\phi+\kappa}\mathcal{K}_{\Dg_\phi+\kappa}\left(J_1^2\mU_0\mU_2+J_2^2v_2\mU_0\mU_1\mU_2+J_3^2\mU_0\mU_1 \right).
\end{equation}
The action of the permutation $\si = (13)(64)$ on cross-ratios follows from its action on the two-point invariants, 
$\si:X_{ij}\mapsto X_{\si(i)\si(j)}$. The result was displayed in eq.~\eqref{eq:sigma}. From there, it is easy to show that 
$\si$ leaves the prefactor $\om(\dutchcal{u}_i,v_i)$ invariant and transforms the remaining cross-ratios in the argument 
of $f_{J_1(J_2,\kappa)J_3}$ as
\begin{equation}\label{eq:sigma}
\si: (\mathcal{U}_0,v_2,\mathcal{U}_1,\mathcal{U}_2)\mapsto (v_2,\mathcal{U}_0,\mathcal{U}_2,\mathcal{U}_1). 
\end{equation}
At the level of the GFF-normalized conformal blocks $f_{J_1(J_2,\kappa)J_3}(\mathcal{U}_0,v_2;\mathcal{U}_1,\mathcal{U}_2)$, 
this permutation of cross-ratios is realized by the following integral identity:
\begin{equation}
\int_0^\infty \dd J_1' \, W_{J_1 J_1'}^{(\Dg_\phi)}(J_2,\kappa) \int_0^\infty \dd J_3'\, 
W_{J_3 J_3'}^{(\Dg_\phi)}(J_2,\kappa) f_{J_1'(J_2,\kappa)J_3'}(\mathcal{U}_0,v_2;
\mU_1,\mU_2)=f_{J_1(J_2,\kappa)J_3}(v_2,\mathcal{U}_0;\mU_2,\mU_1),
\label{hankel_6pt}
\end{equation}
where $W_{k\ell}^{(\Dg)}(J,\kappa)$ are given by eq.~\eqref{Wkappa}. This integral identity is derived in 
appendix~\ref{app:crossing_kernel}. We conclude that the product of Racah coefficients coincides with the 
six-point crossing kernel for the permutation $\si=(13)(64)$.  In particular, since 
$\mathcal{H}_\nu \circ \mathcal{H}_\nu=1$, our expression for the crossing kernel provides an orthogonal 
representation of the permutation $\si$, which is a product of disjoint two-cycles that squares to the 
identity. More specifically:
\begin{equation}
 \int_0^\infty \dd \ell  \, W_{k \ell}^{(\Dg_\phi)}(J_2,\kappa) W_{\ell m}^{(\Dg_\phi)}(J_2,\kappa) = \dg(k-m).
\end{equation}
Let us note that the integral identity~\eqref{hankel_6pt} only determines $W_{kl}^{(\Dg_\phi)}(J_2,\kappa)$ 
up to a sign. As a result, the non-analytic prefactor $(-1)^{J_2}$ in our formula \eqref{Wkappa} for the large 
spin limit of the Racah coefficients is invisible to the six-point crossing property we studied here. 
 
\subsubsection{Lightcone limits and large-spin OPE coefficients}
\label{sssec:pgff_ls}
To find the limiting form of the GFF OPE coefficients in eq.~\eqref{opec_PiWPi} relevant to the lightcone bootstrap, it is useful to review how it arises from the lightcone limit of the GFF six-point function.

Subleading contributions to the lightcone limit of the correlator correspond \emph{either} to individual subleading contributions to the crossed-channel large-spin OPE coefficients, \emph{or} to families of individually leading terms whose overall asymptotics are damped by relative phases upon summation. This fact is well known from the four-point bootstrap (see e.g.~\cite[App.~B]{Alday:2015eya}), where the sign factor $(-1)^J$ relates the contributions of $t$-channel identity exchange and $u$-channel identity exchange in the $s$-channel OPE coefficients. While the latter subtlety is avoided in four-point functions of identical operators by the constraint that $J$ has to be an even integer,  there is no simple analog of this constraint in the case of six-point functions\footnote{The analogous constraint could only be expressed in an orthonormal eigenbasis of the $S_3$ projection operator in eq.~\eqref{projectors_J1J3}, of which we do not know any examples that are analytically tractable.}. 

To see the damping effect of relative phases explicitly, let us now discuss the lightcone limit of the GFF six-point function of the defining scalar field in more detail. 
The correlator consists of a sum over permutations $\si\in S_6$ of Wick contractions $\left(X_{\si(1)\si(2)} X_{\si(3)\si(4)} X_{\si(5)\si(6)}\right)^{-\Dg_\phi}$.  By dividing out the prefactor $\Om_\CC$ in eq.~\eqref{eq:Omega_definition} and applying the lightcone limit $\bar z_2\rightarrow 0$,  we observe at leading order that triple-twist operators are only exchanged in the crossed-channel decompositions of the six Wick contractions $\left(X_{\si(1)6} X_{\si(2)5} X_{\si(3)4}\right)^{-\Dg_\phi}$ with $\si\in S_3$. To obtain the OPE coefficients that reproduce these terms in the crossed channel, recall that the triple-twist two-point function is normalized by 
\begin{equation}
\langle [[\phi\phi]_{0,k} \phi]_{0,J_2,\kappa}(P_1^\star,Z_1^\star,W_1^\star) 
[\phi[\phi\phi]_{0,\ell}]_{0,J_2,\kappa}(P_2^\star,Z_2^\star,W_2^\star)\rangle =\mathbb{S}_{k\ell},
\end{equation}
where $\mathbb{S}$ is the $S_3$ projection 
operator defined in eq.~\eqref{projectors_J1J3}.
This implies that the product of OPE coefficients in the crossed-channel decomposition of the correlator  corresponding to $\tau_2=3\Dg_\phi+J_2+\kappa$ exchanges is given by
\begin{align}
p_{J_1(J_2,\kappa)J_3}=\sum_{k,\ell} \pgff_{J_1}^k \mathbb{S}_{k\ell}\, \pgff_{J_3}^\ell &= 6 \, \norm{g_{J_1,J_2,\kappa}^{(12)}}^{-1}
\norm{g_{J_3,J_2,\kappa}^{(12)}}^{-1}  \mathbb{S}_{J_1J_3}.
\label{p6pt_to_p4pt}
\end{align}
By tracking the contribution of separate Wick contractions (or by comparison with the case of three non-identical scalars), we can decompose the six-point OPE coefficients into a sum of terms
\begin{equation}
p_{J_1(J_2,\kappa)J_3}=\sum_{\si\in S_3}p_{J_1(J_2,\kappa)J_3}^\si,\quad p_{J_1(J_2,\kappa)J_3}^\si := \norm{g_{J_1,J_2,\kappa}^{(12)}}^{-1}
\norm{g_{J_3,J_2,\kappa}^{(12)}}^{-1} \, W_{J_1J_3}^\si,
\label{pgff_sigma}
\end{equation}
 such that 
\begin{equation*}
 \lim_{\bar z_2\rightarrow 0}\!\sum_{J_1,J_2,\kappa,J_3}\! \pgff_{J_1(J_2,\kappa)J_3}^\si 
 \frac{g^\CC_{(h_i,J_i,\kappa;J_1-\kappa,J_3-\kappa)}
 (\bar z_1,z_1,\bar z_2,z_2,\Upsilon_0,\bar z_3,z_3,w_1,w_2)}
 {(X_{12}X_{34}X_{56})^{\Dg_\phi} (\bar z_2 z_2)^{\Dg_\phi/2}} 
= (X_{\si(1)6}X_{\si(2)5}X_{\si(3)4})^{-\Dg_\phi}.
\end{equation*}
With this equation, the stage is set to make the damping effect manifest. 
To this end, let us now take the full lightcone limit $\LCL^{(16,34)}$. The only leading Wick contraction in this limit is $(X_{16}X_{25}X_{34})^{-\Dg_\phi}$. Thus, only the contribution $p_{J_1(J_2,\kappa)J_3}^{(1)}$ (proportional to $W^{(1)}_{J_1J_3}=\dg_{J_1J_3}$) survives in the decomposition~\eqref{pgff_sigma}.
But note that the other contributions $p_{J_1(J_2,\kappa)J_3}^{\sigma}$ only differ from $p_{J_1(J_2,\kappa)J_3}^{(1)}$ by $\orm(1)$ factors, namely components of the orthogonal matrices $W^\sigma$! Hence, at the level of individual summands, they should be just as leading as $p_{J_1(J_2,\kappa)J_3}^{(1)}$. The only explanation for their absence in the limit is therefore the cancellation of these individual contributions between themselves, which is the effect that we set out to highlight at the beginning of this section.

Let us now determine the concrete asymptotic form of the contribution that survives the damping effect. Following the analysis of Section~\ref{ssec:Casimir_singularity},
the crossed-channel sum localizes according to the large-spin limit~\eqref{eq:Jscaling}. The asymptotics of the coefficients $p_{J_1(J_2,\kappa)J_3}^{(1)}$ in this limit follow from the large-spin asymptotics of the norm:
\begin{equation}
\norm{g_{J_1,J_2,\kappa}^{(12)}}^2 \stackrel{J_1,J_2 \rightarrow \infty}{\sim} \Gamma(\Dg_\phi)^3 \, 
\kappa! \, \pi^{-\frac{1}{2}} 2^{2 J_2+4J_1+\kappa+7\Dg_\phi-4}\,e^{-2J_1} J_2^{1-2J_1-\kappa-
3\Dg_\phi}J_1^{1+2J_1-2\kappa}.
\label{ls_norm2}
\end{equation}
Note that, from the six-point OPE coefficients $p_{J_1(J_2,\kappa)J_3}^{(1)}$, we can also deduce the limiting form of the four-point OPE coefficients $p_{J_1}^\ell(J_2,\kappa)$ implied by the lightcone limit $\LCL^{(16,34)}$. 
Since relation~\eqref{p6pt_to_p4pt} reduces to $\sum_\ell p_{J_1}^\ell p_{J_3}^\ell = p_{J_1(J_2,\kappa)J_3}$, 
we deduce that $p_{J_i}^\ell$ must be proportional to the matrix elements of a square-root of the identity matrix. 
Moreover, the six-point crossing kernel $W_{J_1J_1'}^{(\Dg_\phi)}(J_2,\kappa)W_{J_3J_3'}^{(\Dg_\phi)}(J_2,\kappa)$ must map the $\CC$ OPE coefficients $p_{J_i}^\ell$ to the $\CC'$ OPE coefficients $(p')_{J'_i}^{\ell'}$ while preserving the six-point OPE coefficients $p_{J_1(J_2,\kappa)J_3}$. 
As a result, the triple-twist OPE coefficients of the crossed channel $\CC$ and the dual crossed channel $\CC'$ reduce to
\begin{equation}
p_{J_i}^\ell(J_2,\kappa) \sim \norm{g_{J_i,J_2,\kappa}^{(12)}}^{-1} \dg(J_i-\ell)\ ,\qquad (p')_{J'_i}^{\ell'}(J_2,\kappa) \sim \norm{g_{J_i',J_2,\kappa}^{(12)}}^{-1} W^{(\Dg_\phi)}_{J_i\ell}(J_2,\kappa),
\label{pgff_4pt_ls}
\end{equation}
where the large-spin limit of the norm is given by eq.~\eqref{ls_norm2} and the limiting behavior of the Racah 
coefficient can be found in eq.~\eqref{Wkappa}. 

\section{Triple-twist CFT Data from Six-Point Crossing Equation}
\label{sec:cft_data_3twist}

We are finally prepared to study triple-twist operators in general CFTs, following our exposition in 
Section \ref{sec:definitions}. Concretely, we begin in Section \ref{sec:leadingordersolution} by analyzing 
the leading direct-channel contribution that stems 
from the exchange of two identity fields. We will be able to reconstruct this term through crossed-channel 
contributions. As in the four-point lightcone bootstrap, this analysis puts strong constraints on the large-spin 
limit of the OPE coefficients, essentially implying that the large-spin limit of 
triple-twist OPE coefficients approaches their GFF counterparts. After completing and discussing the analysis
of the leading term, we turn our attention to the two subleading terms in which one identity field gets 
exchanged in the direct channel. These subleading terms contain logarithms of two cross-ratios which we 
shall interpret in terms of triple-twist anomalous dimensions. Since triple-twist operators are highly 
degenerate (infinitely degenerate in the large-spin limit), their anomalous dimensions are encoded in a 
matrix (resp.\ operator in the large-spin limit). 

We determine the relevant operator in Section \ref{sec:CrossingatNLO}, see eq.~\eqref{adim_master}. This 
formula is the main result of our work. Some special eigenfunctions 
and eigenvalues are discussed at the end of this section, leaving a more systematic study of the 
eigenvalue problem for a future publication. 

The direct-channel contributions in the crossing equation we are about to study were spelled out in 
Section~\ref{sec:leadingtwistexpansionDC}, see eq.~\eqref{eq:leading_DC}. Our goal is to match these by summing up crossed-channel contributions, 
\begin{align}
 \frac{1+ \frac{C_{\phi\phi\oo_\star}^2}
{B_{\bar h_\star}}(\mathcal{U}_0^{h_\star}+ v_2^{h_\star})\log \dutchcal{u}_2^{-1} +
\orm(\ep_{16}^{h_\star}\ep_{34}^{h_\star})}{(v_2\mathcal{U}_0\mathcal{U}_1\mathcal{U}_2)^{\Dg_\phi}} 
 = \hspace{-0.25 cm} \sum_{\substack{h_i,J_i,n_i \\ \kappa,\ell}} \hspace{-0.25 cm}\, P_{\oo_1\oo_2^\ell}^{(n_1)}
P_{\oo_3\oo_2^\ell}^{(n_2)}\, \om^{-h_\phi}g_{(h_1,J_1)(h_2,J_2,\kappa)(h_3,J_3);n_1n_2}^{(12)3(4(56))}.
\label{CE_to_subsubleading1} 
\end{align}
In comparison to the earlier expression of the direct-channel terms, we have now divided by the 
factor $\om^{h_\phi}:=(\dutchcal{u}_1^2\dutchcal{u}_2^3\dutchcal{u}_3^2 v_1v_2^2 v_3)^{h_\phi}$ and 
expressed the ratio of the leg factors $\Omega$ through cross-ratios using eq.~\eqref{eq:curlycrossratios}. Furthermore, 
we also split the OPE coefficients as 
\begin{equation}
P_{\oo_1\oo_2^\ell\oo_3}^{(n_1,n_2)}= \sum_{\ell=0}^{J_2} P_{\oo_1\oo_2^\ell}^{(n_1)}P_{\oo_3\oo_2^\ell}^{(n_2)},\qquad 
P_{\oo_i\oo_2^\ell}^{(n)}:= C_{\phi\phi\oo_i} C_{\oo_i\phi\oo_2^\ell}^{(n)}\,,
\label{P6pt_to_P4pt}
\end{equation}
where $\ell$ denotes an extra label for the degeneracy of operators $\oo_2^\ell$ with the same twist 
and spin $(h_2,J_2,\kappa)$. 

In Section~\ref{ssec:Casimir_singularity}, we established that, to the 
order we consider here and in all relevant lightcone limits $\LCL^{(16,34)}$, $\LCL^{(16)}$, or 
$\LCL^{(34)}$, the action of the Casimir and vertex differential operators on the direct-channel 
contributions implies that the crossed-channel conformal block decomposition indeed localizes to 
exchanges of the double-twist operators $(\oo_1,\oo_3)=\left([\phi\phi]_{0,J_1},[\phi\phi]_{0,J_3}
\right)$ and some triple-twist operators $\oo_2=[\phi\phi\phi]_{0,J_2,\kappa}^{\Psi}$. At the same 
time, the tensor structures are fixed to the values  $n_1=J_1-\kappa$, $n_2=J_3-\kappa$ and large 
spins. 

Note that at this stage, the degeneracy label $\Psi$ of the triple-twist operator is not yet 
determined, as it cannot be measured by any of the conformally invariant operators we considered. However, we saw in Section \ref{sec:param_3twist} that the degenerate triple-twist operators can be 
expanded in the basis of twofold double twists $[\phi[\phi\phi]_{0,\ell}]_{0,J_2,\kappa}$. We now use this basis to parameterize the operator exchanges of the middle leg in the 
crossed channel and determine the anomalous dimension matrix that lifts the degeneracy.

\subsection{Leading order solution to crossing: triple-twist OPE coefficients}
\label{sec:leadingordersolution}
Focusing on the most leading contribution $f_{\mathds{1},\phi,\mathds{1}}$ which involves two identity 
exchanges, we see that the constraint $J_1=J_3$ imposed by~\eqref{eq:equal_spins_two_identity} implies 
that the product \eqref{P6pt_to_P4pt} of OPE coefficients at leading order has to be of the form  
\begin{equation}
    P_{\oo_1\oo_2\oo_3}^{(n_1,n_2)}\sim\dg\!\left(J_1-J_3\right)\, 
    P_{\oo_1\oo_2\oo_1}^{(J_1-\kappa,J_1-\kappa)}.
\end{equation}
At leading order, the crossing equation 
therefore reduces to
\begin{align}\label{eq:CElead}
\frac{1}{(v_2 \mU^6)^{\Dg_\phi}} = \sum_{\kappa=0}^\infty \int\dd J_1\dd J_2& \dd J_3 \,\dg\!\left(J_1-J_3\right)
P_{\oo_1\oo_2\oo_1}^{(J_1-\kappa,J_1-\kappa)}\,\mathcal{N}_{J_1(J_2,\kappa)J_1} \times \\ 
&(\mU_1+\mU_2-1)^\kappa \left(\frac{\mU^6}{\mU_1\mU_2}\right)^{\Dg_\phi+\kappa} 
\mathcal{K}_{\Dg_\phi+\kappa}\left(\left[J_1^2(\mU_1^{-1}+\mU_2^{-1})+J_2^2 v_2 \right]\mU^6\right). 
\nonumber
\end{align}
Here, we inserted our formula \eqref{gtilde_to_besselK} for the CC lightcone blocks. The homogeneity of 
the direct-channel contribution on the left-hand side imposes that the OPE coefficients at large spin 
take the form
\begin{equation}\label{P2OPE}
P_{\oo_1\oo_2\oo_1}^{(J_1-\kappa,J_1-\kappa)} \mathcal{N}_{J_1(J_2,\kappa)J_1}=  
J_1^{2(\Dg_\phi+\kappa)-1} J_2^{2\Dg_\phi-1} b_\kappa,
\end{equation}
for some sequence $(b_\kappa)_{\kappa=0}^\infty$ that still needs to be determined.  We can then 
perform the double-integral over $(J_1,J_2)$ with the help of the following integral identity, 
\begin{equation}
    \int_{\Rs_+^2} \frac{\dd x \dd y}{x y} x^{\Dg_\phi+\kappa} y^{\Dg_\phi} 
    \mathcal{K}_{\Dg_\phi+\kappa}(x+y)=\frac12 \Gamma(\Dg_\phi)^2\,\Gamma(\Dg_\phi+\kappa).
\end{equation}
After multiplying both sides of the crossing equation~\eqref{eq:CElead} by $(v_2\mU^6)^{\Dg_\phi}$,
we obtain the following condition on the unknown coefficients $b_k$:
\begin{equation}
1 = (\mU_1+\mU_2)^{-\Dg_\phi} \sum_{\kappa=0}^\infty \left(1- \frac{1}{\mU_1+\mU_2}\right)^{\kappa} 
\Gamma(\Dg_\phi)^2\,\Gamma(\Dg_\phi+\kappa)\frac{b_\kappa}{8} . 
\end{equation}
Assuming $\abs{\mU_1+\mU_2}>1$, this equation can be solved by the binomial series, provided we 
choose $b_\kappa$ as 
\begin{equation} 
 b_\kappa = 8\,  \frac{(\Dg_\phi)_\kappa}{\kappa!\, \Gamma(\Dg_\phi)^2\,\Gamma(\Dg_\phi+\kappa)}\ . 
\end{equation} 
For these values of the coefficients $b_\kappa$, the OPE coefficients \eqref{P2OPE} must then take the form 
\begin{equation}
P_{\oo_1\oo_2\oo_3}^{(J_1-\kappa,J_3-\kappa)} \sim \frac{J_1^{2(\Dg_\phi+\kappa)-1}J_2^{2\Dg_\phi-1}}{\kappa!} 
\frac{8}{\Gamma(\Dg_\phi)^3} 
\frac{\dg\!\left(J_1-J_3\right)}{\mathcal{N}_{J_1(J_2,\kappa)J_1}}. 
\label{PJ1J2J3_from_lcb}
\end{equation}
This formula coincides with the asymptotics of the GFF OPE coefficients $p^{\si=(1)}_{J_1(J_2,\kappa)J_3}$ defined in eq.~\eqref{pgff_sigma} if and only if
\begin{equation}
\lim_{J_1\rightarrow\infty}\lim_{J_2\rightarrow\infty} \frac{J_1^{2(\Dg_\phi+\kappa)-1}J_2^{2\Dg_\phi-1}}{\kappa!} 
\frac{8}{\Gamma_{\Dg_\phi}^3} \frac{\norm{g_{J_1,J_2,\kappa}^{(12)}}^2}{\mathcal{N}_{J_1(J_2,\kappa)J_1}}=1.
\end{equation}
Using the explicit formulas for the denominator $\mathcal{N}$ in eq.~\eqref{norm_6ptblock} and the 
numerator $\norm{g}^2$ in eq.~\eqref{ls_norm2}, one can check that this identity is indeed satisfied. 
Repeating the derivation of Section~\ref{sssec:pgff_ls}, we can then deduce the asymptotics of the 
four-point OPE coefficients. In summary, at leading order in the lightcone limit $\LCL^{(16),(34)}$, 
the solution to crossing coincides with the GFF triple-twist OPE coefficients~\eqref{pgff_4pt_ls}, 
i.e.
\begin{equation}
P^{(n)}_{\oo_1\oo_2^\ell} = C_{\phi\phi[\phi\phi]_{0,J_1}} C^{(n)}_{[\phi\phi]_{0,J_1}\phi [\phi[\phi\phi]_{0,\ell}]_{0,J_2}} \ \sim \ \dg_{n,J_1-\kappa} \,p_{J_1}^\ell(J_2,\kappa).
\end{equation}
To retrieve the full GFF OPE coefficients~\eqref{pgff_4pt_ls} from the above asymptotics induced by the lightcone limit, we must symmetrize by the $S_3$ projector $\mathbb{S}_{k\ell}$ of eq.~\eqref{projectors_J1J3}. More generally, we conjecture that a decomposition of the form $P=\sum_\si P^\si$ with $P^\si = P^{(1)} \, W^\si$ continues to hold for general CFTs at finite spin. This would be a natural 
generalization of double-twist OPE coefficients for the four-point function, where the decomposition $P_\oo = 
P^t_\oo + (-1)^{J_\oo} P^u_\oo$ of $s$-channel OPE coefficients can be defined non-perturbatively in terms of double-discontinuities of the four-point function \cite{Caron-Huot:2017vep}.

\subsection{Crossing at NLO: triple-twist anomalous dimensions}\label{sec:CrossingatNLO}

We now analyze the subleading contributions of the direct channel, see eq.~\eqref{eq:CElead}.
Note that the two terms that arise from the exchange of a single identity 
field in the direct channel both contain a factor of $\log \dutchcal{u}_2$. The same type of logarithms 
also appears in the subleading terms of the four-point lightcone bootstrap where they were 
traced back to anomalous contributions to the conformal weights of the intermediate fields. The 
same is true for the logarithmic contributions in the direct channel of the six-point lightcone 
bootstrap. We shall first discuss this for the term of the form $\mathcal{U}_0^{h_\ast}
\log \dutchcal{u}_2^{-1}$ in the first part of the analysis before we turn to the second logarithmic 
term in the second part. 

\subsubsection{Diagonal contributions to the anomalous dimension matrix} 

In the crossed channel $\CC$, the direct-channel term proportional to $\mU_0^{h_\star} \log 
\dutchcal{u}_2^{-1}$ arises from the anomalous dimensions of the double-twist operators $\oo_\ell=
[\phi\phi]_{0,\ell}$ that constitute the triple-twist operators $\oo_2^\ell = [\phi\oo_\ell]_{0,J_2,
\kappa}$.  Indeed, recall that the large-spin expansion of double-twist anomalous dimensions from the 
four-point bootstrap takes the form 
\begin{equation}
h_{\oo_\ell} = \Dg_\phi +\frac{\gamma_0}{2}\ell^{-2h_\star}+\orm(\ell^{-2h_\star-1})+\orm(\ell^{-2h_>})\,, 
\end{equation}
where the constant $\gamma_0$ is given by 
\begin{equation} \label{eq:gamma0}
 \gamma_0 := -2 \frac{C_{\phi\phi\oo_\star}^2}{\mathrm{B}_{\bar h_\star}} 
\frac{\Gamma_{\Dg_\phi}^2}{\Gamma_{\Dg_\phi-h_\star}^2}\,,
\end{equation} 
and $2h_>$ is the twist of the next leading-twist operator in the OPE of $\phi$ with itself.  If 
we then insert this expansion of $h_{\oo_l}$ into the middle-leg twist $h_2=h_\phi+h_{\oo_\ell}+\dots$ 
of the crossed channel $\CC$, we reproduce the direct channel by virtue of the relation   
\begin{equation}
\frac{C_{\phi\phi\oo_\star}^2}{B_{\bar h_\star}} \frac{U^{h_\star}}{(U VX_1X_2)^{\Dg_\phi}} = 
\sum_{\kappa=0}^\infty \int \dd^3J\frac{8}{\Gamma_{\Dg_\phi}^3} J_2^{2\Dg_\phi-1}
\left(\frac{\gamma_0}{2 J_1^{2h_\star}} \right)   \dg(J_1-J_3)   f_{J_1J_2J_3}(U,V;X_1,X_2),
\label{adim_4ptlike}
\end{equation}
where we used the shorthand $(U,V,X_1,X_2)=(\mU_0,v_2,\mU_1,\mU_2)$. With the help of eq.~\eqref{Ppsi_to_f}, 
we can translate the integrand on the right-hand side of eq.~\eqref{adim_4ptlike} into a matrix product 
\begin{equation} I = \int \dd k\, \dd \ell\, p_{J_1}^k \gamma_{k\ell}\, p_{J_3}^\ell\ . 
\label{eq:matrixproduct}
\end{equation}  
of the anomalous dimension operator $\gamma$ with the GFF OPE coefficients $p_\ell^{J_i}$ of eq.~\eqref{pgff_4pt_ls}. We thereby deduce the 
following contributions to the triple-twist anomalous dimension operator:
\begin{align} \label{eq:gamma1} 
&\DC^{(1)}:\mU_0^{h_\star} \log\dutchcal{u}_2^{-1} \quad \leftrightarrow 
\quad \CC: \gamma_{k\ell}^{(1)} =\gamma_0 \frac{\dg(k-\ell)}{\ell^{2h_\star}} 
\end{align}
We have placed the superscript $(1)$ in order to show that this is only part of the anomalous 
dimension operator, namely the part that is needed in order to reproduce the direct-channel 
term $\mU_0^{h_\star} \log\dutchcal{u}_2^{-1}$. It still remains to analyze the second term.

\subsubsection{Non-diagonal contributions to the anomalous dimension matrix} 

Let us now address the term that involves the product $v_2^{h_\star} \log \dutchcal{u}_2^{-1}$. It is not difficult to 
understand how to recover this term from the crossed channel, based on the following crucial 
observation: the two logarithmic terms in the direct channel are actually mapped to each other 
by a simple exchange of the external insertion points via the permutation $\sigma = (13)(46)$. 
Indeed, we have already computed the action of $\sigma$ on our cross-ratios in eq.~\eqref{eq:sigma}.
From these formulas, it is evident that $\sigma$ maps $\mU_0$ to $v_2$ while leaving the product 
$v_2\mU_0\mU_1\mU_2$ invariant. Hence, the second logarithmic term in the direct channel possesses
a similar interpretation in terms of the crossed-channel expansion as the first one, as long as 
we expand in terms of the crossed channel $\CC' = \CC \circ \sigma$. Hence, if we apply the 
identity \eqref{adim_4ptlike} with the choice $(U,V,X_1,X_2)=(v_2, \mU_0,\mU_2,\mU_1)$ we 
can rewrite the second term in terms of lightcone blocks for the crossed channel $\CC'$. 
Before we can read off the contribution to the anomalous dimension matrix $\gamma$, we have 
to transform back from the lightcone blocks in the $\CC'$ channel to those in the crossed 
channel $\CC$. But this is exactly what we discussed in Section \ref{sec:sixptcrossingkernel}, see in particular 
the integral relation~\eqref{hankel_6pt}. In conclusion, we deduce that 
\begin{eqnarray}
& & \hspace*{-15mm} \frac{C_{\phi\phi\oo_\star}^2}{B_{\bar h_\star}} 
\frac{v_2^{h_\star}}{(\mU_0 v_2\mU_1\mU_2)^{\Dg_\phi}} 
= \label{adim_hankel} \\[2mm]  
& = & \sum_{\kappa=0}^\infty \int \dd^3 J  \frac{8}{\Gamma_{\Dg_\phi}^3} J_2^{2\Dg_\phi-1} 
\left\{ \int \dd J_1' W^{(\Dg_\phi)}_{J_1'J_1} \frac{\gamma_0}{2 J_1'^{2h_\star}} 
W^{(\Delta_\phi)} _{J_1'J_3} \right\}  
\dg(J_1-J_3)   f_{J_1J_2J_3}(\mU_0,v_2;\mU_1,\mU_2).\nonumber 
\end{eqnarray}
After translating the integrand into the matrix product of the form \eqref{eq:matrixproduct} by 
means of eq.~\eqref{Ppsi_to_f}, we obtain the following expression for the second contribution 
to the anomalous dimension matrix 
\begin{align} \label{eq:gamma2}
&\DC:v_2^{h_\star} \log\dutchcal{u}_2^{-1} \quad \leftrightarrow 
\quad \CC: \gamma_{k\ell}^{(2)} =\int \dd k' \dd \ell '\,W^{(\Dg_\phi)}_{kk'} \gamma_{k'\ell'}^{(1)} W^{(\Dg_\phi)}_{\ell' \ell}
\end{align}
\paragraph{The full anomalous dimension matrix.} In summary, we have found that crossing symmetry 
entails the following anomalous dimension matrix for triple-twist operators 
$[\phi[\phi\phi]_{0,\ell}]_{0,J,\kappa}$:
\begin{equation}
\gamma_{k\ell}(J,\kappa)= \gamma_0\frac{\dg(k-\ell)}{\ell^{2h_\star}} + \gamma_0\int_0^\infty 
\dd m \frac{W_{k m}^{(\Dg_\phi)}(J,\kappa)W_{m\ell}^{(\Dg_\phi)}(J,\kappa)}{m^{2h_\star}} + \orm(J^{-2h_\star}).
\label{adim_master}
\end{equation}
The second term can be evaluated explicitly using \cite[eq.~(10.22.58)]{dlmf}: 
\begin{align}
& \int_0^\infty \dd  m \frac{W_{k m}^{(\Dg_\phi)}(J,\kappa) W^{(\Dg_\phi)}_{m\ell}(J,\kappa)}{m^{2h_\star}} 
= \label{eq:nondiagonal_adim_integral}\\[2mm]  
& \hspace*{10mm} =\frac{2}{\sqrt{k\ell}} \frac{\Gamma_{\Dg_\phi+\kappa-h_\star}}{\Gamma_{\Dg_\phi+\kappa}\Gamma_{h_\star}} \, 
\left(\frac{k\ell}{k^2 +\ell^2}\right)^{\Dg_\phi+\kappa} \left(\frac{k^2+\ell^2}{J^2}\right)^{h_\star} 
\hypg{\frac{\Dg_\phi+\kappa-h_\star}{2}}{\frac{\Dg_\phi+\kappa+1-h_\star}{2}}{\Dg_\phi+\kappa}
\left(\frac{4 k^2\ell^2}{(k^2+\ell^2)^2}\right). \nonumber 
\end{align}
Equations \eqref{adim_master} and \eqref{eq:nondiagonal_adim_integral} constitute the main result of 
this work. In the large-spin limit $1 \ll k,\ell, J$ with $k^2,\ell^2=\orm(J)$, they provide the leading correction to the anomalous dimension of triple-twist operators in the double-twist 
basis. 

This result generalizes straightforwardly to the triple-twist primaries $[\phi_1\phi_2\phi_3]$ exchanged in the six-point function $\langle \phi_1\phi_2\phi_3\phi_3\phi_2\phi_1\rangle$ of non-identical scalars. In the direct channel, the four-point functions $\langle \phi_1\phi_1\phi_2\phi_2\rangle$ and $\langle \phi_2\phi_2\phi_3\phi_3\rangle$ exchange $\oo_{\star 1}$ and $\oo_{\star 2}$ at leading twist respectively, while in the crossed channel, the large-spin Racah coefficients for the $(13)$ permutation is modified to $W_{k\ell}^{(\Dg_2)}$. As a result, the first term in eq.~\eqref{adim_master} is modified via 
\begin{equation*}
   h_{\star}\rightarrow h_{\star 1}, \quad  \gamma_0\rightarrow -\frac{2}{\mathrm{B}_{\bar h_{\star 1}}}C_{\phi_1\phi_1\oo_{\star 1}} C_{\phi_2 \phi_2\oo_{\star 1}}\frac{\Gamma_{\Dg_1}}{\Gamma_{\Dg_1-h_{\star 1}}}\frac{\Gamma_{\Dg_2}}{\Gamma_{\Dg_2-h_{\star 1}}} ,
\end{equation*}
while the second term is modified via
\begin{equation*}
   h_{\star}\rightarrow h_{\star 2}, \quad  \gamma_0\rightarrow -\frac{2}{\mathrm{B}_{\bar h_{\star 2}}}C_{\phi_2\phi_2\oo_{\star 2}} C_{\phi_3 \phi_3\oo_{\star 2}}\frac{\Gamma_{\Dg_2}}{\Gamma_{\Dg_2-h_{\star 2}}}\frac{\Gamma_{\Dg_3}}{\Gamma_{\Dg_3-h_{\star 2}}},\quad W^{(\Dg_\phi)}(J,\kappa)\rightarrow W^{(\Dg_2)}(J,\kappa).
\end{equation*}
In section~\ref{sssec:twist4}, we examine a specific case where $\phi_1=\phi_2=\phi$ and $\phi_3=\phi^2$ in perturbation theory.

\subsection{Double-twist limit and first correction}
\label{sec:dtlim_and_corr}
We do not intend to give a complete analysis of the eigenvalue equation $\gamma \cdot 
\Psi_\la = \la \,\Psi_\la$ for the anomalous dimensions matrix $\gamma$, but would like 
to study the eigenvalues near a limiting regime that recovers the behavior of triple-twist 
operators expected from the four-point bootstrap. To this end, we reintroduce the independent 
scaling of the degeneracy labels $k,\ell$ (which behave like $J_1,J_3$) and the triple-twist 
spin $J$ (which behaves like $J_2$) that followed from the six-point lightcone bootstrap, 
i.e.\ we shall assume that $k^2,\ell^2=\orm(\ep_{16}^{-1})$, while $J^2 = \orm(\ep_{16}^{-1}
\ep_{34}^{-1})$. For this scaling behavior, we observe that the two terms $\gamma^{(1)}$ 
and $\gamma^{(2)}$ that contribute to $\gamma$ scale differently. Considering the term 
$\gamma^{(1)}$ first, it is obvious from its definition \eqref{eq:gamma1} that 
\begin{equation}
\gamma_{k\ell}^{(1)}(J,\kappa) := \frac{\dg(k-\ell)}{k^{2h_\star}} = \orm(\ell^{-2h_\star-1}).
\end{equation}
The non-diagonal term $\gamma^{(2)}$, on the other hand, can be seen from eq.\ 
\eqref{eq:nondiagonal_adim_integral} above to scale as 
\begin{equation}
\gamma_{k\ell}^{(2)}(J,\kappa):= \int_0^\infty \dd m \frac{W_{k m}^{(\Dg_\phi)}(J,\kappa)
W_{m\ell}^{(\Dg_\phi)}(J,\kappa)}{m^{2h_\star}}=\orm(\ell^{2h_\star-1}J^{-2h_\star}).
\end{equation}
Thus, the non-diagonal term is suppressed relative to the diagonal term in the regime 
$\ell^2\ll J$, i.e.\ for $ \ep_{16} \gg \ep_{34}$. In this case, the anomalous dimension 
matrix is diagonalized by wave functions $\Psi_{\ell_\la,J,\kappa}^{(12)}$ with eigenvalue 
$\gamma_0 \, \ell_\la^{-2h_\star}$ at leading order. The wave functions correspond to the 
triple-twist operators $[\phi[\phi\phi]_{0,\ell_\la}]_{0,J,\kappa}$,  while the eigenvalues 
coincide with the leading anomalous dimensions of the constituent double-twist operators 
$[\phi\phi]_{0,\ell_\la}$.  We therefore call this regime the ``double-twist" limit. Let us stress that the limiting behavior that arises from the leading diagonal term 
$\gamma^{(1)}$ is expected from a bootstrap analysis of the four-point function 
$\langle \oo_\ell(X_a,Z_a)\phi(X_3)\phi(X_4)\oo_\ell(X_c,Z_c)\rangle$, where 
$\oo_\ell=[\phi\phi]_{0,\ell}$. 

To compute the first correction to the triple-twist anomalous dimensions away from the 
double-twist limit, we consider the perturbation theory of the free Hamiltonian 
$\gamma_{k\ell}^{(1)}$ by the small perturbation $\gamma_{k\ell}^{(2)}$.  In this 
case, the first correction to the eigenvalue is given by
\begin{align}
\gamma([\phi[\phi\phi]_{0,\ell_\la}]_{0,J,\kappa})& \stackrel{\ell_\la^2 \ll J}{\sim} 
\gamma_0 \left( \ell_\la^{-2h_\star} +\ell_\la  \,\gamma_{\ell_\la\ell_\la}^{(2)}(J,\kappa)
+\dots\right) \nonumber  \\[2mm] 
&=\frac{\gamma_0}{\ell_\la^{2h_\star}} +  2\gamma_0\left(\frac{\ell_\la}{J}\right)^{2h_\star} 
\frac{\Gamma(\Dg_\phi+\kappa-h_\star)}{\Gamma(\Dg_\phi+\kappa+h_\star-1)} 
\frac{\Gamma(2h_\star-1)}{\Gamma(h_\star)^2} +\dots 
\label{eq:adim_diagonal_limit}
\end{align}
In the second line,  we evaluated the matrix elements $\gamma_{\ell\ell}^{(2)}(J,\kappa)$ 
directly using \cite[eq.~(10.22.57)]{dlmf}.  

\paragraph{The dual double-twist limit.} We can also study a dual limit $\ell^2 \gg J$ (i.e.\ $\ep_{16}\ll \ep_{34}$), where $\gamma_{k\ell}^{(2)}$ is the leading contribution and $\gamma_{k\ell}^{(1)}$ is the small perturbation.  To diagonalize the operator in this limit,  recall that the two matrices $\gamma_{k\ell}^{(1)}$ and $\gamma_{k\ell}^{(2)}$ are conjugate to one another via the Racah matrix $W_{k\ell}^{(\Dg_\phi)}$,  which realizes the change of basis $\Psi_{\ell,J,\kappa}^{(12)}\mapsto \Psi_{\ell,J,\kappa}^{(32)}$.  As a result, the anomalous dimension operator is diagonalized by $\Psi^{(32)}_{\ell'_\la,J,\kappa}$ at leading order, with eigenvalues $\gamma_0\, \ell_\la'^{-2h_\star}$ and first correction of order $(J/\ell'_\la)^{-2h_\star}$.  This means that the dual limit again localizes to the double-twist basis,  and one can move from one regime to the other via the identification $\ell'_\la = J/\ell_\la$.

\section{Comparison to \texorpdfstring{$\epsilon$-Expansion in Scalar Field Theories}{epsilon-Expansion in Scalar Field Theories}}\label{sec:checks_and_appl}
 
In this section, we show that our results for the anomalous dimension~\eqref{adim_master} of triple-twist 
operators in the large-spin limit of eq.~\eqref{lim_Jsimell2} agree with the anomalous dimension operator 
of certain fixed points that can be studied perturbatively through the $\epsilon$-expansion. More specifically, 
we will compute the triple-twist anomalous dimensions of $\phi^{3}$ theory in $d=6-2\ep$ dimensions at one loop (order $\ep$) and of $\phi^4$ theory in $d= 4-2\ep$ dimensions at two loops (order $\ep^2$) at finite spin following the approach of Derkachov 
and Manashov, which first appeared in \cite{Derkachov:1995zr} and was later reviewed in \cite{Derkachov:2010zza}. 
The first subsection is devoted to the fixed point of scalar $\phi^3$ theory. We shall first review the 
results of Derkachov and Manashov on the anomalous dimensions of triple-twist operators with vanishing MST 
spin $\kappa=0$. These are stated in eqs.~\eqref{gammaDM} and \eqref{Hdef} and then evaluated in the large-spin limit, see eq.~\eqref{eq:Gammalargespin} where we find perfect agreement with our formula~\eqref{adim_master}. 
We also extend the one-loop analysis to triple-twist operators with non-zero MST spin $\kappa$. Once again the 
large-spin limit is shown to agree with the results of the lightcone bootstrap. In the second subsection, we 
then turn our attention to the Wilson-Fisher fixed points of scalar $\phi^4$ theory. We shall show that the 
bootstrap result~\eqref{adim_master} makes a prediction for the two-loop anomalous dimensions in this case. 
The latter are then computed perturbatively through the evaluation of the relevant Feynman diagrams. The 
calculation mimics the approach Manashov and Derkachov used for $\phi^3$ theory. The resulting formulas for 
the two-loop anomalous dimensions of triple-twist operators at the Wilson-Fisher fixed point are new. In 
the large-spin limit, they are shown to reproduce eq.~\eqref{adim_master}. Finally. we shall conclude with 
a few comments on the relation of our bootstrap analysis with a recent study of one-loop twist-four 
anomalous dimensions in \cite{Henriksson:2023cnh}.

\begin{figure}[htp]
    \centering
    \includegraphics[scale=0.13]{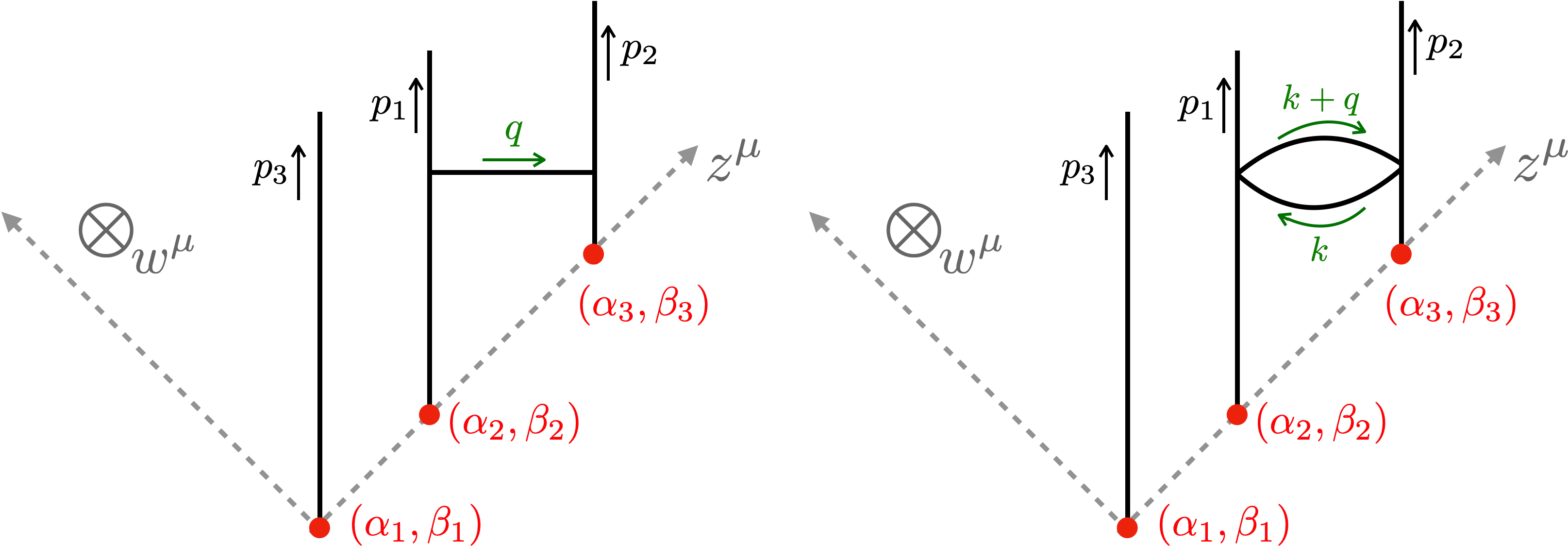}
    \caption{First Feynman diagrams in the $\epsilon$-expansion of $\phi^3$ theory (left) and $\phi^4$ theory (right) that contribute to the anomalous dimensions of triple-twist operators in the large-spin limit $1\ll k^2,\ell^2=\orm(J)$. Starting from a disconnected contribution to the six-point function, we Fourier transform three fields to $\tilde{\phi}(p_i)$ and take the normal-ordered product $\mathbb{O}(\ag_i;\bg_i)$ of the three others, inserted along the null directions spanned by the orthogonal polarization vectors $z,w$. The poles of these diagrams in the $\ep$-expansion are then renormalized by the anomalous dimensions of the triple-twist operators that appear in the lightcone OPE of the normal-ordered product  $\mathbb{O}(\ag_i;\bg_i)$.}
    \label{fig:feyndiag}
\end{figure}

\subsection{\texorpdfstring{$\phi^3$}{ϕ\^3} theory in \texorpdfstring{$6-2\epsilon$}{6-2ε} dimensions}
\label{phi3app}

In this section, we compare the bootstrap results we obtained in Section~\ref{sec:cft_data_3twist} with 
known results for $\phi^3$ theory in $d= 6-2\epsilon$ dimensions. The fixed point of a scalar field 
with interaction $g\phi^3$ was studied to leading order in the epsilon expansion in  \cite{Derkachov:1997uh,
Derkachov:2010zza}.  Following \cite[eq.~(4.11)]{Derkachov:2010zza}, the beta function of the $\phi^3$ model 
and the anomalous dimension of its fundamental scalar $\phi$ are given by
\begin{equation}
\bg(g) = -g\left(\ep+ \frac{3}{256\pi^3} g^2+\orm(g^4)\right), 
\quad \Dg_\phi(g)=2-\ep +\frac{g^2}{768\pi^3}+\orm(g^4).
\end{equation}
Therefore, at this order in the epsilon expansion, the coupling and scaling dimension at the fixed point are 
given by \cite{Gracey2015,deAlcantaraBonfim:1980pe}
\begin{equation}
g_\star^2 = -\frac{256\pi^3}{3} \ep + \orm(\ep^2),\quad \Dg_\phi(g_\star)=2-\frac{10}{9}\ep+\orm(\ep^2).
\end{equation}
These values determine the substractions we need to perform in passing from the triple-twist conformal 
dimensions to the anomalous contributions,  
\begin{equation}\label{gammaDM}
\gamma_{[\phi\phi\phi]_{0,J}}= \Dg_{[\phi\phi\phi]_{0,J}}-3\Dg_\phi(g_\star)-J = -\frac{g_\star^2}{(4\pi)^3} 
\mathbb{H}+\orm(g_\star^4) =\frac{4\ep}{3}\mathbb{H}+\orm(\ep^2),
\end{equation}
where $\mathbb{H}= \orm(\ep^0)$ denotes some operator on the space of triple-twist wave functions. In general, 
these wave functions depend on six variables $\alpha_i, \beta_i, i=1,2,3$. Derkachov and Manashov computed the 
leading contributions to the triple-twist anomalous dimensions at the fixed point only for STT triple-twist 
operators, i.e.\ for operators with $\kappa=0$. Their approach is based on the study of the correlation function 
$\langle \phi(x_1)\phi(x_2)\phi(x_3)\mathbb{O}(\ag_1,\ag_2,\ag_3;\bg_1,\bg_2,\bg_3)\rangle$, where
$\mathbb{O}=:\mathrel{\phi\phi\phi}:$ is the normal-ordered product of three fundamental fields at lightlike 
separation (see Fig.~\ref{fig:feyndiag}), which can then be expanded into a linear combination of triple-twist 
operators following the methods of Section~\ref{sec:param_3twist}. In perturbation theory, the six-point Feynman 
diagrams that contribute to this correlator exhibit collinear singularities when the three points are lightlike
separated. The corresponding $1/\ep$ pole is then renormalized by the anomalous dimensions of the triple-twist 
operators that span $\mathbb{O}$, and these anomalous dimensions are expressed in terms of an integral operator 
acting on the space of triple-twist wave functions $\Psi_{J,\kappa}$. By acting on the wave functions
$\Psi_{\ell,J,\kappa}^{(12)}$, we can then compare the resulting anomalous dimension in the double-twist basis 
with our prediction from the lightcone bootstrap at large spin.

\subsubsection{Anomalous dimensions for vanishing MST spin}

In the case $\kappa=0$ the wave functions only depend on the three variables $\alpha_i$. So, in order to state 
the results of Derkachov and Manashov for $\gamma$ we shall drop the variables $\beta_i$.
Recall from eq.~\eqref{3t_cb_dec} that the triple-twist wave functions $\Psi(\alpha_1,\alpha_2,\alpha_3)$ can 
be expanded in the basis of double-twist wave functions~\eqref{dtPsi_sym}. On these wave functions, the 
Hamiltonian $\mathbb{H}$ takes the following nice form \cite[eqs.~(5.13),(5.17a)]{Derkachov:2010zza}
\begin{equation}\label{Hdef}
 \mathbb{H}=\sum_{1\leq i<k\leq 3}  \mathbb{H}_{ik},\qquad  \mathbb{H}_{ik}= \left( \frac{1}{\Cs_{ik}^2} 
 -\frac{1}{24} \dg_{\Cs_{ik}^2,2}\right).
\end{equation}
Here, the operators $\mathbb{C}^2_{ik}$ denote (quadratic) Casimir operators of $\mathfrak{su}(1,1)$ with 
generators given by sums $S_i+S_k$ of operators that act on $\alpha_i$ and $\alpha_k$, respectively, see 
eqs.\ (\ref{eq:Sp}-\ref{eq:Sm}) for concrete expressions. The action of the resulting Casimir operators 
on the triple-twist wave functions $\Psi_{\ell,J}^{(ik)}$ is given by 
\begin{equation}\label{eq:Casik} 
\Cs_{ik}^2 \Psi_{\ell,J}^{(ik)}(\ag_1,\ag_2,\ag_3) := 
(\Dg_\phi+\ell)(\Dg_\phi+\ell-1)\Psi_{\ell,J}^{(ik)}(\ag_1,\ag_2,\ag_3),
\end{equation}
Below we shall explain that the same formulas~\eqref{Hdef} and \eqref{gammaDM} also give the anomalous 
dimensions of triple-twist primaries in MST representations, i.e.\ with $\kappa \neq 0$, provided one replaces 
the Casimir operators $\mathbb{C}^2$ of $\mathfrak{su}(1,1)$ by those of the Lie algebra $\mathfrak{su}(1,2)$. 
But before we do so, let us first compare the original result of Derkachov and Manashov for fields with 
$\kappa=0$ with the outcome of our bootstrap analysis. 
\medskip 

To compare formulas~\eqref{Hdef} and \eqref{gammaDM} with our result \eqref{adim_master} for the 
triple-twist anomalous dimensions in the large-spin limit $J\gg 1$, $\ell^2=\orm(J)$, we study the 
matrix elements of \eqref{gammaDM} in the $\psi_\ell^{(12)}$ basis, that is 
\begin{equation}\label{eq:Gamma}
\Gamma_{k\ell}(J,\kappa=0) = \frac{4}{3} \ep\, \langle \psi_k^{(12)},\mathbb{H}\psi_\ell^{(12)}\rangle. 
\end{equation}
We want to evaluate the asymptotics of these matrix elements in the limit~\eqref{lim_Jsimell2} with 
$k^2\sim\ell^2\sim J$ and show that these coincide with eq.\ \eqref{adim_master}. To begin with, let us
look at the two terms in \eqref{Hdef} that involve the Casimir operators $(\Cs_{12}^2)^{-1}$ and $(\Cs_{23}^2)^{-1}$. 
The first term is diagonal in the $(12)$ double-twist basis, see eq.\ \eqref{eq:Casik}, and the 
asymptotic behavior of its matrix elements is obtained from 
\begin{equation} \label{eq:Cas12asym} 
\langle \psi_{k,J}^{(12)},\Cs^2_{12}\psi_{\ell,J}^{(12)}\rangle = (\ell^2+\orm(\ell)) \dg_{k\ell}\ . 
\end{equation} 
The second term that involves the Casimir $\Cs^2_{23}$ may be expressed in terms of the $\mathfrak{su}(1,1)$ 
Racah coefficients $W^{(\Dg_\phi)}_{k\ell}(J,0)$ that provide the transformation from the $(12)$ to the $(23)$
double-twist basis, see eq.~\eqref{Wracah_def}. 
The large-spin asymptotics of the Racah coefficients was given in eq.~\eqref{Wkappa}. From there we see that  
the matrix elements of the Casimir elements $\Cs_{23}^2$ behave as 
\begin{equation} \label{eq:Cas23asym}
\langle \psi_{k,J}^{(12)},\Cs_{23}^2 \psi_{\ell,J}^{(12)}\rangle = \orm(J^2/\ell^2).
\end{equation}
Consequently, in the limiting regime in which we want to compare the results of Derkachov and Manashov 
with our expression for anomalous dimensions, the terms $(\Cs_{12}^2)^{-1}$ and $(\Cs_{23}^2)^{-1}$ contribute at order $1/J$.  
There is one more Casimir operator to look at, namely the operator $\Cs_{13}^2$. Since the latter operator is quadratic in the generators $S_i$, it can be rewritten in terms of the Casimir operator $\Cs^2_{123}$ that is built 
from the sum $S_1+S_2+S_3$ as follows:
\begin{equation} \label{eq:Cas13asym}
\Cs_{13}^2=\Cs_{123}^2-\Cs_{12}^2-\Cs_{23}^2-3h_\phi(h_\phi-1)=J^2+\orm(\ell^2,J^2/\ell^2).
\end{equation}
The statement on the asymptotic behavior of $\Cs_{13}^2$ then follows from the fact that the 
Casimir operator $\Cs_{123}^2$ is constant and takes the value $(3h_\phi+J)(3h_\phi+J-1)$. As a result, the contribution $(\Cs_{13}^2)^{-1}$ to the operator $\mathbb{H}$ is 
subleading in the limit $J= \orm(\ell^2)\gg 1$. Finally, for none of the pairs $(i,k)\in 
\{(1,2),(2,3),(1,3)\}$ does the Casimir operator $\Cs_{ik}^2$ possess eigenvectors with eigenvalue 
$\lambda=2$ in this regime. Indeed, the spectrum of the Casimir operator of $\mathfrak{su}(1,1)$ is given by 
$\lambda = (\Dg_\phi+\ell)(\Dg_\phi+\ell-1)$ and hence, given that $\Dg_\phi = 2+\orm(\ep)$, the 
eigenvalue $\lambda =2$ is only assumed for $\ell=0$. Since the constant term in $\mathbb{H}$ 
is proportional to $\delta_{\Cs_{ik}^2,2}$, this term can never contribute. Putting all this 
together we have now shown that the large-spin limit of the matrix elements $\Gamma_{k\ell}$ 
defined in eq.~\eqref{eq:Gamma} reduces to
\begin{equation}
\frac{3}{4\ep}\Gamma_{k\ell}(J,0) =\frac{\dg_{k\ell}}{(\Dg_\phi+\ell)(\Dg_\phi+\ell-1)}  
+ \langle W^{(2)}(J,0) \psi_{k,J}^{(23)}, \frac{1}{\Cs_{23}^2} W^{(2)}(J,0)
\psi_{\ell,J}^{(23)}\rangle +\orm(J^{-2}).
\end{equation}
Here we have used that $\Dg_\phi = 2+\orm(\ep)$. In the large-spin limit, we can then approximate the sum over double-twist basis elements in the second term by an integral, and replace the Racah coefficients $W^{(\Dg_\phi=2)}_{k\ell}(J,\kappa=0)$ by their large-spin limit~\eqref{Wkappa}. In conclusion, the large-spin limit of Derkachov and Manashov's one-loop anomalous dimension operator is given by
\begin{equation} \label{eq:Gammalargespin}
\Gamma_{k\ell}(J,0) = \frac{4}{3}\ep\left(\frac{\dg(k-\ell)}{\ell^2} + \int_0^\infty\dd m 
\frac{W^{(\Dg_\phi)}_{km} W^{(\Dg_\phi)}_{m\ell}}{m^2} \right) + \orm(J^{-2}).
\end{equation}
 In order to compare the large-spin limit of the 
one-loop anomalous dimensions with our formula \eqref{adim_master}, we still need to determine 
the two parameters $h_\star$ and $\gamma_0$ that appear in eq.\ \eqref{adim_master}. In scalar 
$\phi^3$ theory, the leading-twist operator in the OPE of $\phi$ with itself is $\oo_\ast = 
\phi$ and hence $h_\star = \Dg_\phi/2=1+\orm(\ep)$. Moreover, for the coefficient $\gamma_0$
that was defined in \eqref{eq:gamma0} we obtain the following value, 
\begin{equation}
\gamma_0=-2 \frac{ \Gamma_{\Dg_\phi}^3}{\Gamma_{\Dg_\phi/2}^4} C_{\phi\phi\phi}^2 = 
-2C_{\phi\phi\phi}^2 = 4\ep/3+\orm(\ep^2) 
\end{equation}
In the last step we have inserted the well-known value of the OPE coefficient 
$C_{\phi\phi\phi}^2=-2\ep/3+\orm(\ep^2)$ in scalar $\phi^3$ theory, see e.g.\ \cite{Hasegawa_2017,Gopakumar:2016cpb}. In conclusion, we have established that the large-spin 
limit \eqref{eq:Gammalargespin} of the one-loop anomalous dimension matrix $\Gamma$ found by 
Derkachov and Manashov indeed coincides with our general result \eqref{adim_master} for 
$\gamma$ once we plug in the parameters of scalar $\phi^3$ theory and set the MST spin 
$\kappa$ to $\kappa=0$. 

\subsubsection{Generalization to non-zero MST spin}\label{genmst}

In the previous subsection, we have evaluated our general formula \eqref{adim_master} for the 
large-spin behavior of the anomalous dimension matrix for scalar $\phi^3$ theory and compared 
it with the existing one-loop calculation of the same quantity. Derkachov and Manashov performed 
their calculation for triple-twist operators in STT representations only. In this subsection, we 
want to briefly describe how their formula (\ref{gammaDM},\ref{Hdef}) can be extended to 
triple-twist primaries with $\kappa\neq 0$ and we shall sketch how the resulting formula 
for $\mathbb{H}$ compares with our bootstrap analysis. 

It is actually not difficult to generalize the original derivation of one-loop, $M$-twist anomalous 
dimensions in \cite[Sec.~4]{Derkachov:2010zza} to non-zero MST spin. This only requires to add 
an additional orthogonal null vector $w \in \Cs^d$ such that $z^2=w^2=z\cdot w = 0$. The 
normal-ordered product of $M$ operators then takes the form\footnote{Note that our notation 
differs from \cite{Derkachov:2010zza} in that $u^{\mathrm{DM}}=z^{\mathrm{ours}}$ and 
$z_i^{\mathrm{DM}} = \ag_i^{\mathrm{ours}}$.}
\begin{equation}
\mathbb{O}^{(2)}(\ag_i;\bg_i) := :\mathrel{\phi(x+\ag_1 z+\bg_1 w)\dots \phi(x+\ag_M z + \bg_M w)}:.
\end{equation}
Among the two diagrams in \cite[Fig.~2]{Derkachov:2010zza} that contribute to $\langle \phi\phi\phi\mathbb{O}\rangle$ at one loop, the rightmost diagram can only produce anomalous dimensions for $[\phi\,\phi^2]_{0,J}$, which therefore do not exist for non-zero MST spin $\kappa>0$. The former property follows from the fact that the diagram connects two external legs to the same vertex, as explained at the beginning of section~\ref{sssec:phi4_feynds}. For the remaining diagram, with the labeling of external legs and arguments of $\mathbb{O}$ given on the left-hand-side of Fig.~\ref{fig:feyndiag}, one can check that the $1/\ep$ divergence is renormalized by the anomalous dimension operator $(4\ep/3) \mathbb{H}_{12}$, where
\begin{equation}  \label{eq:Hint}
\mathbb{H}_{12} \Psi_{J,\kappa}(\ag_i;\bg_i) = \prod_{i=1}^3 \int_0^1 \dd t_i \, 
\dg(t_1+t_2+t_3-1) \Psi_{J,\kappa}(\ag_{12}^{t_1},\ag_{21}^{t_2},\ag_3;\bg_{12}^{t_1},
\bg_{21}^{t_2},\bg_3),
\end{equation}
and $\alpha_{12}^t := (1-t)\alpha_1+ t \alpha_2$.  Using the highest-weight 
condition~\eqref{hw_PsiJkappa}, we can then express the $\bg_i$-dependence of the MST 
wave function as in eq.~\eqref{eq:Psiababstar}. In addition, one can easily check that the 
factor $\om$ defined in eq.\ \eqref{eq:Psiababstar} obeys 
\begin{equation} 
\om(\ag_{12}^{t_1},\ag_{21}^{t_2},\ag_3;\bg_{12}^{t_1},\bg_{21}^{t_2},\bg_3) = \om(\ag_i;\bg_i)\ . 
\end{equation} 
Hence, we have 
\begin{equation}
\mathbb{H}_{12} \, \om(\ag_i;\bg_i)^{\kappa}\,  \Psi_{J,\kappa}(\ag_i,\bg_i^\star)= 
\om(\ag_i;\bg_i)^{\kappa}\, \mathbb{H}_{12} \Psi_{J,\kappa}(\ag_i,\bg_i^\star). 
\end{equation}
Using the property \eqref{eq:gred} of the (non-normalized) double-twist basis $g^{(12)}_{\ell,J,\kappa}
(\ag_i,\bg_i)$ we can finally determine the action of the operator $\mathbb{H}_{12}$ on the function 
by taking the limit $\ag_1\rightarrow\ag_2$. At leading order, we find 
\begin{equation}
\mathbb{H}_{12}g_{\ell,J}^{(12)}(\ag_i)\stackrel{\ag_1\rightarrow \ag_2}{\sim} \ag_{13}^{J-\ell} 
\mathbb{H}_{12}\ag_{12}^\ell =\frac{\ag_{12}^\ell\ag_{13}^{J-\ell}}{ (2+\ell)(1+\ell)}.
\end{equation}
Since $(2+\ell)(1+\ell)$ is the eigenvalue of the second-order Casimir operator $\Cs_{12}^2$ for the 
$\mathfrak{su}(1,2)$ generators $S_1+S_2$ with corresponding eigenvector $\Psi^{(12)}_{\ell,J,\kappa}
(\ag_i;\bg_i)$, we deduce that $\mathbb{H}_{12}=(\Cs_{12}^2)^{-1}$ even for $\kappa>0$. After summing 
over permutations of the external legs on the leftmost diagram of Fig.~\ref{fig:feyndiag} that produced $\mathbb{H}_{12}$, we then find that the one-loop 
anomalous dimension operator at $\kappa>0$ is given by
\begin{equation}\label{H12result}
\mathbb{H}= \sum_{1\leq i<k\leq 3}  \frac{1}{\Cs^2_{ik}}.
\end{equation}
While formally identical to the expression we reviewed before in eq.~\eqref{Hdef}, there is an important difference: in eq.\ \eqref{Hdef}, the symbol $\mathbb{C}^2$ denotes the quadratic Casimir elements 
of the Lie algebra $\mathfrak{su}(1,1)$. Instead, the extension to operators with $\kappa \neq 0$ we derived here involves the quadratic Casimir element of the Lie algebra $\mathfrak{su}(1,2)$. 
\medskip 

As in the previous subsection for triple-twist primaries with $\kappa=0$ we can now calculate the matrix 
elements of the operator~\eqref{H12result} for $\kappa \neq 0$ in the large-spin limit and then compare 
with with our result \eqref{adim_master} from the lightcone bootstrap. At large spin and finite $\kappa$, 
where $J\sim J-\kappa$, we thus retrieve the same shift relation $\Dg_\phi \mapsto \Dg_\phi+\kappa$ for 
$\Dg_\phi=2$ as seen from the property \eqref{Wkappa} of Racah coefficients.  Note that the expression 
\eqref{H12result} does not have the Kronecker delta contributions like \eqref{Hdef}. Those terms are 
important only for $\ell=0$ of the basis vectors $\psi^{12}_{\ell}$, and hence can be neglected in the 
comparison which is done in the limiting regime where $\ell^2\sim J \gg 1$. The conclusion generalized 
that of the previous subsection: when applied to triple-twist operators in scalar $\phi^3$ theory, our 
bootstrap result \eqref{adim_master} coincides with the large-spin limit of the one-loop anomalous 
dimension matrix \eqref{H12result} even for operators with non-vanishing MST spin $\kappa$. 
\smallskip 

Before we conclude our discussion of $\phi^3$ theory, we would like to note one additional outcome 
of our new formula for the one-loop anomalous dimension of operators with $\kappa\neq 0$. It concerns 
the special case in which $\kappa =J$. For this maximal value of $\kappa$, the MST triple-twist families 
are non-degenerate, with a unique wave function 
$$\Psi_{\kappa,\kappa}(\ag_i;\bg_i) =\left(\ag_{21}\bg_{31}-\ag_{31}\bg_{21}\right)^\kappa$$ 
up to normalization. We can insert this simple expression for the wave function directly into our 
integral formula \eqref{eq:Hint} to obtain 
\begin{equation}
\gamma_{[\phi\phi\phi]_{0,J=\kappa,\kappa}} = \frac{4}{3} \ep \prod_{i=1}^3\int_0^1 \dd t_i\, 
\dg(t_1+t_2+t_3-1)\left(t_3^\kappa+t_1 ^\kappa+t_2^\kappa\right) =\frac{4\ep}{(\kappa+1)(\kappa+2)}.
\end{equation}
This is a new result for the anomalous dimension of triple-twist operators with maximal $\kappa = J$ in $\phi^3$ theory. Note that it is a corollary of our perturbative analysis, so its validity 
does not require any large-spin limit.

\subsection{\texorpdfstring{$\phi^4$}{ϕ\^4} theory in \texorpdfstring{$4-2\epsilon$}{4-2ε} dimensions}

In this subsection, we will consider the two-loop (second order in $\ep$) anomalous dimensions of triple-twist operators with minimal twist in the $O(1)$ Wilson-Fisher CFT, that is to say the fixed point of a single-scalar $\phi^4$ theory in the $\epsilon$-expansion. Kehrein studied a very similar problem in the multi-scalar $O(n)$ Wilson-Fisher CFT, for triple-twist operators with vanishing MST spin that transform in a STT representation of the $O(n)$ global symmetry group \cite{Kehrein:1995ia}. While Kehrein's methods can be extended to the $n=1$ case, the resulting anomalous dimension operator would be expressed as an operator on the whole conformal multiplets rather than the primaries. This realization of the anomalous dimension as an infinite-dimensional matrix complicates its diagonalization, and its large-spin limit in this form is also not obvious. More recently, the two-loop anomalous dimension of the $[\phi\phi^2]_{0,J}$ subsector of triple-twist operators was computed in \cite[Sec.~5]{Bertucci:2022ptt}. This computation, based on the Lorentzian inversion formula, relied on the fact that $[\phi\phi^2]_{0,J}$ is the unique operator at fixed $J$ to acquire an anomalous dimension at first order in $\ep$. The computation in this section is complementary to previous works, as we derive the part of the two-loop anomalous dimension operator that acts non-trivially on the complement, spanned by $[\phi[\phi\phi]_{0,\ell>0}]_{0,J,\kappa}$. This operator, acting on the finite-dimensional space of primaries, is defined by eqs.~\eqref{eq:gammaWF}, \eqref{eq:HWF}. Together with the anomalous dimension of $[\phi\phi^2]_{0,J}$ in \cite[Sec.~5]{Bertucci:2022ptt}, its eigenvalues provide the whole spectrum of triple-twist anomalous dimensions of $\phi^4$ theory at two loops. Our derivation relies again on Derkachov and Manashov's parameterization, explained in e.g.~\cite{Derkachov:2010zza}. The large-spin limit of this operator provides another check for the results predicted by the six-point lightcone bootstrap.

\subsubsection{Prediction from six-point lightcone bootstrap}
\label{sec:wf_bstrp_analysis}

For Wilson-Fisher theory in $d=4-2\ep$ dimensions, the large-spin triple-twist anomalous dimensions 
predicted by the lightcone bootstrap first appear at second order in the $\epsilon$ expansion. To 
understand why this is the case, recall the scaling dimensions of the two lowest-lying scalars 
$\phi$ and $\phi^2$ \cite{WILSON197475}:
\begin{equation}
\Dg_\phi= 1-\ep+\frac{\ep^2}{27}+\orm(\ep^3),\quad \Dg_{\phi^2}=2-\frac{4}{3}\ep+\orm(\ep^2)= 
2\Dg_\phi+\frac{2}{3}\ep+\orm(\ep^2).
\label{wf_dimensions}
\end{equation}
We will use the general formula \eqref{adim_master} to compute the triple-twist anomalous dimension 
operator in the limit $J = \orm(\ell^2)\gg 1$. In the limit $\epsilon\to 0$, the leading-twist 
operators in the $\phi\times\phi$ OPE are $\phi^2$ and the higher-spin currents $[\phi\phi]_{0,\ell\geq 2}$, 
all appearing with twist two. Now, if we set $\oo_\star \equiv \phi^2$ as the leading-twist operator 
for nonzero epsilon in our formula, then the triple-twist anomalous dimension is proportional to
\begin{equation}
\gamma_0 \Bigl\vert_{\oo_\star=\phi^2} =
- \frac{4+\orm(\ep)}{ \Gamma\left(\Dg_\phi-\Dg_{\phi^2}/2\right)^2} =-\frac{4}{9}\ep^2+\orm(\ep^3). 
\end{equation}
Moreover, while the stress tensor and broken higher-spin currents $[\phi\phi]_{0,\ell\geq 2}$ have 
lower twist than $\phi^2$,  their anomalous dimensions first appear at second order with  
\cite{WILSON197475,Derkachov1998}
\begin{equation}
\tau_{[\phi\phi]_{0,\ell}} = 2-2\ep+\frac{2}{27}\left(1-\frac{6}{\ell(\ell+1)}\right)\ep^2+\orm(\ep^3).
\end{equation}
This absence of anomalous dimensions at leading order implies a further suppression of $\gamma_0$:
\begin{equation}\label{higherspincont}
\gamma_0 \Bigl\vert_{\oo_\star=[\phi\phi]_{0,\ell}} = -\frac{\mathrm{B}_{1+\ell}^{-1}
C_{\phi\phi[\phi\phi]_{0,\ell}}^2+\orm(\ep)}{\Gamma\left(\Dg_\phi-\tau_{[\phi\phi]_{0,\ell}}/2\right)^2} 
=\frac{4}{729}\left(-\frac{6}{\ell(\ell+1)}\right)^2\mathrm{B}_{1+\ell}^{-1}
C_{\phi\phi[\phi\phi]_{0,\ell}}^2 \ep^4+\orm(\ep^5).
\end{equation}
Accordingly, if we first expand both sides of the crossing equation up to order $\ep^2$,  then only 
$\phi^2$ appears at leading twist in the lightcone OPEs of the direct channel.  We conclude that our 
formula \eqref{adim_master} for the universal large-spin anomalous dimension operator with $\oo_\star 
= \phi^2$ and $\gamma_0=-4\ep^2/9$ is a prediction for the two-loop anomalous dimension of triple-twist 
operators in Wilson-Fisher theory. 

Let us comment on why we are allowed to consider an infinite family of minimal twist operators $\oo_\star$ 
with a vanishing twist gap. First, it is important to note that the sum over $\ell$ in eq.~\eqref{higherspincont} 
converges, allowing us to make sense of an order-by-order 
analysis in $\epsilon$. In this case, what we are really doing is isolating the deviation of the Wilson-Fisher theory 
from the free field theory at a given order of $\epsilon$. At first order, this isolates the contribution 
of $\phi^2$. Moreover, while $\phi^2$ has a higher twist than the family of currents with $\ell\ge 2$ for $\ep>0$, the opposite is true for $\epsilon <0$ (i.e.\ $d>4$), where the interacting CFT still exists (albeit not as a stable IR 
fixed point). The OPE data could then be analytically continued to negative values of $\epsilon$ where this applies. Finally, note that a similar analysis of anomalous dimensions for double-twist operators in 
\cite{Dey:2017oim} was shown to give correct results up to order $\orm(\epsilon^5)$.

\subsubsection{Explicit check with Feynman diagrams}
\label{sssec:phi4_feynds}
\begin{figure}[htp]
    \centering
    \includegraphics[scale=0.13]{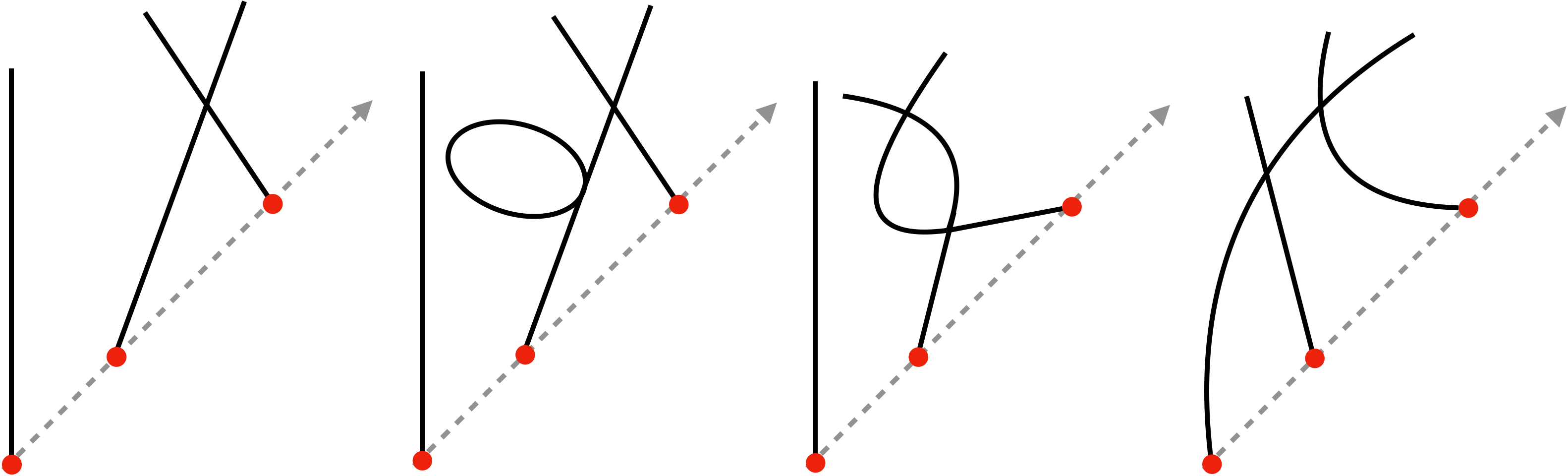}
    \caption{Four of the five Feynman diagrams that contribute to the $\langle \phi(x_1)
    \phi(x_2)\phi(x_3)\mathbb{O}(\ag_i;\bg_i)\rangle$ four-point function up to two loops. The fifth diagram is depicted on the right 
    of Fig.~\ref{fig:feyndiag}. Contrary to the latter, these four diagrams shown here share the property that two 
    of the external legs are connected to the same vertex. It follows that all their poles in $\ep$ are renormalized 
    by the anomalous dimension of $[\phi\phi^2]_{0,J}$.}
    \label{fig:trivial_feyndiags}
\end{figure}

In order to verify our bootstrap results in Wilson-Fisher theory, we now need to compute the anomalous 
dimension matrix of triple-twist operators to order $\ep^2$ in the epsilon expansion. This can be done 
with the methods of \cite{Derkachov:2010zza}, i.e.\ by computing the UV divergences of the correlation 
function $\langle\mathbb{O}(\ag_1,\ag_2,\ag_3;\bg_1,\bg_2,\bg_3)\tilde{\phi}(p_1)\tilde{\phi}(p_2)
\tilde{\phi}(p_3)\rangle$, where $\tilde{\phi}$ is the Fourier transform of $\phi$ and $\mathbb{O}$ 
denotes the normal-ordered product of fundamental fields at lightlike separation
\begin{equation}
\mathbb{O}(\ag_i;\bg_i) = :\mathrel{\prod_i \phi(y_i)}:,\quad y_i := \ag_i z+\bg_i w.
\end{equation}
There are five diagrams that can contribute up to two loops. One of them is shown on the right of 
Fig.~\ref{fig:feyndiag} while the remaining ones are displayed in Fig.~\ref{fig:trivial_feyndiags}.  

Out of the five diagrams that contribute to the connected part\footnote{There is a sixth diagram at one loop, labeled by ``6" in \cite{Kehrein:1995ia} and ``(b)" in \cite{Derkachov:1995zr}, which contributes to the disconnected parts $\langle\phi\phi\rangle\langle\phi\mathbb{O}\rangle$. From bootstrap perspective, this clearly cannot contribute to the triple-twist anomalous dimension, as it involves identity exchange in one of the OPEs. Moreover, Derkachov and Manashov showed that the anomalous dimension operator generated by this diagram exchanges three $\phi$'s for one single-twist operator $\ds^2 \phi$. Since there is no contribution that maps $\ds^2\phi$ to triple-twist operators, the anomalous dimension operator then acquires the structure of a block triangular matrix, whose eigenvalue problem reduces to that of the triple-twist-triple-twist block.} of this correlation function, only the diagram on the right-hand side of 
Fig.~\ref{fig:feyndiag} is responsible for the anomalous dimension of triple-twist operators in the large-spin 
limit~\eqref{lim_Jsimell2}, while the four remaining diagrams in Fig.~\ref{fig:trivial_feyndiags} can only produce 
anomalous dimensions for $[\phi \phi^2]_{0,J}$.  To see why this is the case, note that all four diagrams in
Fig.~\ref{fig:trivial_feyndiags} share the property that two external legs are connected to a single vertex.  
Without loss of generality, we can label each of these external legs by a momentum $p_1$ or $p_2$.  In momentum 
space,  this property implies that the Feynman integral only depends on the sum $p_1+p_2$ of the two momenta.  
After Fourier transforming back to position space, the corresponding diagram must then be localized to coincident 
points $x_1=x_2$.  Consequently,  the $1/\ep$ divergences of such a diagram can only be renormalized by anomalous 
dimension operators that localize the normal-ordered product $\mathbb{O}(\ag_i,\bg_i)$ to a subset with two 
coincident points $(\ag_{\si(1)},\bg_{\si(1)})=(\ag_{\si(2)},\bg_{\si(2)})$,  for some permutation $\si\in S_3$.  
Indeed, it is only in these cases that the condition $x_1=x_2$ can be reproduced by Wick contraction.  
Finally, in expanding the normal ordered product $\mathbb{O}(\ag_i;\bg_i)$ into a linear combination of primaries,  
the restriction to coincident points imposes that only the double-twist operators $[\phi\phi^2]_{0,J}$ (i.e.\ 
the $\ell=0$ elements in the double-twist basis) can appear. Note that the same mechanism occurs for triple-twist 
anomalous at one loop, where only the leftmost diagram of Fig.~\ref{fig:trivial_feyndiags} contributes to the 
correlator, reproducing the result of Kehrein-Pismak-Wegner \cite{Kehrein:1992fn} and Derkachov-Manashov \cite{Derkachov:1995zr} that $[\phi\phi^2]_{0,J}$ are the 
only triple-twist operators to acquire an anomalous dimension at one loop in Wilson-Fisher theory.

Having established that only the right diagram of Fig.~\ref{fig:feyndiag} contributes to the anomalous dimension 
of triple-twist operators outside of $[\phi \phi^2]_{0,J}$,  we can isolate it and express the two-loop correlator 
as
\begin{equation*}
\langle\mathbb{O}(\ag_i;\bg_i)\tilde{\phi}(p_1)\tilde{\phi}(p_2)\tilde{\phi}(p_3)\rangle = \dots 
+ \frac{g^2}{2}\sum_{1\leq i\neq j \neq k\leq 3} I(y_i,y_j;p_i,p_j) \langle \phi(y_k)\tilde{\phi}(p_k)\rangle+\orm(g^3),
\end{equation*}
where the ellipsis denotes the four diagrams of Fig.~\ref{fig:trivial_feyndiags}, while
\begin{equation*}
I(y_1,y_2;p_1,p_2) = e^{\mathrm{i}(p_1 y_1+p_2 y_2)} \int \frac{\dd^d q}{(2\pi)^d} 
\frac{e^{\mathrm{i} q(y_1-y_2)}}{(q+p_1)^2(q-p_2)^2} I_{11}(q),\quad I_{11}(q):=\int \frac{\dd^d k}{(2\pi)^d} \frac{1}{k^2(k+q)^2}
\end{equation*}
corresponds to the connected part of the diagram on the right-hand side of Fig.~\ref{fig:feyndiag}.  We can now 
proceed to compute $I(y_1,y_2;p_1,p_2)$ using
\begin{equation}
I_{11}(q) = \frac{\Gamma\left(2-\frac{d}{2}\right)\Gamma\left(\frac{d-2}{2}\right)^2}{\Gamma(d-2)} (q^2)^{\frac{d-4}{2}}.
\end{equation}
After this, we introduce Feynman parameters to help carry out the remaining loop momentum integral, i.e.
\begin{equation}
\frac{1}{q^{2\ep} (q+p_1)^{2}(q+p_2)^{2}} =\frac{\Gamma(2+\ep)}{\Gamma(\ep)} \prod_{i=1}^3 \int_0^1\dd t_i \,
\dg\left(\sum_{i=1}^3t_i-1\right) \frac{t_3^{\ep-1}}{(t_1t_2)^{2\ep}} (Q^2+M^2)^{-2-\ep}, \label{param_feyn}
\end{equation} 
where $Q:= q+t_1p_1-t_2p_2$ and $M^2:=t_1 t_2s_{12}:=t_1t_2(p_1+p_2)^2$.  Apart from the $(Q^2+M^2)^{-2-\ep}$ in 
eq.~\eqref{param_feyn},  the remaining dependence in the loop momentum comes from the phase $e^{\mathrm{i}Q\cdot 
y_{12}}$.  However, after expanding the latter into a power series in $y_{12}^\mu$,  it follows from Lorentz 
invariance that any higher order term in $y_{12}$ will be proportional to $y_{12}^2$.  Since $y_{12} \in 
\mathrm{Span} (z,w)$, where $z,w$ are mutually orthogonal null vectors, these terms must then vanish\footnote{The same argument can found in \cite{Derkachov:2010zza}, below eq.~(4.19),  for the one-loop 
diagram of $\phi^3$ theory on the left-hand side of Fig.~\ref{fig:feyndiag}.}.  We can therefore drop this 
phase and explicitly integrate over $Q$ to obtain
\begin{equation}
G(y_1,y_2;p_1,p_2) =  \frac{\Gamma(1-\ep)^2\Gamma(2\ep)}{(4\pi)^{2-\ep} \Gamma(2-2\ep)s_{12}^{2\ep}}
\prod_{i=1}^3\int_0^1\dd t_i \frac{t_3^{\ep-1}}{(t_1t_2)^{2\ep}}\dg\left(\sum_{i=1}^3t_i-1\right)\, 
e^{\mathrm{i}\left(p_1 y_{12}^{t_1}+p_2 y_{21}^{t_2} \right)},
\end{equation}
where $y_{12}^t := (1-t)y_1+t y_2$.  The $\Gamma(2\ep)$ produces a simple pole, which is eliminated by adding 
a two-loop anomalous dimension to the composite operator $\mathbb{O}$.  Following the conventions of 
\cite[Sec.~4]{Derkachov:2010zza}, it takes the form
\begin{equation} \label{eq:gammaWF}
\gamma = -\frac{4\ep^2}{9} \left(\mathbb{H}_{12}+\mathbb{H}_{23}+\mathbb{H}_{13}\right),
\end{equation}
where
\begin{equation} \label{eq:HWF} 
\mathbb{H}_{12} \Psi^{(2)}(\ag_i;\bg_i) = \prod_{i=1}^3 \int_0^1 \dd t_i\,\dg(t_1+t_2+t_3-1)\,\frac{t_3^{\ep-1}}{(t_1t_2)^{2\ep}} \Psi^{(2)}(\ag_{12}^{t_1},\ag_{21}^{t_2},\ag_3;\bg_{12}^{t_1},\bg_{21}^{t_2},\bg_3).
\end{equation}
Using once again eq.~\eqref{eq:Psiababstar}, we can repeat the derivation  of the one-loop anomalous dimension 
of $\phi^3$ theory from Subsection~\ref{phi3app} and \ref{genmst} with $[\mathbb{H}_{12},\om^\kappa]=0$ and
\begin{equation}
\mathbb{H}_{12}g_{\ell,J}^{(12)}(\ag_i)\stackrel{\ag_1\rightarrow \ag_2}{\sim} \ag_{13}^{J-\ell} 
\mathbb{H}_{12}\ag_{12}^\ell =\frac{\ag_{12}^\ell\ag_{13}^{J-\ell}}{ (1-\ep+\ell)(\ell-\ep)}.
\end{equation}
Given $\Dg_\phi=1-\ep$ in $d=4-2\ep$, we then equate $\mathbb{H}_{12}$ with the inverse $\su(1,2)$ Casimir operator and 
write $\mathbb{H}=\sum_{i,k}1/\Cs_{ik}^2$. At large spin, we thus retrieve the same result as the lightcone 
bootstrap with $\oo_\star = \phi^2$ and $\gamma_0=-4\ep^2/9$. In conclusion, we have not only computed the 
anomalous dimension \eqref{eq:gammaWF} of triple-twist operators in Wilson-Fisher theory to order $\ep^2$ in 
the epsilon expansion but also shown that the large-spin limit of this two-loop result agrees beautifully 
with the prediction of our bootstrap analysis in formula~\eqref{adim_master}. 

\subsubsection{Two-loop, triple-twist anomalous dimensions at finite spin}
The results from the previous subsection allow us to determine the two-loop anomalous dimensions of the one-loop-degenerate triple-twist primaries\footnote{We thank Johan Henriksson for helping us understand the two-loop mixing problem, and for sharing results \cite{johan_private} on the two-loop anomalous dimensions of one-loop degenerate triple-twist operators at low spin.}. Indeed, the anomalous dimension operator at two loops takes the general form
\begin{equation}
\gamma = c_1\ep\, \mathbb{K}_1 +c_2\ep^2\mathbb{K}_2^{(\ep)}-\frac{4\ep^2}{9}\mathbb{H}^{(\ep)}, \quad \mathbb{H}:=\mathbb{H}_{12}+\mathbb{H}_{23}+\mathbb{H}_{13},
\end{equation}
where $\mathbb{H}_{12}$ was defined by eq.~\eqref{eq:HWF}, while $\mathbb{K}_L$ are operators that originate from the $\ep$-poles of the $L$-loop diagrams in Fig.~\ref{fig:trivial_feyndiags}, where two external legs are attached to the same vertex. As explained in subsection~\ref{sssec:phi4_feynds}, $\mathbb{K}_L$ therefore vanish on the subspace of polynomials that vanish at coincident points: $\Psi(\alpha_1,\alpha_1,\alpha_3;\beta_1,\beta_1,\beta_3)=0.$ Note that we keep an explicit dependence on $\epsilon$ for the operators $\mathbb{K}_2^{(\ep)}$, $\mathbb{H}^{(\ep)}$ because they each exhibit a pole in $\epsilon$ proportional to $\mathbb{K}_1$. However, it was shown in \cite{Kehrein:1995ia} that these poles cancel in the final expression.  For non-zero MST spin $\kappa>0$, all wave functions $\Psi$ vanish at coincident points, such that $\mathbb{H}^{(\ep=0)}$ is finite and the two-loop anomalous dimension operator reduces to $-4\ep^2\mathbb{H}^{(0)}/9$. In particular, there is a unique primary at maximal MST spin $\kappa=J$, whose anomalous dimension we can easily compute from eqs.~\eqref{eq:gammaWF} and \eqref{eq:HWF}. The result is
\begin{equation}
\gamma_{[\phi\phi\phi]_{0,J=\kappa,\kappa}} = -\frac{4\ep^2}{9} \prod_{i=1}^3\int_0^1 \dd t_i\, 
\dg(t_1+t_2+t_3-1)\left(t_3^{\kappa-1}+t_1 ^{\kappa-1}+t_2^{\kappa-1}\right) =-\frac{4\ep^2}
{3\kappa(\kappa+1)}.
\end{equation}
In the case $J>\kappa=0$, the orthogonal complement of the kernel of $\mathbb{K}_1,\mathbb{K}_2$ is spanned by the wave function $\Psi_{\ell=0}$ corresponding to $[\phi\phi^2]_{0,J}$ in the double-twist basis. This primary therefore acquires an anomalous dimension at order $\ep$, and its anomalous dimension at order $\ep^2$ was computed in \cite[Sec.~5]{Bertucci:2022ptt}. Since the remaining anomalous dimensions correspond to polynomial wave functions in the kernel of $\mathbb{K}_1,\mathbb{K}_2$, they can be related to the eigenvalues of $\mathbb{H}$ from the fact that
\begin{equation}
\Psi(\alpha,\alpha,\beta)=0 \Longrightarrow \gamma\cdot  \Psi = -\frac{4\ep^2}{9}\mathbb{H}^{(\ep)}\cdot\Psi \Longrightarrow \gamma_i = -\frac{4\ep^2}{9} \lim_{\ep\rightarrow 0} \la_i^{(\ep)}.
\end{equation}
In practice, the matrix elements of $\mathbb{H}^{(\ep)}$ in the double-twist basis $\Psi_\ell^{(12)}$ can be expressed in terms of the eigenvalues $(1-\ep+\ell)^{-1}(\ell-\ep)^{-1}$ of $\mathbb{H}_{12}$ and the Racah coefficients $W_{k\ell}^{(1-\ep)}(J,0)$. After diagonalization, the spectrum of $\mathbb{H}^{(\ep)}$ has one eigenvalue $\la_0^{(\ep)} = \orm(\ep^{-1})$ that diverges as $\epsilon\rightarrow 0$, while its other eigenvalues $\la_i^{(\ep)}=\orm(1)$ are finite as $\epsilon\rightarrow 0$: the latter correspond to the two-loop anomalous dimensions of one-loop degenerate triple-twist operators. For example, at low spins $J=6,8,9,10,11$, the space of triple-twist operators is two-dimensional, and the kernel of $\mathbb{K}_1,\mathbb{K}_2$ is therefore one-dimensional. Correspondingly, there is a single finite eigenvalue $\la_1(J)$, for which we find
\begin{equation}
\la_1^{(0)}(6) = \la_1^{(0)}(8) = \frac{5}{18},\quad \la_1^{(0)}(7) = \frac{79}{672},\quad \la_1^{(0)}(10) = \frac{23}{84},\quad \la_1^{(0)}(11) = \frac{2599}{21600}.
\end{equation}
It is interesting to note that (the absolute value of) these numbers are the same as in \cite[Tab.~3]{Kehrein:1995ia}, which tabulates the eigenvalues $\alpha_2^{(2lp)}$ of the operator $V_2^{(2lp)}$. The latter provides the two-loop anomalous dimension of one-loop-degenerate triple-twist operators that transform in a STT representation of $O(n)$, and are therefore not immediately related to our family of $n=1$ primaries. However, we expect that the diagram-based arguments of this section also imply the reduction of the one-loop degenerate part to a multiple of $V_2^{(2lp)}$, as claimed in \cite[Sec.~4.3]{Kehrein:1995ia}.

In summary, the $\orm(\ep^0)$ eigenvalues of $-4\mathbb{H}^{(\ep)}/9$ correspond to the two-loop anomalous dimensions of one-loop-degenerate triple-twist operators in $\phi^4$ theory. At vanishing MST spin $\kappa=0$, this confirms the computations of \cite{johan_private}, which were based on Kehrein's approach in \cite{Kehrein:1995ia}. 

\subsubsection{Comment about one-loop, twist-four anomalous dimensions}
\label{sssec:twist4}
There is another test of our bootstrap analysis we can perform in the context of Wilson-Fisher theories. 
Recently, the authors of \cite{Henriksson:2023cnh} studied the one-loop anomalous dimension of the 
twist-four primaries $[\phi\,[\phi \phi^2]_{0,\ell}]_{0,J,0}$. While the analysis we have presented 
was restricted to identical scalars and hence does not quite cover these twist-four operators, it is 
not too difficult to extend our framework accordingly. More specifically, after a minimal adjustment 
of our setup to the six-point function $\langle \phi(X_1)\phi(X_2)\phi^2(X_3)\phi^2(X_3)\phi(X_5)
\phi(X_6)\rangle$ of non-identical scalars, we expect to recover a formula like 
eq.~\eqref{eq:adim_diagonal_limit} that describes the anomalous dimension of the primaries $[\phi\,[\phi \phi^2]_{0,\ell}]_{0,J,\kappa}$ near the double-twist limit $\ell^2\ll J$. However, while the 
$(J/\ell)^{-2h_\star}$ term remains proportional to $\gamma_0\propto C_{\phi\phi\oo_\star}^2$, the 
prefactor of the $\ell^{-2h_\star}$ term will be modified to $\gamma_0\propto C_{\phi^2\phi^2\oo_\star}C_{\phi\phi\oo_\star}$. Given the leading-twist operators $\oo_\star$ of the 
$\phi\times\phi$ OPE listed in Section~\ref{sec:wf_bstrp_analysis}, the latter contribution will 
therefore be subleading in the $\ep$-expansion due to the suppression of the OPE coefficient 
$C_{\phi^2\phi^2\oo_\star}$. Focusing on the former contribution, we can then use the same 
reasoning as in Section~\eqref{sec:wf_bstrp_analysis} to deduce that $\oo_\star=\phi^2$ is the 
dominant exchange in the $\ep$-expansion. Now, while $\gamma_0 \vert_{\oo_\star=\phi^2}=
-\frac{4}{9}\ep^2$ is of second order in $\epsilon$, the pole $\Gamma(\Dg_\phi-h_\star)=\orm(\ep^{-1})$ in the $(J/\ell)^{-2h_\star}$ term of eq.~\eqref{eq:adim_diagonal_limit} implies 
that this anomalous dimension does indeed start at one loop. Finally, after plugging in the 
scaling dimensions $\Dg_\phi$ and $h_\star=\Dg_{\phi^2}/2$ of eq.~\eqref{wf_dimensions}, our 
analysis predicts that the one-loop anomalous dimension of $[\phi\,[\phi \phi^2]_{0,\ell}]_{0,J,0}$ 
in Wilson-Fisher theory in the double-twist limit is given by
\begin{equation}
\gamma\left([\phi\,[\phi \phi^2]_{0,\ell}]_{0,J,0}\right) \sim \stackrel{1\ll \ell^2 \ll J}
{\sim} \frac{8}{3}\, \ep \left(\frac{\ell}{J}\right)^2+\dots
\end{equation}
Quite remarkably, this does indeed coincide with the anomalous dimension correction at the bottom 
of \cite[Tab.~1]{Henriksson:2023cnh}, for even $\ell$ and $k=-m=\ell+1\gg 1$.

\section{Conclusions and Outlook}\label{sec:conclusions}
 
In this paper, we employed a higher-point lightcone bootstrap approach to extract large-spin CFT 
data of triple-twist operators. Just like in the four-point lightcone bootstrap, our analysis of 
the six-point crossing equation in lightcone limits demonstrates that triple-twist CFT data asymptotes 
to generalized free field theory at large spin. The corresponding scalar/double-twist/triple-twist OPE 
coefficients $p_\ell^{J_i}$ are presented in eq.~\eqref{opec_PiWPi}, with an asymptotic form given by
eq.~\eqref{pgff_4pt_ls}.  Beyond this expected behavior, we successfully derived the triple-twist 
anomalous dimension matrix $\gamma_{k\ell}(J,\kappa)$ displayed in~\eqref{adim_master}, and valid 
in the large-spin limit $1 \ll k,\ell, J$ with $k^2,\ell^2=\orm(J)$. Each of these results can describe triple-twist operators in a mixed-symmetry representation with two spin labels, $J$ and $\kappa$. 

To achieve our goal, we considered a six-point comb-to-comb crossing under a triple lightcone limit, 
tailored to project onto lowest-twist contributions for the direct channel and triple-twist exchanges 
in the crossed channel. Following the approach of~\cite{Kaviraj:2022wbw}, we reproduced the finite number of terms of the direct channel from the 
crossed-channel expansion, making crucial use 
of the Casimir and vertex differential operators that can be used to characterize conformal 
blocks~\cite{Buric:2020dyz, Buric:2021ywo, Buric:2021ttm, Buric:2021kgy}. This allowed us to 
determine both the relevant scaling of crossed-channel eigenvalues, cf.~Section~\ref{sec:definitions}, 
and the crossed-channel lightcone blocks corresponding to those eigenvalues in Section~\ref{sec:blocks_cc}. 
The expressions of six-point lightcone conformal blocks~\eqref{gtilde_to_besselK} and their
normalization~\eqref{norm_6ptblock} are new, and their derivation constitutes a significant step 
forward in the higher-point bootstrap program.

To correctly interpret and express our results, it was necessary to discuss the parameterization 
of triple-twist operators and their degeneracy in generalized free field theory. Our discussion in
Section~\ref{sec:param_3twist} addressed this in the language of Derkachov and Manashov~\cite{Derkachov:1995zr,Derkachov:2010zza}, 
and generalized their framework to account for triple-twist operators transforming in mixed-symmetry representations $\kappa>0$. The application of this formalism to the double-twist basis $[\phi[\phi\phi]_{0,\ell}]_{0,J,\kappa}$ was central to the introduction of the Racah coefficients $W_{k\ell}^{(\Delta)}(J,\kappa)$, as well as the derivation of their large-spin limit $1 \ll k=
\orm(J/\ell)$ presented in eq.~\eqref{Wkappa}. These Racah coefficients are an important ingredient in 
the anomalous dimension matrix~\eqref{adim_master} of triple-twist operators.

Let us stress again that our results only apply to triple-twist operators supported on a subset of the entire range $0\leq \ell \leq J$ spanned by the degeneracy parameter $\ell$.\footnote{A triple-twist primary is characterized by its anomalous 
dimension $\la$ and its wave function $\Psi_\lambda$, which admits an expansion in the double-twist basis $\Psi_{\ell,J,\kappa}$. The values of $\ell$ that `contribute significantly' 
to this expansion are referred to as the \textit{support} of the triple-twist operator.} More specifically, our formula~\eqref{adim_master} captures the regime $1 \ll \ell \ll  J$. Near the 
`edges' of this region of validity, i.e.\ when either $1 \ll \ell \ll \sqrt{J}$ or $\sqrt{J} \ll \ell \ll J$, 
the anomalous dimension matrix becomes nearly diagonal in the double-twist basis. In this limit, triple-twist operators localize to elements of the double-twist basis, while the anomalous dimension reduces to that of its corresponding double-twist constituent. For small perturbations away from the $1 \ll \ell \ll  \sqrt{J}$ limit, the first correction to the anomalous dimension is given by eq.~\eqref{eq:adim_diagonal_limit}. But as $\ell$ moves 
further toward the center of the region of validity, where $\ell = \orm(\sqrt{J})$ and the anomalous dimension matrix scales as $J^{-h_\star}$, non-trivial mixing occurs within the double-twist basis. This results in a strong deviation away from double-twist behavior, and therefore a genuine feature of triple-twist dynamics that goes beyond the reach of the four-point lightcone bootstrap. 

It would be particularly interesting to explore the $\ell = \orm(J)$ region, which is furthest away from the double-twist approximation and goes well beyond the domain of validity of formula~\eqref{adim_master}. From explicit examples like $\phi^3$ theory at one loop \cite{Derkachov:1997uh,Derkachov:2010zza},  we expect this regime to 
describe the triple-twist operators with the lowest absolute anomalous dimension at large spin, scaling 
like $J^{-2h_\star}$ instead of $J^{-h_\star}$.  However,  to reach such operators from the six-point crossing equation of this paper,  it 
appears necessary to relax the $\ep_{34}\rightarrow 0$ limit so that all three intermediate spins
$J_a\propto \ep_{16}^{-1/2}$ scale in the same way. In the absence of the lightcone limit on $X_{34}$, we would then need to include all higher-twist exchanges in the $(34)$ OPE of the direct channel. As such, any comprehensive analysis requires at least significant further work on lightcone blocks in both channels.  On the other hand,  the holographic description of triple-twist operators in AdS provides a useful and complementary approach to resolving their dynamics in this regime.  This exploration of triple-twist dynamics based on effective field theory in AdS is the subject of upcoming work \cite{PK_JAM_inprep}.

\medskip 

\begin{figure}[htp]
    \centering
    \includegraphics[scale=0.445]{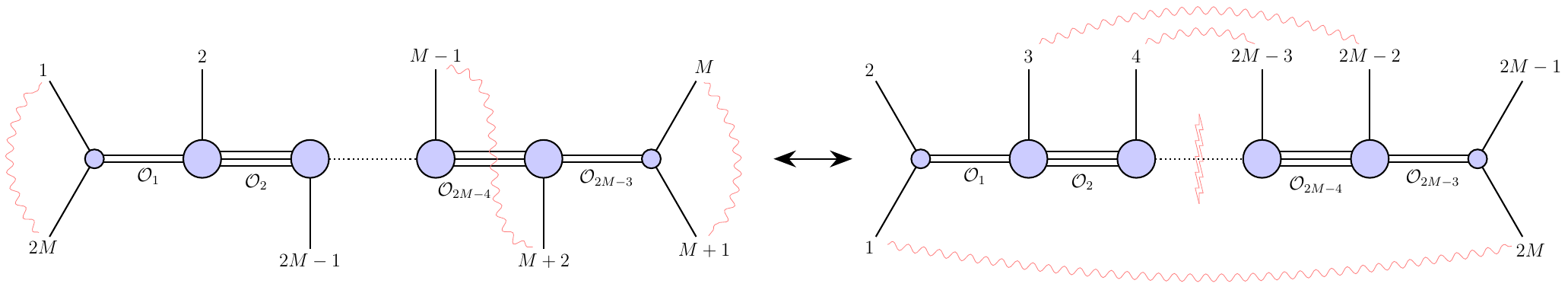}
    \caption{Proposed comb-to-comb duality to entail exchanges of twist-$M$ operators in the middle leg of the crossed channel (right). The red wavy lines connect points whose distances are taken to be null and are meant to project on lowest-twist exchanges on all the internal legs of the direct channel (left). The CC middle-leg leading-twist limit $\dutchcal{u}_{M-1}\rightarrow 0$ is only represented on the right as a red line that cuts the diagram in half.}
    \label{fig:higher-twists}
\end{figure}

In addition to studying higher corrections to our analysis of six-point functions, the extension to correlation functions with more than six operator insertions would certainly be 
relevant. Some parts of our analysis 
seem relatively straightforward to generalize in this direction. Indeed, it should be possible to bootstrap twist-$M$ operators  
of the form $[\phi\dots\phi]_{0,J,\{\kappa\}}$ by studying a comb-channel crossing equation for $2M$
insertions of $\phi$. A natural generalization of the six-point crossing equation we used above to $M > 3$ 
is depicted in Figure~\ref{fig:higher-twists}.  
By taking the lightcone limits $X_{1(2M)}, X_{3(2M-2)},\dots,X_{M(M+1)}\rightarrow 0$ and projecting to leading twist in the crossed-channel middle leg via $\dutchcal{u}_{M-1}\rightarrow 0$, we expect the 
direct-channel contribution with a maximal number of identity exchanges to be reproduced by a 
sequence of twist-$(a+1)$ exchanges $\oo_a = [\phi\dots\phi]_{0,J_a,\{\kappa\}}$ for $a=1,\dots, 
M-1$, with increasingly large spin $J_a \ll J_{a+1}$. At next-to-leading order, the $M-1$ exchanges of 
one single leading-twist operator $\oo_\star$ in the direct channel should then entail a large-spin 
anomalous dimension matrix of the form 
\begin{equation}
\gamma=\gamma^{(1)}+ \sum_{a=3}^M W^{(1a)} \gamma^{(1)} \left(W^{(1a)}\right)^{-1},
\end{equation}
where $\gamma^{(1)}$ has eigenvalues $\gamma_0\,J_1^{-2h_\star}$ and $W^{(1a)}$ realizes the permutation 
$(1a) \in S_M$ in the space of twist-$M$ operators. In fact, the $M-1$ terms we displayed are the natural 
extension of the two terms in eq.~\eqref{adim_master}. To write down the corresponding twist-$M$ OPE coefficients and anomalous dimensions explicitly, we can 
generalize the basis arising from the iterated construction of double-twist operators. In fact, in 
accordance with the connection to conformal blocks observed in this paper, the Derkachov-Manashov wave 
functions for these basis elements should correspond to eigenfunctions of the Gaudin integrable model 
on the $M$-punctured sphere associated with $\mathfrak{su}(1,r)$, where $r-1$ is the number of 
non-vanishing transverse spins, in the same comb-channel limit used for conformal blocks, see
\cite{Buric:2020dyz,Buric:2021ywo}. The only minor difference is that instead of imposing $M$-point 
conformal invariance as in the case of conformal blocks, the diagonal action of $\mathfrak{su}(1,r)$ 
on twist-$M$ wave functions is now determined by the quantum numbers of the twist-$M$ 
primary that exchanged in the middle leg. In the absence of transverse spins, i.e.\ for $r=1$,  the 
method of separation of variables determines the dual wave functions $\tilde{\Psi}$ explicitly in 
terms of a product of $M-1$ Jacobi polynomials. These results should allow us to study 
the anomalous dimension matrix explicitly and diagonalize it in certain limits. 

The analysis of higher multi-twist operators is particularly interesting in connection with recent work on large-charge effective descriptions of 3d CFTs\footnote{We thank Alessio Miscioscia for enjoyable discussions on that topic.} \cite{Cuomo:2022kio} (see also the follow-ups \cite{cuomo2024chiral,cuomo2023numerical}). 
In this context, it would be helpful to obtain a bootstrap understanding of the emergence of the giant vortex phase in the regime where the squared charge $Q^2$ approaches the spin $J$. One could possibly probe this regime by picking special configurations of $Q$ pairs of charge $\pm1$ operator insertions at near-to-lightlike separation, letting the $Q \rightarrow \infty$ limit and the lightcone limit be controlled by the same parameter\footnote{Note that this would differ from the analysis of lower-point correlators of highly charged operators in \cite{Jafferis_2018} that successfully provided a bootstrap perspective on the superfluid phase. Instead it would be more similar to the setup used by the authors of \cite[Sec.~2.2]{Anous:2016kss}.}.
As noted in \cite{Cuomo:2022kio}, the asymptotics of such a regime are determined by a delicate balance between the combinatorics of different contributions to the OPEs in the correlator and the suppression of higher-twist terms by the lightcone limit. 
Concretely, \cite{Cuomo:2022kio} sketches a qualitative description of the situation to argue for a match between the boundaries of the Regge phase and giant vortex phase.
However, this qualitative description is based on the assumption that the twist gap is exactly one. In contrast, a bootstrap analysis could provide important insight into the general dependence on the twist gap and the behavior that arises for generic values of it.
Moreover, the intrinsically model-agnostic approach of the bootstrap could point towards direct generalizations to space-time dimensions other than three.
Finally, a putative bootstrap description of the conjectured Regge-Giant-Vortex phase transition may help elucidate the similarities between the multi-twist Hilbert space on the one hand, and the Hilbert space of fluctuations around the giant vortex on the other hand. 
\bigskip

\textbf{Acknowledgements:} We are grateful to Ant\'onio Antunes, Carlos Bercini, Aleix Gimenez-Grau, Johan Henriksson, Petr Kravchuk, and Alessio Miscioscia
for useful discussions. This project received funding from the German Research 
Foundation DFG under Germany’s Excellence Strategy -- EXC 2121 Quantum Universe -- 
390833306. In the course of this work, J.A.M.\ was supported by the Royal Society under grant URF$\backslash$R1$\backslash$211417 and by the European Research Council (ERC) under the European Union’s Horizon 2020 research and innovation
program – 60 – (grant agreement No.~865075) EXACTC. S.P.H.\ is further supported by the Studienstiftung des Deutschen Volkes.

\appendix
\section{Large-Spin Vertex Operators}\label{app:VertexOperator}
This appendix provides a simple expression for the basis of conformal tensor structures for STT-MST-scalar correlators 
\begin{align}
     \langle \mathcal{O}_1(X_1,Z_1)\mathcal{O}_2(X_2,Z_2,W_2) \mathcal{O}_3(X_3) \rangle
\end{align}
 that diagonalizes the vertex operator
 \begin{align}
     \mathcal{V}=\frac12 \tr \left( \mathcal{T}_1^3 \mathcal{T}_3\right)-\frac{1}{16} \tr \left( \mathcal{T}_1^2\right)\tr \left( \mathcal{T}_2^2\right)+\frac{1}{8} \tr \left( \mathcal{T}_1^2\right)\tr \left( \mathcal{T}_3^2\right)+\frac{d(d-1)}{8}\left[\tr \left( \mathcal{T}_3^2\right)-\tr \left( \mathcal{T}_2^2\right)\right]+\frac{1}{4}\tr \left( \mathcal{T}_1^4\right), 
 \end{align}
in the regime
 \begin{align}
     d,J_1, h_1, h_2, \kappa, \Delta_3 \ll J_2.
 \end{align}
 Concretely, in terms of the standard basis
 \begin{align}
\{ \Omega_3 \mathcal{X}^{n}\}_{n=0}^{J_1 - \kappa}, && \text{where} && \mathcal{X} = \frac{H_{21}X_{23}X_{31}}{J_{1,23}J_{2,13}} && \text{and} &&\Omega_3 = \frac{U_{2,13}^{\kappa}J_{1,23}^{J_1-\kappa_2} J_{2,13}^{J_2-\kappa}}{ X_{21}^{\bar h_{21;3}-\kappa} X_{23}^{\bar h_{23;1}-\kappa+J_1} X_{31}^{\bar h_{3 1;2}+J_2}},
 \end{align}
 we find that a basis that diagonalizes $\mathcal{V}$ is given by
\begin{align}\label{eq:vertexwavefunction}
    \{\Omega_3 g_\nu(\mathcal{X})\}_{\nu=0}^{J_1 - \kappa} && \text{where} && g_\nu(\mathcal{X}) = (1-\mathcal{X})^{J_1 - \kappa -\nu} \mathcal{X}^\nu + \orm (1/J_2).
\end{align}
The corresponding eigenvalues are
\begin{align}
        \mathcal{V} g_\nu(\mathcal{X}) = J_2^2\left[\left(h_1+J_1-\nu\right)\left(2(h_1+J_1-\nu)-d\right)-2\kappa_2 \left(h_1+J_1-\nu-1\right)    \right] g_\nu(\mathcal{X}) + \orm(J_2)\,,
\end{align}
which can be verified to correspond to the parametrization of eigenvalues presented in~\cite[eq.~(4.15)]{Kaviraj:2022wbw} with $N=J_1-\nu$, once one sets $\kappa=0$ and factors in the different shifts by Casimir operators in the definition of the vertex operator.
Note that the basis $\{g_\nu\}_{\nu=0}^{J_1 - \kappa}$ can be expressed as a simple polynomial basis if one redefines the prefactor and cross-ratios in use. Indeed, with 
\begin{align}
    \tilde\Omega_3 := \Omega_3 (1- \mathcal{X})^{J_1 -\kappa} && \tilde{\mathcal{X}} := \frac{\mathcal{X}}{1-\mathcal{X}},
\end{align}
we find
\begin{align}
    \Omega_3 g_\nu(\mathcal{X}) = \tilde{\Omega}_3 \tilde {\mathcal{X}}^\nu.
\end{align}
Moreover, we would like to point out that the $g_\nu$--basis is preferable over the standard basis for a description of vertex operators at finite $J_2$ in the sense that $\mathcal{V}$ is simply a tridiagonal matrix in the $g_\nu$--basis, whereas in the standard basis the second diagonal above the main diagonal is non-vanishing. 

\section{Normalization of Six-point Lightcone Blocks}
\label{app:normalization_6ptblocks}
In Subsection~\ref{ssec:normalization_6ptblocks}, we outlined the main steps in the derivation of the normalization $\mathcal{N}_{J_1(J_2,\kappa)J_3}$ of eq.~\eqref{norm_6ptblock} that enters formula~\eqref{gtilde_to_besselK} for large spin, crossed-channel lightcone blocks. In this appendix, we will provide explicit formulas for the second-order Casimir equations and their solutions that appear in the derivation. 

\subsection{Casimir equations for lightcone blocks.} Given the parameterization~\eqref{psi_to_Ftilde} of normalized lightcone blocks, the second-order Casimir equations with eigenvalues $\la_1(h_1,\bar h_1),\la_2(h_2,\bar h_2,\kappa),\la_3(h_3,\bar h_3)$ can be expressed as
\begin{align}
&(\mathcal{D}_{12}^2-\la_1) g_{\oo_i;n_s} \stackrel{\Upsilon_0,\bar z\rightarrow 0}{\sim}\prod_{i=1}^3 \bar z_i^{h_i}z_i^{\bar h_i}\prod_{s=1}^2 (1-w_s)^{n_s} \Upsilon_0^\kappa\ \,\mathcal{D}_1 \cdot\tilde{F}_{(h_i,\bar h_i,\kappa;n_s)}(z_1,z_2,z_3,w_1,w_2), \\
&(\mathcal{D}_{456}^2-\la_2) g_{\oo_i;n_s} \stackrel{\Upsilon_0,\bar z\rightarrow 0}{\sim}  \prod_{i=1}^3 \bar z_i^{h_i}z_i^{\bar h_i}\prod_{s=1}^2 (1-w_s)^{n_s} \Upsilon_0^\kappa\, \mathcal{D}_2 \cdot \tilde{F}_{(h_i,\bar h_i,\kappa;n_s)}(z_1,z_2,z_3,w_1,w_2),\\
&(\mathcal{D}_{56}^2-\la_3) g_{\oo_i;n_s} \stackrel{\Upsilon_0,\bar z\rightarrow 0}{\sim}  \prod_{i=1}^3 \bar z_i^{h_i}z_i^{\bar h_i}\prod_{s=1}^2 (1-w_s)^{n_s} \Upsilon_0^\kappa\, \mathcal{D}_3 \cdot \tilde{F}_{(h_i,\bar h_i,\kappa;n_s)}(z_1,z_2,z_3,w_1,w_2),
\end{align}
where, using the notation $j_i:=J_i-\kappa$, the differential operators on the right-hand side are given by
\begin{align}
\mathcal{D}_1 = \vt_{z_1}(\vt_{z_1}+2\bar h_1-1) - z_1\vt_{z_1} (\bar h_{12;\phi}-\kappa+\vt_{z_1}+\vt_{z_2}-\vt_{w_1}+w_1(n_1-j_2-\vt_{z_2}+\vt_{w_1})), \label{D1_lc} \\
\mathcal{D}_3 = \vt_{z_3}(\vt_{z_3}+2\bar h_3-1) - z_3\vt_{z_3} (\bar h_{23;\phi}-\kappa+\vt_{z_2}+\vt_{z_3}-\vt_{w_2}+w_2(n_2-j_2-\vt_{z_2}+\vt_{w_2})), \label{D3_lc}
\end{align}
and
\begin{align}
\mathcal{D}_2 = \vt_{z_2}(\vt_{z_2}+2\bar h_2-\kappa-1) - z_2&(\bar h_{12;\phi}-\kappa+\vt_{z_1}+\vt_{z_2}-\vt_{w_1}+w_1(n_1-j_1-\vt_{z_1}+\vt_{w_1}))\label{D2_lc}  \\
&(\bar h_{23;\phi}-\kappa+\vt_{z_2}+\vt_{z_3}-\vt_{w_2}+w_2(n_2-j_3-\vt_{z_3}+\vt_{w_2})),\nonumber
\end{align}
This system of differential equations $\mathcal{D}_a\tilde{F}=0$ with initial condition $\tilde{F}(0,w_s)=1$ can be solved in terms of a hypergeometric power series in $z_i,w_s$, or expressed as an integral representation.  
\subsection{The limit \texorpdfstring{$\ep_{34}\rightarrow 0$}{ε34->0}.} Now,  to interpolate with the lightcone limit of crossed-channel blocks, we will directly solve the system of equations in the limit $\ep_{34}\rightarrow 0$.  To do this, we will make the change of variables to $v_2=1-z_2=\orm(\ep_{34})$ and $\xx_s=1-w_s=\orm(\ep_{34})$. Then in the limit $\ep_{34}\rightarrow 0$ with $J_2^2=\orm(\ep_{34}^{-1})$, we find at leading order
\begin{align}
\mathcal{D}_{1,3} = \mathcal{D}_{\bar h_{1,3},h_{2\phi}+n_{1,2}}(z_{1,3},\vt_{z_{1,3}}; \xx_{1,2},\vt_{\xx_{1,2}};-\ds_{v_2}) + \orm(\ep_{14}),  \label{D13_X34lim} \\ 
\mathcal{D}_2 = \ds_{v_2}(\vt_{v_2}+h_{1\phi}+h_{3\phi}+\kappa+\vt_{\xx_1}+\vt_{\xx_2})-J_2^2+\orm(\ep_{34}^0),
\end{align}
where the limiting form of the outer-leg Casimirs in eq.~\eqref{D13_X34lim} are given by
\begin{equation}
\mathcal{D}_{\bar h,\ag}(z,\vt_z;x,\vt_x;\tau) :=\vt_z(2\bar h+\vt_z-1) - z(\bar h+\vt_z)(\bar h+\ag+\vt_x+\vt_z)-zx  \tau(\bar h+\vt_z).
\end{equation}
We assume that the solution takes the form
\begin{align}
\tilde{F}_{(h_i,\bar h_i,\kappa;n_s)}&(z_1,1-v_2,z_3,1-\xx_1,1-\xx_2) \stackrel{v_2,\xx_s\rightarrow 0}{\sim} \nonumber\\
&\mathcal{N}^{\mathrm{4pt}}_{(h_i,\bar h_i,\kappa,n_s)} \,G_1(z_1,-z_1\xx_1\ds_{v_2})G_3(z_3,-z_3\xx_2\ds_{v_2}) \mathcal{K}_{\ag_2} (J_2^2 v_2),
\label{Ftilde_x34}
\end{align}
where $\ag_2:= h_{1\phi}+h_{3\phi}+n_1+n_2+\kappa$ and for $a=1,3$:
\begin{equation}
G_a(z_a,y_a) = \sum_{m,\nu=0}^\infty \frac{(\bar h_a+\ag_a+\nu)_m}{m!} \frac{1}{\nu!} \frac{(\bar h_a)_{m+\nu}}{(2\bar h_a)_{m+\nu}} z_a^m y_a^\nu,\quad (\ag_1,\ag_3):=(h_{2\phi}+n_1+\kappa,h_{2\phi}+n_2+\kappa).
\label{Ga_of_zaya}
\end{equation}
To prove this ansatz and determine $\mathcal{N}^{\mathrm{4pt}}$, we compare the solution of the quadratic Casimir equation $\mathcal{D}_2\tilde{F}_{\oo_i;n_s}(z_1=0,z_2,z_3=0,w_s=1)=0$, which is ${}_2 F_1\! \left(\bar h_2\!+\!h_{1\phi}\!+\!n_1\!+\!\kappa,\bar h_2\!+\!h_{3\phi}\!+\!n_2\!+\!\kappa;2\bar h_2\!-\!\kappa;z_2 \right)$ when normalized by the OPE limit, to the expression~\eqref{Ftilde_x34} evaluated at $z_1,z_3,\xx_1,\xx_2=0$. Using the identity
\begin{equation}
\lim_{\bar h\rightarrow \infty} \frac{\Gamma(\bar h+c-a)\Gamma(\bar h+c-b)}{2\, \Gamma(2\bar h+c)}\, \hypg{\bar h+a}{\bar h+b}{2\bar h+c}\left(1-\bar h^{-2}x\right) = \mathcal{K}_{a+b-c}(x),
\end{equation}
we deduce that $\mathcal{N}^{\mathrm{4pt}}$ is given by eq.~\eqref{norm_34}.
\subsection{The limit \texorpdfstring{$\ep_{16}\rightarrow 0$}{ε16->0}} 
To reach the kinematical limit of crossed-channel lightcone blocks, we change variables to
\begin{equation}
\xx_1=\frac{(1-z_1)v_2}{z_1}x_1,\quad \xx_2 =\frac{v_2(1-z_3)}{z_3}x_2,\quad (x_1,x_2):=(1-\mU_1-\mU^6,\mU_1).
\end{equation}
Next, we rewrite the functions $G_a$ as
\begin{align}
G_a(y_a,z_a) &= \frac{1}{(1-z_a)^{\alpha_a}} \sum_{\nu_a=0}^\infty \frac{1}{\nu!} \frac{(\bar h_a)_\nu}{(2\bar h_a)_\nu} \left(\frac{y_a}{1-z_a}\right)^{\nu} \hypg{\bar h_a}{\bar h_a-\ag_a}{2\bar h_a-\nu}(z_a)\\
&=\frac{1}{(1-z_a)^{\alpha_a}}\sum_{m_a=0}^\infty \frac{z_a^{m_a}}{m_a!}\frac{(\bar h_a)_{m_a}(\bar h_a-\ag_a)_{m_a}}{(2\bar h_a)_{m_a}}  \,_1 F_1\begin{bmatrix}\bar h_a\\2\bar h_a+m_a\end{bmatrix}\left(\frac{y_a}{1-z_a}\right). \label{app:tildeGa}
\end{align}
After applying the identity $ y^{-\ag}\mathcal{K}_{\ag}(y) = \mathcal{K}_{-\ag}(y)$ and inserting the integral representation of $\mathcal{K}_{-\ag}$, we then obtain the following expression for $F(v_i,x_s) \equiv v_1^{\ag_1}v_2^{\ag_2}v_3^{\ag_3} \tilde{F}(1-v_i,1-\xx_s(v_s,v_{s+1},x_s))$:
\begin{align}
F_{(h_i,\bar h_i,\kappa:n_s)}(v_i,x_s) \stackrel{\ep_{34}\rightarrow 0}{\sim}\frac{\mathcal{N}_{(h_i,\bar h_i,\kappa;n_s)}^{\mathrm{4pt}}}{2J_2^{2\ag_2}} \int_0^\infty& \frac{\dd t}{t^{1-\ag_2}} e^{-\left(t+\frac{v_2J_2^2}{t}\right)} \label{F_to_1F1}\\&\sum_{m_a=0}^\infty \frac{z_a^{m_a}}{m_a!}\frac{(\bar h_a)_{m_a}(h_{a2;\phi}-\dg n_a)_{m_a}}{(2\bar h_a)_{m_a}}  \,_1 F_1\begin{bmatrix}\bar h_a\\2\bar h_a+m_a\end{bmatrix}\left(t x_{[a]}\right)  \nonumber,
\end{align}
where $(x_{[1]},x_{[3]}):=(x_1,x_2)$ and  $(\dg n_1,\dg n_3):=(J_1-n_1,J_3-n_2)$. We now define the lightcone limit $X_{16}=\orm(\ep_{16})$, $\ep_{16}\rightarrow 0$ where the cross-ratios and quantum numbers scale with $\ep_{16}$ as follows:
\begin{equation}
\left(v_a,\mU_1,\mU^6;J_2,J_a,\dg n_a\right) \propto (1,1,\ep_{12};\ep_{12}^{-1/2},\ep_{12}^{-1/2},1).
\end{equation}
To retrieve the limiting form~\eqref{gtilde_to_besselK} of blocks in this regime, we must first recall the prefactor that relates $F=\prod_a v_a^{\ag_a} \tilde{F}$, defined by eq.~\eqref{psi_to_Ftilde} to $\tilde g$, defined by eq.~\eqref{eq:rescaled_blocks_omega}:
\begin{equation}
\tilde g = x_0^{\kappa} \prod_{a=1,3} z_a^{\dg n_a} x_{[a]}^{j_a-\dg n_a} F,\quad x_0:=\frac{z_1 z_2 z_3 \Upsilon_0}{v_1v_2v_3}.
\end{equation}
We then write the expansion~\eqref{F_to_1F1} as
\begin{equation}
x_0^{-\kappa} \tilde{g} = \mathcal{N}_{(h_i,\bar h_i,\kappa;n_s)}^{\mathrm{4pt}}\prod_{a=1,3} x_{[a]}^{-(h_a+\dg n_a+\kappa)}\sum_{m_a=0}^{\infty}\frac{z_a^{m_a}}{m_a!}\frac{(\bar h_a)_{m_a}(h_{a2;\phi}-\dg n_a)_{m_a}}{(2\bar h_a)_{m_a}} I^{m_1m_3}_{\bar h_1\bar h_3},
\label{appB:gtilde_exp}
\end{equation}
where
\begin{equation}
I^{m_1m_3}_{\bar h_1\bar h_3} := \frac{1}{2} J_2^{-2\ag_0}\int_0^\infty \frac{\dd k}{k^{1+\ag_0}} e^{-\left(k J_2^2+\frac{v_2}{k}\right)} \prod_{a=1,3} (k J_2^2 x_{[a]})^{\bar h_a}\,_1F_1\begin{bmatrix}
    \bar h_a\\ 2\bar h_a+m_a
\end{bmatrix}(k J_2^2 x_a),
\end{equation}
and $\ag_0:=2h_{\phi}+\kappa+\dg n_1+\dg n_2$. In the limit $J_2^2=\orm(\ep_{16}^{-1})$, we can expand the confluent hypergeometric functions at large argument using \cite[eq.~(13.7.1)]{dlmf}, which yields
\begin{equation*}
(k J_2^2 x_{[a]})^{\bar h_a}\,_1F_1\begin{bmatrix}
    \bar h_a\\ 2\bar h_a
\end{bmatrix}(k J_2^2 x_{[a]}) = \frac{\Gamma(2\bar h_a)}{\Gamma(\bar h_a)}(k J_2^2 x_{[a]})^{-m_a} \exp(k J_2^2 x_{[a]}-\frac{\bar h_a^2}{k J_2^2 x_{[a]}}) \left(1+\orm(\ep_{12}^{1/2})\right).
\end{equation*}
Given $x_1+x_2=1-\mathcal{U}^6$ and $x_2=\mathcal{U}_1$, we then obtain 
\begin{equation}
I^{m_1m_3}_{\bar h_1\bar h_3}\stackrel{\ep_{16}\rightarrow 0}{\sim} J_2^{-2(m_1+m_3)}\left(\mU^6\right)^{\ag_0+m_1+m_2} \mathcal{K}_{\ag_0+m_1+m_2}\left(\left[\frac{\bar h_1^2}{1-\mU_1}+ J_2^2 v_2+\frac{\bar h_3^2}{\mU_1}\right]\mU^6\right).
\end{equation}
Inserting this leading form back into eq.~\eqref{appB:gtilde_exp}, one finds that $I^{m_1m_3}$ is suppressed by a factor $J_2^{-2(m_1+m_3)}$ relative to $I^{00}$. We then take the leading term $m_a=0$ to retrieve the asymptotics of crossed-channel lightcone blocks:
\begin{multline}
\tilde{g}_{(h_i,h_i+J_i,\kappa;J_{[s]}-\dg n_{[s]})}\stackrel{\ep_{16}\rightarrow 0}{\sim}
\frac{\mathcal{N}_{(h_i,h_i+J_i,\kappa;J_{[s]}-\dg n_{[s]})}^{\mathrm{4pt}}}{J_2^{2(h_1+h_2+J_1+J_3-2h_\phi-\kappa-\dg n_1-\dg n_2)}} \times\\
\times \prod_{a=1,3}\frac{\Gamma(2\bar h_a)}{\Gamma(\bar h_a)} G^{h_1-2h_\phi,h_3-2h_\phi}_{(J_1,J_2,\kappa,J_3);\dg n_1\dg n_3}(v_2,\mU_1,1-\mU_1,\mU^6).
\end{multline}
In the case $(h_1,h_2,h_3;\dg n_1,\dg n_3)=(2 h_\phi,3h_\phi,2h_\phi;0,0)$, one deduces formula~\eqref{norm_6ptblock} for the normalization of six-point blocks with GFF twists exchanged.
\section{Leading-Twist GFF Correlators from Representation Theory}
\label{app:rep_theory}
In this appendix, we address the derivation of the GFF correlator of a triple-twist operator and three scalars displayed in~\eqref{dt_wf}. After introducing the necessary background on the highest-weight representation theory of $\su(1,2)$, we then derive the generalization of the latter, displayed in~\eqref{dtPsi_sym_mst}, to triple-twist operators in MST representations.
\subsection{STT sector: derivation of eq.~\texorpdfstring{\eqref{dt_wf}}{(4.40)}}
Performing the Wick contractions in the GFF correlator that defines $\Psi_{\ell,J}$ leads to
\begin{align}
    \Psi_{\ell,J} = \check{\Psi}_{\ell,J}(\partial_{\bar\alpha_i})  \sum\limits_{\sigma \in S_3}\prod\limits_{i=1}^3 X_{ib}^{\Delta_\phi}  (X_{\sigma(i)b}+\bar\alpha_i Z X_{\sigma(i)})^{-\Delta_\phi}.
\end{align}
Inserting \eqref{dt_norm2} for $\check{\psi}$ in eq.~\eqref{eq:defPsicheck} furthermore yields
\begin{equation}
\check{\Psi}_{L,0}(\ds_{\bar\ag_1},\ds_{\bar\ag_2}) = \cgff_{J-\ell} \cgff_\ell \sum_{n=0}^{J-\ell} \sum\limits_{k=0}^n\sum\limits_{j=0}^\ell\frac{(\ell-J)_n}{n!} \binom{n}{k} \frac{(-\ell)_j}{j!}\frac{\ds_{\bar\ag_1}^{k+j}\ds_{\bar\ag_2}^{\ell+n-k-j}\ds_{\bar\ag_3}^{J-\ell-n}}{(2\Delta_\phi+2\ell)_n(\Delta_\phi)_{J-\ell-n}(\Delta_\phi)_{j}(\Delta_\phi)_{\ell-j}}.
\end{equation}
Therefore,
\begin{equation}
\Psi_{\ell,J}(\ag_1,\ag_2,\ag_3) = \norm{\ag_{12}^\ell}_{\Dg_\phi,\Dg_\phi}^{-1}\norm{\ag_{12}^{J-\ell}}_{\Dg_\phi,2\Dg_\phi+2\ell}^{-1}\sum_{\si\in S_3} g_{\ell,J}^{(12)}(\ag_{\si(1)},\ag_{\si(2)},\ag_{\si(3)}), 
\end{equation}
with 
\begin{align}\label{eq:g_unpractical_frame}
    g_{\ell,J}^{(12)}=\sum_{n=0}^{J-\ell} \sum\limits_{k=0}^n\sum\limits_{j=0}^\ell\frac{(\ell-J)_n}{n!} \binom{n}{k} \frac{(-\ell)_j}{j!}\frac{\alpha_{1}^{k+j}\ag_{2}^{\ell+n-k-j}\ag_{3}^{J-\ell-n}(\Delta_\phi)_{k+j}(\Delta_\phi)_{\ell+n-k-j}}{(2\Delta_\phi+2\ell)_n(\Delta_\phi)_{j}(\Delta_\phi)_{\ell-j}(-1)^J}.
\end{align}
On the one hand, by translation symmetry, we obtain the same function by evaluating $\alpha_1 = 0$ while setting $\alpha_2 = - \alpha_{12}$ and $\alpha_3 = - \alpha_{13}$. On the other hand, the condition $\alpha_1 = 0$ directly trivialises two of the sums in eq.~\ref{eq:g_unpractical_frame} and we conclude that
\begin{align}
    g_{\ell,J}^{(12)}=\sum_{n=0}^{J-\ell} \frac{(\ell-J)_n}{n!} \frac{\ag_{2}^{\ell+n}\ag_{3}^{J-\ell-n}(\Delta_\phi)_{\ell+n}}{(2\Delta_\phi+2\ell)_n(\Delta_\phi)_{\ell}(-1)^J} = \alpha_{12}^\ell \alpha_{13}^{J-\ell}\sum_{n=0}^{J-\ell} \frac{(\ell-J)_n (\Dg_\phi+\ell)_n}{(2\Dg_\phi+2\ell)_n n!} \left(\frac{\ag_{12}}{\ag_{13}}\right)^{n}.
\end{align}

\subsection{MST sector and representation theory of \texorpdfstring{$\mathfrak{su}(1,2)$}{su(1,2)}}
\label{app:sl3}
\subsubsection{Generators and action on polynomials}
The generators of $\sl(3)$ in a lowest-weight representation are given in \cite[Sec.~3.1]{Derkachov:2005hw} as first order differential operators $\mathrm{T}_{ij}(x,y,z,\ds_x,\ds_y,\ds_z;m,n)$ and $\mathrm{H}_i(x,y,z,\ds_x,\ds_y,\ds_z;m,n)$ acting on power series in three variables $x,y,z$. In this realization, the unit-normalized highest-weight vector is $1$. The relation to generators in our notation is then given by
\begin{align}
S_{ij}(\ag,\bg,\cg,\ds_\ag,\ds_\bg,\ds_\cg) := \mathrm{T}_{ij}(\ag,\bg,\cg,-\ds_\ag,-\ds_\bg,\ds_\cg;J-\kappa,\kappa-\bar\tau),  \quad 1\leq i\neq j\leq 3, \\
S_{0i}(\ag,\bg,\cg,\ds_\ag,\ds_\bg,\ds_\cg) := \frac{1}{2}\mathrm{H}_{i}(\ag,\bg,\cg,\ds_\ag,\ds_\bg,\ds_\cg;J-\kappa,\kappa-\bar\tau), \quad i=1,2
\end{align}
The lowest-weight vector in this realization is $S_{ki}\cdot 1=0$, $k>i$, with weights $(S_{01},S_{02})\cdot 1= (\bar\tau-\kappa,\kappa-J)$.  The Hilbert space is then spanned by vectors $\psi_{r,s,t}(\ag,\bg,\cg)=S_{12}^r S_{23}^s S_{13}^t \cdot 1$. When $J-\kappa>0$ is an integer, the corresponding basis is restricted to $s\leq J-\kappa$.  
\subsubsection{Duality, scalar product and norms}
Define $(\dg,j):=(\bar\tau-\kappa,J-\kappa)$. The action of the triangular subalgebra spanned by $(S_{12},S_{23},S_{13})$ is then given by
\begin{align}
&\left(e^{\bar\ag S_{12}}\cdot \psi\right)(\ag,\bg,\cg) = (1-\bar\ag \ag)^{-\dg} \psi\left(\frac{\ag}{1-\bar\ag \ag},\frac{\bg}{1-\bar\ag \ag},\cg-\bar\ag(\ag\cg-\bg)\right), \label{S12}\\
&\left(e^{\bar\cg S_{23}}\cdot \psi\right)(\ag,\bg,\cg) = (1+\bar\cg \cg)^{j} \psi\left(\ag-\bar\cg \bg,\bg,\frac{\cg}{1+\bar\cg \cg}\right), \label{S23}\\
&\left(e^{\bar\bg S_{13}}\cdot \psi\right)(\ag,\bg,\cg) =\frac{ (1+\bar\bg (\ag\cg-\bg))^{j}}{(1-\bar\bg \bg)^{\dg}} \psi\left(\frac{\ag}{1-\bar\bg \bg},\frac{\bg}{1-\bar\bg \bg},\frac{\cg}{1+\bar\bg (\cg\ag-\bg)}\right).\label{S13}
\end{align}
From these properties, it is easy to show that the duality $\check{\psi}\mapsto \psi$ in eq.~\eqref{psi_to_psihat_mst} satisfies
\begin{align}
\psi(\ag,\bg,\cg)&= \check{\psi}(\ds_{\bar\ag},\ds_{\bar\bg},\ds_{\bar\cg}) \frac{\left[1+\bar\cg\cg-(\bar\ag\bar\cg-\bar\bg)(\ag\cg-\bg)\right]^{j}}{(1-\bar\ag\ag-\bar\bg\bg)^{\dg}} \Bigl\vert_0 \label{genfunc_mst} \\
&= \check{\psi}(\ds_{\bar\ag},\ds_{\bar\bg},\ds_{\bar\cg})  e^{\bar\ag S_{12}} e^{\bar\cg S_{23}} e^{\bar\bg S_{13}} \cdot 1\\
&= \mathrm{L}\left\{\mathrel{\check{\psi}\left(S_{12},S_{23},S_{13}\right)}\right\} \cdot 1,
\end{align}
where $\mathrm{L}\{\}$ now denotes left-ordering of $S_{12}$ with respect to $S_{23}$ (both of which commute with $S_{13}$).  The inverse map $\psi \mapsto \check{\psi}$ is complicated by the fact that $S_{12}^rS_{23}^sS_{13}^t\cdot 1$ is no longer a monomial in $\ag,\bg,\cg$. This, in turn, reflects the fact that the monomials $\bg$ and $\ag\cg$ have the same homogeneity degree with respect to the Cartan subalgebra of $\mathfrak{sl}(3)$, such that the overlaps of states $(r,s,t)\neq (r',s',t')$ need not vanish for $(r-s,s+t)=(r'-s',s'+t')$.  Fortunately,  we need only compute norms that come from the two following subsets of states:
\begin{enumerate}
\item $\psi \in \mathrm{Span}(S_{12}^rS_{13}^t\cdot 1)$.  Scalar products between these states are obtained by applying eq.~\eqref{psi_to_psihat_mst} with $\bar\cg=0=\cg$, and for monomials we obtain
\begin{equation} \label{eq:normab}
\ds_\ag^r\ds_\bg^t \longleftrightarrow (\dg)_r(\dg-j+r)_t\, \ag^r \bg^t, \quad \langle \ag^r\bg^t, \ag^{r'}\bg^{t'}\rangle_{\dg,j} =  \frac{\dg_{rr'}\dg_{tt'} r! t!}{ (\dg)_r(\dg-j+r)_t}.
\end{equation}
In the specific case $j=0$,  it will be useful to instead rewrite the monomial norms in terms of shifted norms for the $\mathfrak{su}(1,1)$ subgroup, $\norm{\ag^r}^2_{\bar\tau}= r!/(\bar\tau)_r$, such that
\begin{equation}
\norm{\psi(\ag)\bg^t}^2_{\dg,0} = t!\, \norm{\psi(\ag)}^2_{\bar\tau+t}.
\end{equation}

\item $\psi \in \mathrm{Span}(S_{12}^rS_{23}^s\cdot 1)$.  Scalar products between these states are obtained by applying eq.~\eqref{psi_to_psihat_mst} with $\bar\bg=0=\bg$, and for monomials we obtain
\begin{equation} 
\ds_\ag^r\ds_\cg^s \longleftrightarrow (\dg-s)_r(-j)_s (-1)^s \ag^r \cg^s, \quad \langle \ag^r\cg^s, \ag^{r'}\cg^{s'}\rangle_{\dg,j} =  \frac{\dg_{rr'}\dg_{ss'} r! s!}{ (\bar\tau-\kappa-s)_r(\kappa-J)_s (-1)^s}.
\label{norm_YtoX}
\end{equation}
It is again useful to rewrite these norms in terms of the $\mathfrak{su}(1,1)$ norms $\norm{\ag^r}^2_{\bar\tau}= r!/(\bar\tau)_r$, such that
\begin{equation}
\norm{\psi(\ag)\cg^s}^2_{\dg,j} = \binom{j}{s}^{-1} \norm{\psi(\ag)}^2_{\dg-s}.
\label{norm_ZtoX}
\end{equation}
\end{enumerate}

\subsubsection{Norm and dual of the double-twist wave function}
Up to normalization, the unique lowest-weight vector of weight $(\bar\tau_1+\bar\tau_2+J+\kappa,J-\kappa)$ in the tensor product of two $\mathfrak{su}(1,2)$ representations of weights $(\bar\tau_1,0)$ and $(\bar\tau_2,\ell)$ is given by
\begin{align}
\frac{\psi_{J-\ell,\kappa,0}(\ag_a,\bg_a,\cg_a,\ag_3,\bg_3)}{\psi_{J-\ell,\kappa,0}(0,0,0,1,1)} &= \ag_{a3}^{J-\ell} \left(\bg_{a3}-\cg_a\ag_{a3}\right)^\kappa \\
&= \ag_3^{J-\ell}\bg_3^\kappa (1-\ag_3^{-1}\ag_{a})^{J-\ell} \left[1-\bg_3^{-1}\ag_3 \cg_a+\bg_3^{-1}(\cg_a\ag_a-\bg_a)\right]^\kappa.
\end{align}
In the last line, we expressed the dependence on $\ag_a,\bg_a,\cg_a$ in the same way as eq.~\eqref{genfunc_mst} with $\bar\ag_a=\ag_3^{-1}$, $\bar\cg_a=-\bg_3^{-1}\ag_3$, and $\bar\bg_a=0$. This implies that $\psi_{J-\ell,\kappa,0}\in \mathrm{Span}(S_{12}^r S_{13}^t\cdot 1) \otimes \mathrm{Span}(S_{12}^r S_{23}^s\cdot 1)$, and that the norm can be computed in terms of scalar products of vectors in this subspace. In particular, we can set $\bg_2=0$ and expand the norm in terms of the norms of eq.~\eqref{norm_YtoX},\eqref{norm_ZtoX} as follows:
\begin{align}
\norm{\ag_{a3}^L \left(\bg_{a3}-\cg_a\ag_{a3}\right)^\kappa}^2_{\bar\tau_a,\ell,\bar\tau_3}  &= \norm{\ag_{a 3}^L(\cg_a\ag_{a3}+\bg_3)^\kappa}^2_{\bar\tau_a,\ell,\bar\tau_3}  \\
&= \sum_{q=0}^\kappa \binom{\kappa}{q}^2 \norm{\bg_1^{\kappa-q} \cg_2^q \ag_{12}^{L+q}}^2_{\bar\tau_a,\ell,\bar\tau_3} \\
&= \frac{\kappa!}{\ell!} \sum_{q=0}^\kappa \binom{\kappa}{q} \frac{(\ell-q)!}{(\bar\tau_3)_{\kappa-q}} \norm{\ag_{a3}^{L+q}}^2_{\bar\tau_a-q,\bar\tau_3+\kappa-q}.\label{dtnorm2_mst_finsum}
\end{align}
It also follows from this analysis that the dual wave function for $\ds_{\bar\bg_a}=0$ is given by
\begin{equation}
\check{\psi}(x_a,0,z_a,x_3,y_3)= \frac{\psi_{J-\ell,\kappa,0}(0,0,0,1,1)}{\psi_{J-\ell,0}(0,1)} \sum_{q=0}^\kappa \binom{\kappa}{q} \frac{z_a^q}{(-1)^q (-\ell)_q} \frac{y_3^{\kappa-q}}{(\bar\tau_3)_{\kappa-q}} \check{\psi}_{L+q,0}^{(\bar\tau_a-q,\bar\tau_3+\kappa-q)}(x_a,x_3).
\label{checkpsi_specialkin}
\end{equation}
\subsubsection{Derivation of eq.~\texorpdfstring{\eqref{dtPsi_sym_mst}}{(4.61)}}
The wave function $\Psi_{\ell,J,\kappa}(\ag_i,\bg_i)$ corresponding to $[\phi\oo_\ell]_{0,J,\kappa}$,  $\oo_\ell:=[\phi\phi]_{0,\ell}$ is given by
\begin{equation}
\Psi_{\ell,J,\kappa}(\ag_i,\bg_i) = \sum_{\si \in S_3} \Psi^{(12)}_{\ell,J,\kappa}(\ag_{\si(i)},\bg_{\si(i)}),
\end{equation}
where each permutation amounts to a Wick contraction in the six-point function of $\phi$.  The (non-symmetric) wave function $\Psi^{(12)}$ corresponding to a single permutation takes the form
\begin{align}
\Psi^{(12)}_{\ell,J,\kappa}(\ag_i,\bg_i) = \check{\psi}_{J-\ell,\kappa,0}(\ds_{\bar\ag_a},\ds_{\bar\bg_a},\ds_{\bar\cg_a},\ds_{\bar\ag_3},\ds_{\bar\bg_3}) \check{\psi}_{\ell,0}(\ds_{\bar\ag_1},\ds_{\bar\ag_2})\prod_{i=1}^3 (1-\xi_i\ag_i-\eta_i\bg_i)^{-\Dg_\phi},
\end{align}
where
\begin{align}
&(\xi_1,\xi_2,\xi_3) := (\bar\ag_1+\bar\ag_a,\bar\ag_2+\bar\ag_a,\bar\ag_3) \\
& (\eta_1,\eta_2,\eta_3):= (\bar\bg_a-\bar\cg_a\bar\ag_1,\bar\bg_a-\bar\cg_a\bar\ag_2,\bar\bg_3).
\end{align}
The lowest-weight condition uniquely defines $\Psi^{(12)}_{\ell,J,\kappa}(\ag_i,\bg_i)$ in terms of $\Psi^{(12)}_{\ell,J,\kappa}(\ag_1,0,\ag_2,0,\ag_3,1)$, which reduces to 
\begin{align*}
\Psi^{(12)}_{\ell,J,\kappa}(\ag_1,0,\ag_2,0,\ag_3,1) = \check{\psi}_{J-\ell,\kappa,0}\vert_{\ds_{\bar\bg_a},\ds_{\bar\cg_a}=0} \check{\psi}_{\ell,0}(1\!-\!\bar\ag_1\ag_1)^{-\Dg_\phi}(1\!-\!\bar\ag_2\ag_2)^{-\Dg_\phi}(1\!-\!\bar\ag_3\ag_3\!-\!\bar\bg_3\bg_3)^{-\Dg_\phi} \vert_0.
\end{align*}
The dual polynomial $\check{\psi}_{J-\ell,\kappa,0}(x_a,0,0,x_3,y_3)$ is a special case of eq.~\eqref{checkpsi_specialkin},  which after insertion leads to 
\begin{align*}
\Psi^{(12)}_{\ell,J,\kappa}(\ag_1,0,\ag_2,0,\ag_3,1) = C\,& \check{\psi}_{J-\ell,0}^{(\bar\tau_a,\Dg_\phi+\kappa)}(\ds_{\bar\ag_1}+\ds_{\bar\ag_2},\ds_{\bar\ag_3})\check{\psi}_{\ell,0}^{(\Dg_\phi,\Dg_\phi)}(\ds_{\bar\ag_1}+\ds_{\bar\ag_2},\ds_{\bar\ag_3})\frac{ \ds_{\bg_3}^\kappa}{(\Dg_\phi)_\kappa} \\
\times &(1-\bar\ag_1\ag_1)^{-\Dg_\phi}(1-\bar\ag_2\ag_2)^{-\Dg_\phi}(1-\bar\ag_3\ag_3-\bar\bg_3\bg_3)^{-\Dg_\phi} \vert_{\bar\ag_i,\bar\bg_3=0},
\end{align*}
where $\bar\tau_a:=2\Dg_\phi+2\ell$ and $C$ is the ratio of $\psi_{J-\ell,\kappa,0}(0,0,0,1,1)$ by $\psi_{J-\ell,0}(0,1)$. The action of $(\Dg_\phi)_\kappa^{-1}\ds_{\bar\bg_3}^\kappa\vert_{\bar\bg_3=0}$ has the effect of shifting the exponent of $(1-\bar\ag_3\ag_3)$ by $-\kappa$.  The remaining expression is equivalent to the double-twist basis wave function $g_{\ell,J}^{(12)}$ at $\kappa=0$, but with shifted arguments $(J,\ell,\Dg_\phi)\rightarrow (J-\kappa,\ell-\kappa,\Dg_\phi+\kappa)$.  Since the $g_{\ell,J}^{(12)}$ as defined in eq.~\eqref{dt_wf} is invariant under this shift, we retrieve the formula 
\begin{equation}
\Psi^{(12)}_{\ell,J,\kappa}(\ag_1,0,\ag_2,0,\ag_3,1) = C\,\psi_{\ell,J}^{(12)}(\ag_1,\ag_2,\ag_3).
\end{equation}
To retrieve the full result $\Psi_{\ell,J,,\kappa}$, we first reintroduce the full dependence on the $\bg_i$ by virtue of the lowest-weight condition:
\begin{equation}
\Psi^{(12)}_{\ell,J,\kappa}(\ag_i,\bg_i) = \frac{(\ag_{12}\bg_{13}-\ag_{13}\bg_{12})^\kappa}{\ag_{12}^\kappa}\Psi^{(12)}_{\ell,J,\kappa}(\ag_1,0,\ag_2,0,\ag_3,1).
\end{equation}
The final result then reduces to 
\begin{equation}
\Psi_{\ell,J,\kappa}(\ag_i,\bg_i) = C\sum_{\si \in S_3}(\ag_{\si(1)\si(2)}\bg_{\si(1)\si(3)}-\ag_{\si(1)\si(3)}\bg_{\si(1)\si(2)})^\kappa\ag_{\si(1)\si(2)}^{-\kappa} \psi_{\ell,J}^{(12)}(\ag_{\si(i)}).
\end{equation}
Using the fact that $\psi_{\ell,J}^{(12)}$ is the ratio of $g_{\ell,J}^{(12)}$ by its norm, and that $C$ is the ratio of $\psi_{J-\ell,\kappa,0}(0,0,0,1,1)$ by $\psi_{J-\ell,0}(0,1)$, one can then retrieve eq.~\eqref{dtPsi_sym_mst}.

\section{OPE Limits to Lower-Point Functions}
\label{app:6pt_to_4pt}

The scope of this appendix is to use OPE limits from six-point conformal blocks to four-point ones to work out some of the results relevant for Section~\ref{sec:GFFOPECoefficients}. In Appendix~\ref{app:form_blocks_GFF}, we provide a derivation of the GFF four-point blocks presented in~\eqref{eq:4ptlightconeblock}. In the second part, Appendix~\ref{app:crossratios_prefactor_from_OPElimit}, we both derive the relation between cross-ratios and the coordinates $\alpha_i$ presented in~\eqref{eq:z1_z2_w_as_alphas}, as well as deriving the multiplicative prefactor used in~\eqref{3t_cb_dec} to express the GFF four-point correlator $\langle \phi(X_1)\phi(X_2)\phi(X_3)[\phi\phi\phi]_{0,J,\kappa}(X_b,Z_b)\rangle$ in terms of linear combinations of four-point lightcone blocks $g_{\oo_1;n_1}(z_1,w_1)$.

\subsection{GFF four-point blocks from six-point OPE limit}
\label{app:form_blocks_GFF}
The spinning four-point lightcone blocks can be obtained from an OPE limit of six-point lightcone blocks. For this reason, we consider again the conformal block decomposition of the six-point function in OPE cross-ratios:
\begin{equation}
\langle \phi(X_1)\dots\phi(X_6)\rangle = \sum_{\oo_1,\oo_2,\oo_3;n_1,n_2} P_{\oo_1\oo_2\oo_3}^{(n_1n_2)}
\frac{g_{\oo_1\oo_2\oo_3;n_1n_2}(\bar z_1,z_1,\bar z_2,z_2,\Upsilon_0,\bar z_3,z_3,w_1,w_2)}
{(X_{12}X_{34}X_{56})^{\Dg_\phi} (\bar z_2 z_2)^{\Dg_\phi/2}}.
\tag{\ref{eq:sixpointblockdecomp}}
\end{equation}
Taking the OPE limit in the second and third leg along with the lightcone limit $\bar z_1\rightarrow 0$, we obtain the lightcone blocks for the four-point function in eq.~\eqref{eq:defPsiJ} (and eq.~\eqref{eq:defPsiJ_mst} for $\kappa>0$):
\begin{equation}
g_{\oo_1\oo_2\oo_3;n_1n_2} \sim \bar z_1^{h_1}g_{\oo_1;n_1}(z_1,w_1) (\bar z_2 z_2)^{h_2} z_2^{J_2+\kappa} \Upsilon_0^\kappa (1-w_2)^{n_2} (\bar z_3 z_3)^{h_3} z_3^{J_3}.
\end{equation}
The multiplicative prefactor relating linear combinations of $\psi_{\oo_1;n_1}$ to the four-point correlator itself is derived in Appendix~\ref{app:crossratios_prefactor_from_OPElimit}.  After setting 
\begin{equation}
    g_{\oo_1;n_1}(z_1,w_1) = z_1^{\bar h_1} (1-w_1)^{n_1} f_{\oo_1;n_1}(z_1,w_1)
    \label{eq:fourptasf_appendix}
\end{equation},  the second-order Casimir equation for the spinning lightcone block reduces to $\mathcal{D}_1 f_{\oo_1;n_1}=0$, where 
\begin{equation}
\mathcal{D}_1 = \vt_{z_1}(\vt_{z_1}+2\bar h_1-1) - z_1\vt_{z_1} (\bar h_{12;\phi}-\kappa+\vt_{z_1}+\vt_{z_2}-\vt_{w_1}+w_1(n_1-j_2-\vt_{z_2}+\vt_{w_1})).
\end{equation}
The corresponding solution satisfying the OPE-limit normalization $f_{\oo_1;n_1}(0,w_1)=1$ is given by
\begin{equation}
f_{\oo_1;n_1}(z_1,w_1) = F_1(\bar h_1; \bar h_{12;\phi}-\kappa,n_1+\kappa-J;2\bar h_1;z_1,w_1 z_1).
\end{equation}
In comparing the generic conformal block decomposition of a spinning four-point function with the GFF correlator of eq.~\eqref{eq:defPsiJ} (\eqref{eq:defPsiJ_mst} for $\kappa>0$),  one notices an apparent discrepancy in the number of degrees of freedom: two cross-ratios $z_1,w_1$ in the conformal block decomposition,  but one cross-ratio $\frac{\ag_1-\ag_2}{\ag_1-\ag_3}$ for the homogeneous, gauge-invariant polynomial $\Psi_{J_2,\kappa}(\ag_i,\bg_i)$.  This discrepancy is resolved by a special property of GFF OPE coefficients derived above eq.~\eqref{gff_dt_opec}: all OPE coefficients with tensor structure $n_1+\kappa<J_1(<J_2)$ vanish.  For the single non-vanishing tensor structure $n_1=J_1-\kappa$ and the GFF quantum numbers
\begin{equation}
(h_1,\bar h_1) = (\Dg_\phi,\Dg_\phi+J_1),\quad (h_2,\bar h_2)=\left(\frac{3}{2}\Dg_\phi,\frac{3}{2}\Dg_\phi+J_2+\kappa \right),
\end{equation}
the four-point lightcone block further reduces to a one-variable hypergeometric function that is polynomial of order $J_2-J_1$:
\begin{equation}
F_1\left(\Dg_\phi+J_1;2\Dg_\phi+J_2+J_1,J_1-J_2;2\Dg_\phi+2J_1;z_1,z_1w_1\right)=\frac{\hypg{\Dg_\phi+J_1}{J_1-J_2}{2\Dg_\phi+2J_1}\left(\frac{z_1(w_1-1)}{1-z_1}\right)}{\left(1-z_1\right)^{\Dg_\phi+J_1}}.
\end{equation}
Once this last expression is plugged in~\eqref{eq:fourptasf_appendix}, we recover the expression of GFF four-point lightcone blocks displayed in eq.~\eqref{eq:4ptlightconeblock}.

\subsection{From triple-twist wave functions to conformal blocks}
\label{app:crossratios_prefactor_from_OPElimit}
The scope of this appendix is, on the one hand, to provide a derivation of equation~\eqref{eq:z1_z2_w_as_alphas} relating the cross-ratios $z_1$, $z_2$, and $w_1$ to the coordinates $\alpha_i$ and, on the other hand, to derive the multiplicative prefactor used to express the GFF four-point correlator $\langle \phi(X_1)\phi(X_2)\phi(X_3)[\phi\phi\phi]_{0,J,\kappa}(X_b,Z_b)\rangle$ in terms of linear combinations of four-point lightcone blocks $g_{\oo_1;n_1}(z_1,w_1)$.

We will use the results of~\cite[Section~3.2]{Buric:2021kgy}, which allow the computation of OPE limits directly in embedding space, and provide a direct construction of the lower-point degrees of freedom in terms of the higher-point ones. This can be directly applied to two consecutive OPE limits for the six-point conformal blocks $g_{\oo_1\oo_2\oo_3;n_1n_2}(z_i,\bar{z_i},w_1,w_2,\Upsilon_0)$ to recover the four-point degrees of freedom associated with three external scalars and one external MST. 

Since we are ultimately interested in four-point lightcone blocks, and since the Lorentzian OPE limits can be taken by making points approach each other along a null direction, we can start by considering the lightcone limits
\begin{equation}
   X_{12},X_{56}, X_{46}, X_{45} \rightarrow 0 \qquad \text{with} \quad \frac{X_{45}}{X_{46}}=\mathrm{const}
\end{equation}
under which three cross-ratios are constrained to $\bar z_1,\bar z_2, \bar z_3=0$ while the remaining ones are determined in terms of embedding space positions via~\eqref{eq:OPE_CR_as_us} and~\eqref{eq:curlycrossratios}. With these limits taken, we can then perform the first OPE limit between the fields at position $X_5$ and $X_6$ by first enforcing 
\begin{align}
X_6 = X_c && X_5 = X_c + \epsilon_c Z_c   
\end{align} 
with $Z_c$ being a null embedding space vector due to $X_{56}=0$, to then take $\epsilon \to 0$. Due to the projective nature of embedding space vectors, $Z_c$ remains a finite and non-trivial degree of freedom after the limit, and its normalization can be chosen such that it becomes a standard polarization vector associated with the field $\oo_{c}(X_c,Z_c)$ produced in this first OPE limit, where $X_c=X_5=X_6$. Using~\eqref{eq:OPE_CR_as_us} and~\eqref{eq:curlycrossratios}, one can readily verify that this limit constrains
\begin{align}
  z_3= \ep_c \frac{X_{3} \cdot Z_c}{X_3 \cdot X_c} + \dots  
\end{align}
and leaves us with the finite cross-ratios $z_1,z_2,w_1,w_2,\Upsilon_0$ that remain the degrees of freedom of a spinning five-point correlator under two lightcone limits. 

The second OPE limit can now be taken in two steps that are similar to what we just performed. We first require 
\begin{align}
    X_4 = X_c+\epsilon_b Z_{b}
\end{align}
$\epsilon_b \to 0$ such that we produce a first polarization vector, and then we take 
\begin{align}
    Z_c = Z_b  - \tilde{\epsilon}_b W_{b}
\end{align} with $\tilde{\epsilon}_b \to 0$, producing a second polarization vector associated with the MST spin. Naming $X_b=X_4=X_c$ after the limit, we have this way constructed the degrees of freedom of an MST field $\oo_b(X_b,Z_b,W_b)$ produced in the $\phi(X_4)\times \oo_c(X_c,Z_c)$ OPE. 

Performing these limits in~\eqref{eq:OPE_CR_as_us} once the cross-ratios are expressed in terms of embedding space vectors via~\eqref{eq:curlycrossratios}, we see that this finally constrains 
\begin{align}\label{eq:w2adnY}
    z_3 (1-w_2) = \epsilon_c \left( \frac{1}{\epsilon_b} + \frac{X_c \cdot Z_c}{X_3 \cdot X_c}\right) + \dots, && \frac{\Upsilon_0}{(1-w_1)(1-w_2)} = \left(\frac{\beta_{21}}{\alpha_{21}}-\frac{\beta_{23}}{\alpha_23}\right) \tilde{\ep}_b + \dots, 
\end{align} and leaves us with the two finite cross-ratios
\begin{equation}
    z_1=1-\frac{X_{1b}X_{23}}{X_{13}X_{2b}}\,, \qquad w_1 = 1+ \frac{J_{b,12}}{J_{b,23}}\frac{X_{23}X_{3b}}{\left(X_1 \otimes X_2\right)\cdot \left(X_b\wedge X_3\right)}.
\end{equation}

To understand how this reflects in the conformal block expansion of $\Psi(\ag_1,\ag_2,\ag_3)$, we can take one final OPE limit $X_1\equiv X_a$, $X_2=X_a+\ep_a Z_a$ with $\epsilon_a \to 0$ to relate the $z_1,w_1$ cross-ratios in the full OPE limit to the $\ag_i$'s. In this case, we find
\begin{equation}
z_1= 1-v_1 = \frac{(X_1\otimes X_2)(X_3\wedge X_b)}{X_{13}X_{2b}} = \frac{J_{a,3b}}{X_{a3}X_{ab}} \epsilon_a+\dots, \quad 1-w_1 = \xx = \frac{H_{ab} X_{a3} X_{b3}}{J_{a,b3} J_{b,3a}} +\dots
\end{equation}
and
\begin{align}\label{eq:z2a2a3}
    z_2 = 1 - v_2 = 1 - \frac{X_{2c}X_{34}}{X_{24}X_{3c}} = \epsilon_b\left( \frac{X_2 Z_b}{X_{2c}} - \frac{X_3 Z_b}{X_{3c}}  \right) + \dots = \epsilon_b(\alpha_2 - \alpha_3)+ \dots.
\end{align}
On the other hand,
\begin{align}
&\ag_2-\ag_1 = \frac{Z_b X_2}{X_b X_2} - \frac{Z_b X_1}{X_b X_1} = \frac{(X_a X_b)(Z_a Z_b)-(X_a Z_b)(X_b Z_a)}{X_{ab}^2} \epsilon_a+\dots =  \frac{H_{ab}}{X_{ab}^2} \epsilon_a + \orm(\epsilon^2)\\
&\ag_i-\ag_3 = \frac{(X_3 \otimes X_i)(X_b\wedge Z_b)}{X_{ib}X_{3b}} = \frac{J_{b,3a}}{X_{ab}X_{3b}}+\dots, \quad \forall i=1,2,
\end{align}
which implies
\begin{align}\label{eq:z1(w1-1)}
    \frac{z_1 (w_1 - 1)}{1 - z_1} = - \epsilon_a \frac{J_{a,3b}}{X_{a3}X_{ab}} \frac{H_{ab} X_{a3} X_{b3}}{J_{a,b3}J_{b,3a}} = \frac{H_{ab}}{X_{ab}^2}\epsilon_a \frac{X_{ab} X_{3b}}{J_{b,3a}} = \frac{\alpha_2 - \alpha_1}{\alpha_2 - \alpha_3}.
\end{align}
Since the comb is symmetric under the exchange of $X_1$ and $X_2$, eqs. \eqref{eq:z2a2a3} and \eqref{eq:z1(w1-1)} provide the relation between OPE cross-ratios and $\alpha$s sought for in eq.~\eqref{eq:z1_z2_w_as_alphas}.

With these identities, the relation between four-point and six-point wave functions stated 
in eq.~\eqref{3t_cb_dec} follows immediately. In order to recall which lightcone blocks 
are relevant, let us reproduce eq.~\eqref{eq:sixpointblockdecomp} from the main text. It 
reads
\begin{align}
\langle \phi(X_1)\dots\phi(X_6)\rangle & = \sum P_{\oo_1\oo_2\oo_3}^{(n_1n_2)}G_{\oo_1\oo_2\oo_3}^{(n_1n_2)}, 
\\[2mm] \textrm{with} \quad \quad & G_{\oo_1\oo_2\oo_3}^{(n_1n_2)}=\frac{\psi_{\oo_1\oo_2\oo_3;n_1n_2}(\bar z_1,z_1,\bar z_2,z_2,\Upsilon_0,\bar z_3,z_3,w_1,w_2)}{(X_{12}X_{34}X_{56})^{\Dg_\phi} (\bar z_2 z_2)^{\Dg_\phi/2}}.
\end{align}
In Appendix \ref{app:form_blocks_GFF}, we computed the lightcone limit of $G_{\oo_1\oo_2\oo_3}^{(n_1n_2)}$ that is given in Section \ref{sec:blocksGFFtwist} as
\begin{align}
\label{eq:G2n3O}
G_{\oo_1\oo_2\oo_3}^{(n_1n_2)} = \frac{\Upsilon_0^\kappa(z_1 \bar z_1 z_2 \bar z_2 z_3 \bar z_3)^{\Delta_\phi}\hypg{\Dg_\phi+J_1}{J_1-J_2}{2\Dg_\phi+2J_1}\left(\frac{z_1(w_1-1)}{1-z_1}\right)}{(X_{12}X_{34}X_{56})^{\Dg_\phi}} \frac{ z_1^{J_1}z_2^{J_2+\kappa}z_3^{J_3}(1-z_1)^{\Delta_\phi + J_1}}{(1-w_1)^{\kappa-J_1}(1-w_2)^{\kappa-J_3}}.
\end{align}
If we use that 
\begin{align}
    \frac{(z_1 \bar z_1 z_2 \bar z_2 z_3 \bar z_3)^{\Delta_\phi}}{(X_{12}X_{34}X_{56})^{\Dg_\phi}} = \frac{1}{X_{3b}^{\Delta_\phi} X_{ab}^{2\Delta_\phi}} + \dots
\end{align}
along with equation \eqref{eq:z1(w1-1)}, our expression \eqref{eq:G2n3O} simplifies to
\begin{align}
    G_{\oo_1\oo_2\oo_3}^{(n_1n_2)} =\frac{\Upsilon_0^\kappa \hypg{\Dg_\phi+J_1}{J_1-J_2}{2\Dg_\phi+2J_1}\left(\frac{\alpha_{21}}{\alpha_{23}}\right)}{(X_{3b}X_{ab}^2)^{\Dg_\phi} }(1-w_1)^{-\kappa}(1-w_2)^{J_3-\kappa} \left(\frac{\alpha_{12}}{\alpha_{23}}\right)^{J_1}z_2^{J_2+\kappa}z_3^{J_3}.
\end{align}
With the help of equations \eqref{eq:w2adnY} and \eqref{eq:z2a2a3} we deduce the following equivalent 
formula 
\begin{align}
       G_{\oo_1\oo_2\oo_3}^{(n_1n_2)} =\frac{ (\alpha_{23} \epsilon_b)^{J_2+\kappa}\left(\frac{\alpha_{12}}{\alpha_{23}}\right)^{J_1} \hypg{\Dg_\phi+J_1}{J_1-J_2}{2\Dg_\phi+2J_1}\left(\frac{\alpha_{21}}{\alpha_{23}}\right)}{(X_{3b}X_{ab}^2)^{\Dg_\phi} }\left(\left(\frac{\beta_{21}}{\alpha_{21}}-\frac{\beta_{23}}{\alpha_{23}}\right) \tilde{\ep}_b\right)^{\kappa}(\epsilon_c/\epsilon_b)^{J_3}.
\end{align}
Removing the $\ep$ parameters that we used for bookkeeping and using the $1 \leftrightarrow 2$ symmetry of the comb channel, we conclude 
\begin{align}
       G_{\oo_1\oo_2\oo_3}^{(n_1n_2)} =\prod\limits_{i=1}^3 \langle \phi(X_i)\phi(X_b)\rangle \alpha_{13}^{J_2}\alpha_{21}^{J_1} \hypg{\Dg_\phi+J_1}{J_1-J_2}{2\Dg_\phi+2J_1}\left(\frac{\alpha_{12}}{\alpha_{13}}\right) \frac{\left(\ag_{13}\bg_{12}-\ag_{12}\bg_{13}\right)^\kappa}{\ag_{12}^\kappa}
\end{align}
This indeed confirms the claim on the relation between four-point lightcone blocks and the 
triple-twist wave functions in GFF that we stated at the end of Section \ref{sec:blocksGFFtwist}.

\section{Six-point crossing kernel}
\label{app:crossing_kernel}
To derive eq.~\eqref{hankel_6pt}, we begin by expressing $f_{J_1'(J_2,\kappa) J_3'}$ in terms of the integral representation of the modified Bessel function:
\begin{equation}
f_{J_1'(J_2,\kappa)J_3'}(v_2,\mathcal{U}_0,\mathcal{U}_1,\mathcal{U}_2) = \frac{1}{2} \int_0^\infty \frac{\dd t}{t^{1+\Dg}} e^{-t \mathcal{U}_0-\frac{J_2^2}{t} v_2 \mathcal{U}_1 \mathcal{U}_2} \left( J_1'^{\Dg-\frac{1}{2}} e^{- \frac{\mathcal{U}_1}{t}J_1'^2 }\right) \left( J_3'^{\Dg-\frac{1}{2}} e^{- \frac{\mathcal{U}_2}{t}J_3'^2 }\right).
\end{equation}
Next, we use the following identity from \cite[eq.~(10.22.51)]{dlmf}:
\begin{equation}
\int_0^\infty \dd a \, a^{\nu+1} J_\nu(a b) e^{-p^2 a^2}=\frac{b^\nu}{(2 p^2)^{\nu+1}} e^{-\frac{b^2}{4p^2}},
\end{equation}
which translates to an identity for the Hankel transform of a Gaussian:
\begin{equation}
\int_0^\infty \dd J_1' W_{J_1J_1'}\, J_1'^{\Dg-\frac{1}{2}} e^{- \frac{\mathcal{U}_1}{t}J_1'^2 } = k^{\Dg-\frac{1}{2}} \left(\frac{t}{J_2\mathcal{U}_1}\right)^\Dg e^{-\frac{J_1^2 t}{J_2^2 \mathcal{U}_1}}.
\end{equation}
The same applies for the integral over $J_1$, so the LHS of eq.~\eqref{hankel_6pt} can be expressed as
\begin{equation}
\mathrm{LHS}=\frac{(J_1 J_3)^{\Dg-\frac{1}{2}}}{(J_2^2 \mathcal{U}_1\mathcal{U}_2)^\Dg} \int_0^\infty \frac{1}{2} \frac{\dd t}{t} t^\Dg e^{-\frac{J_2^2 \mathcal{U}_1\mathcal{U}_2}{t} v_2 - t \mathcal{U}_0 - \frac{t}{J_2^2 \mathcal{U}_1} J_1^2-\frac{t}{J_2^2 \mathcal{U}_2} J_3^2}.
\end{equation}
After the change of variables 
\begin{equation}
\tilde{t}:= \frac{J_2^2 \mathcal{U}_1\mathcal{U}_2}{t},
\end{equation}
we obtain explicitly the integral transform of the modified Bessel function with permuted arguments, i.e.
\begin{equation}
\mathrm{LHS}= (J_1J_3)^{\Dg-\frac{1}{2}} v_2^\Dg \mathcal{K}_\Dg\left(J_1^2 v_2 \mathcal{U}_2+ J_2^2 \mathcal{U}_0 v_2 \mathcal{U}_1 \mathcal{U}_2 +J_3^2 v_2 \mathcal{U}_1\right).
\end{equation}


\begin{thebibliography}{10}
\providecommand{\arxivref}[2]{\href{http://arxiv.org/abs/#1}{#2}}
\providecommand{\doiref}[2]{\href{http://dx.doi.org/#1}{#2}}
\providecommand{\nbbstauthor}[1]{#1}
\providecommand{\nbbstjournal}[1]{\textsf{#1}}
\providecommand{\nbbsttitle}[1]{\textit{``#1''}}
\providecommand{\nbbsturl}[1]{\texttt{#1}}
\providecommand{\nbbsteprint}[1]{\texttt{#1}}
\providecommand{\nbbststyle}{\raggedright\small\parskip0pt}
\nbbststyle

\bibitem{El-Showk:2012cjh}
\nbbstauthor{S.~El-Showk, M.~F.~Paulos, D.~Poland, S.~Rychkov, D.~Simmons-Duffin and A.~Vichi},
\nbbsttitle{{Solving the 3D Ising Model with the Conformal Bootstrap}},
\nbbstjournal{\doiref{10.1103/PhysRevD.86.025022}{Phys.~Rev.~D~86,~025022~(2012)}},
\nbbsteprint{\arxivref{1203.6064}{arxiv:1203.6064}}.

\bibitem{El-Showk:2014dwa}
\nbbstauthor{S.~El-Showk, M.~F.~Paulos, D.~Poland, S.~Rychkov, D.~Simmons-Duffin and A.~Vichi},
\nbbsttitle{{Solving the 3d Ising Model with the Conformal Bootstrap II. c-Minimization and Precise Critical Exponents}},
\nbbstjournal{\doiref{10.1007/s10955-014-1042-7}{J.~Stat.~Phys.~157,~869~(2014)}},
\nbbsteprint{\arxivref{1403.4545}{arxiv:1403.4545}}.

\bibitem{Kos:2014bka}
\nbbstauthor{F.~Kos, D.~Poland and D.~Simmons-Duffin},
\nbbsttitle{{Bootstrapping Mixed Correlators in the 3D Ising Model}},
\nbbstjournal{\doiref{10.1007/JHEP11(2014)109}{JHEP~1411,~109~(2014)}},
\nbbsteprint{\arxivref{1406.4858}{arxiv:1406.4858}}.

\bibitem{Simmons-Duffin:2016wlq}
\nbbstauthor{D.~Simmons-Duffin},
\nbbsttitle{{The Lightcone Bootstrap and the Spectrum of the 3d Ising CFT}},
\nbbstjournal{\doiref{10.1007/JHEP03(2017)086}{JHEP~1703,~086~(2017)}},
\nbbsteprint{\arxivref{1612.08471}{arxiv:1612.08471}},
\href{https://link.springer.com/article/10.1007/JHEP03(2017)086}{\nbbsturl{https://link.springer.com/article/10.1007/JHEP03(2017)086}}.

\bibitem{Caron-Huot:2020adz}
\nbbstauthor{S.~Caron-Huot, D.~Mazac, L.~Rastelli and D.~Simmons-Duffin},
\nbbsttitle{{Dispersive CFT Sum Rules}},
\nbbstjournal{\doiref{10.1007/JHEP05(2021)243}{JHEP~2105,~243~(2021)}},
\nbbsteprint{\arxivref{2008.04931}{arxiv:2008.04931}}.

\bibitem{Liu:2020tpf}
\nbbstauthor{J.~Liu, D.~Meltzer, D.~Poland and D.~Simmons-Duffin},
\nbbsttitle{{The Lorentzian inversion formula and the spectrum of the 3d $\mathrm{O}(2)$ CFT}},
\nbbstjournal{\doiref{10.1007/JHEP09(2020)115}{JHEP~2009,~115~(2020)}},
\nbbsteprint{\arxivref{2007.07914}{arxiv:2007.07914}}.

\bibitem{Su:2022xnj}
\nbbstauthor{N.~Su},
\nbbsttitle{{The Hybrid Bootstrap}},
\nbbsteprint{\arxivref{2202.07607}{arxiv:2202.07607}}.

\bibitem{Kos:2013tga}
\nbbstauthor{F.~Kos, D.~Poland and D.~Simmons-Duffin},
\nbbsttitle{{Bootstrapping the $O(N)$ vector models}},
\nbbstjournal{\doiref{10.1007/JHEP06(2014)091}{JHEP~1406,~091~(2014)}},
\nbbsteprint{\arxivref{1307.6856}{arxiv:1307.6856}}.

\bibitem{Kos:2015mba}
\nbbstauthor{F.~Kos, D.~Poland, D.~Simmons-Duffin and A.~Vichi},
\nbbsttitle{{Bootstrapping the O(N) Archipelago}},
\nbbstjournal{\doiref{10.1007/JHEP11(2015)106}{JHEP~1511,~106~(2015)}},
\nbbsteprint{\arxivref{1504.07997}{arxiv:1504.07997}}.

\bibitem{Chester:2019ifh}
\nbbstauthor{S.~M.~Chester, W.~Landry, J.~Liu, D.~Poland, D.~Simmons-Duffin, N.~Su and A.~Vichi},
\nbbsttitle{{Carving out OPE space and precise $O(2)$ model critical exponents}},
\nbbstjournal{\doiref{10.1007/JHEP06(2020)142}{JHEP~2006,~142~(2020)}},
\nbbsteprint{\arxivref{1912.03324}{arxiv:1912.03324}}.

\bibitem{Erramilli:2022kgp}
\nbbstauthor{R.~S.~Erramilli, L.~V.~Iliesiu, P.~Kravchuk, A.~Liu, D.~Poland and D.~Simmons-Duffin},
\nbbsttitle{{The Gross-Neveu-Yukawa archipelago}},
\nbbstjournal{\doiref{10.1007/JHEP02(2023)036}{JHEP~2302,~036~(2023)}},
\nbbsteprint{\arxivref{2210.02492}{arxiv:2210.02492}}.

\bibitem{Rosenhaus_2019}
\nbbstauthor{V.~Rosenhaus},
\nbbsttitle{Multipoint conformal blocks in the comb channel},
\nbbstjournal{\doiref{10.1007/jhep02(2019)142}{Journal~of~High~Energy~Physics~2019,~39~(2019)}},
\href{http://dx.doi.org/10.1007/JHEP02(2019)142}{\nbbsturl{http://dx.doi.org/10.1007/JHEP02(2019)142}}.

\bibitem{Fortin:2020zxw}
\nbbstauthor{J.-F.~Fortin, W.-J.~Ma and W.~Skiba},
\nbbsttitle{{All Global One- and Two-Dimensional Higher-Point Conformal Blocks}},
\nbbsteprint{\arxivref{2009.07674}{arxiv:2009.07674}}.

\bibitem{Alkalaev:2023axo}
\nbbstauthor{K.~Alkalaev, A.~Kanoda and V.~Khiteev},
\nbbsttitle{{Wilson networks in AdS and global conformal blocks}},
\nbbstjournal{\doiref{10.1016/j.nuclphysb.2023.116413}{Nucl.~Phys.~B~998,~116413~(2024)}},
\nbbsteprint{\arxivref{2307.08395}{arxiv:2307.08395}}.

\bibitem{Fortin:2023xqq}
\nbbstauthor{J.-F.~Fortin, W.-J.~Ma, S.~Parikh, L.~Quintavalle and W.~Skiba},
\nbbsttitle{{One- and two-dimensional higher-point conformal blocks as free-particle wavefunctions in $ {\textrm{AdS}}_3^{\otimes m} $}},
\nbbstjournal{\doiref{10.1007/JHEP01(2024)031}{JHEP~2401,~031~(2024)}},
\nbbsteprint{\arxivref{2310.08632}{arxiv:2310.08632}}.

\bibitem{Gon_alves_2019}
\nbbstauthor{V.~Gon\c{c}alves, R.~Pereira and X.~Zhou},
\nbbsttitle{{$20'$ Five-Point Function from $AdS_5\times S^5$ Supergravity}},
\nbbstjournal{\doiref{10.1007/JHEP10(2019)247}{JHEP~1910,~247~(2019)}},
\nbbsteprint{\arxivref{1906.05305}{arxiv:1906.05305}}.

\bibitem{Fortin:2020bfq}
\nbbstauthor{J.-F.~Fortin, W.-J.~Ma and W.~Skiba},
\nbbsttitle{{Seven-point conformal blocks in the extended snowflake channel and beyond}},
\nbbstjournal{\doiref{10.1103/PhysRevD.102.125007}{Phys.~Rev.~D~102,~125007~(2020)}},
\nbbsteprint{\arxivref{2006.13964}{arxiv:2006.13964}}.

\bibitem{Hoback:2020syd}
\nbbstauthor{S.~Hoback and S.~Parikh},
\nbbsttitle{{Dimensional reduction of higher-point conformal blocks}},
\nbbstjournal{\doiref{10.1007/JHEP03(2021)187}{JHEP~2103,~187~(2021)}},
\nbbsteprint{\arxivref{2009.12904}{arxiv:2009.12904}}.

\bibitem{Poland_2021}
\nbbstauthor{D.~Poland and V.~Prilepina},
\nbbsttitle{{Recursion relations for 5-point conformal blocks}},
\nbbstjournal{\doiref{10.1007/JHEP10(2021)160}{JHEP~2110,~160~(2021)}},
\nbbsteprint{\arxivref{2103.12092}{arxiv:2103.12092}}.

\bibitem{Buric:2020dyz}
\nbbstauthor{I.~Buric, S.~Lacroix, J.~A.~Mann, L.~Quintavalle and V.~Schomerus},
\nbbsttitle{{From Gaudin Integrable Models to $d$-dimensional Multipoint Conformal Blocks}},
\nbbstjournal{\doiref{10.1103/PhysRevLett.126.021602}{Phys.~Rev.~Lett.~126,~021602~(2021)}},
\nbbsteprint{\arxivref{2009.11882}{arxiv:2009.11882}}.

\bibitem{Buric:2021ywo}
\nbbstauthor{I.~Buric, S.~Lacroix, J.~A.~Mann, L.~Quintavalle and V.~Schomerus},
\nbbsttitle{{Gaudin models and multipoint conformal blocks: general theory}},
\nbbstjournal{\doiref{10.1007/JHEP10(2021)139}{JHEP~2110,~139~(2021)}},
\nbbsteprint{\arxivref{2105.00021}{arxiv:2105.00021}}.

\bibitem{Buric:2021ttm}
\nbbstauthor{I.~Buric, S.~Lacroix, J.~A.~Mann, L.~Quintavalle and V.~Schomerus},
\nbbsttitle{{Gaudin models and multipoint conformal blocks. Part II. Comb channel vertices in 3D and 4D}},
\nbbstjournal{\doiref{10.1007/JHEP11(2021)182}{JHEP~2111,~182~(2021)}},
\nbbsteprint{\arxivref{2108.00023}{arxiv:2108.00023}}.

\bibitem{Buric:2021kgy}
\nbbstauthor{I.~Buric, S.~Lacroix, J.~A.~Mann, L.~Quintavalle and V.~Schomerus},
\nbbsttitle{{Gaudin models and multipoint conformal blocks III: comb channel coordinates and OPE factorisation}},
\nbbstjournal{\doiref{10.1007/JHEP06(2022)144}{JHEP~2206,~144~(2022)}},
\nbbsteprint{\arxivref{2112.10827}{arxiv:2112.10827}}.

\bibitem{Fortin:2022grf}
\nbbstauthor{J.-F.~Fortin, S.~Hoback, W.-J.~Ma, S.~Parikh and W.~Skiba},
\nbbsttitle{{Feynman rules for scalar conformal blocks}},
\nbbstjournal{\doiref{10.1007/JHEP10(2022)097}{JHEP~2210,~097~(2022)}},
\nbbsteprint{\arxivref{2204.08909}{arxiv:2204.08909}}.

\bibitem{Poland:2023bny}
\nbbstauthor{D.~Poland, V.~Prilepina and P.~Tadi\'c},
\nbbsttitle{{Improving the five-point bootstrap}},
\nbbsteprint{\arxivref{2312.13344}{arxiv:2312.13344}}.

\bibitem{Poland:2023vpn}
\nbbstauthor{D.~Poland, V.~Prilepina and P.~Tadi\'c},
\nbbsttitle{{The five-point bootstrap}},
\nbbstjournal{\doiref{10.1007/JHEP10(2023)153}{JHEP~2310,~153~(2023)}},
\nbbsteprint{\arxivref{2305.08914}{arxiv:2305.08914}}.

\bibitem{Antunes:2023dlk}
\nbbstauthor{A.~Antunes, S.~Harris, A.~Kaviraj and V.~Schomerus},
\nbbsttitle{{Lining up a Positive Semi-Definite Six-Point Bootstrap}},
\nbbsteprint{\arxivref{2312.11660}{arxiv:2312.11660}}.

\bibitem{Bercini:2020msp}
\nbbstauthor{C.~Bercini, V.~Gon\c{c}alves and P.~Vieira},
\nbbsttitle{{Light-Cone Bootstrap of Higher Point Functions and Wilson Loop Duality}},
\nbbstjournal{\doiref{10.1103/PhysRevLett.126.121603}{Phys.~Rev.~Lett.~126,~121603~(2021)}},
\nbbsteprint{\arxivref{2008.10407}{arxiv:2008.10407}},
\href{https://journals.aps.org/prl/abstract/10.1103/PhysRevLett.126.121603}{\nbbsturl{https://journals.aps.org/prl/abstract/10.1103/PhysRevLett.126.121603}}.

\bibitem{Bercini:2021jti}
\nbbstauthor{C.~Bercini, V.~Gon\c{c}alves, A.~Homrich and P.~Vieira},
\nbbsttitle{{The Wilson Loop - Large Spin OPE Dictionary}},
\nbbsteprint{\arxivref{2110.04364}{arxiv:2110.04364}},
\href{https://arxiv.org/abs/2110.04364}{\nbbsturl{https://arxiv.org/abs/2110.04364}}.

\bibitem{Antunes:2021kmm}
\nbbstauthor{A.~Antunes, M.~S.~Costa, V.~Goncalves and J.~V.~Boas},
\nbbsttitle{{Lightcone bootstrap at higher points}},
\nbbstjournal{\doiref{10.1007/JHEP03(2022)139}{JHEP~2203,~139~(2022)}},
\nbbsteprint{\arxivref{2111.05453}{arxiv:2111.05453}}.

\bibitem{Kaviraj:2022wbw}
\nbbstauthor{A.~Kaviraj, J.~A.~Mann, L.~Quintavalle and V.~Schomerus},
\nbbsttitle{{Multipoint lightcone bootstrap from differential equations}},
\nbbstjournal{\doiref{10.1007/JHEP08(2023)011}{JHEP~2308,~011~(2023)}},
\nbbsteprint{\arxivref{2212.10578}{arxiv:2212.10578}}.

\bibitem{Anous:2021caj}
\nbbstauthor{T.~Anous, A.~Belin, J.~de~Boer and D.~Liska},
\nbbsttitle{{OPE statistics from higher-point crossing}},
\nbbstjournal{\doiref{10.1007/JHEP06(2022)102}{JHEP~2206,~102~(2022)}},
\nbbsteprint{\arxivref{2112.09143}{arxiv:2112.09143}}.

\bibitem{Fitzpatrick:2012yx}
\nbbstauthor{A.~L.~Fitzpatrick, J.~Kaplan, D.~Poland and D.~Simmons-Duffin},
\nbbsttitle{{The Analytic Bootstrap and AdS Superhorizon Locality}},
\nbbstjournal{\doiref{10.1007/JHEP12(2013)004}{JHEP~1312,~004~(2013)}},
\nbbsteprint{\arxivref{1212.3616}{arxiv:1212.3616}}.

\bibitem{Komargodski:2012ek}
\nbbstauthor{Z.~Komargodski and A.~Zhiboedov},
\nbbsttitle{{Convexity and Liberation at Large Spin}},
\nbbstjournal{\doiref{10.1007/JHEP11(2013)140}{JHEP~1311,~140~(2013)}},
\nbbsteprint{\arxivref{1212.4103}{arxiv:1212.4103}}.

\bibitem{Pal:2022vqc}
\nbbstauthor{S.~Pal, J.~Qiao and S.~Rychkov},
\nbbsttitle{{Twist Accumulation in Conformal Field Theory: A Rigorous Approach to the Lightcone Bootstrap}},
\nbbstjournal{\doiref{10.1007/s00220-023-04767-w}{Commun.~Math.~Phys.~402,~2169~(2023)}},
\nbbsteprint{\arxivref{2212.04893}{arxiv:2212.04893}}.

\bibitem{Alday:2015ewa}
\nbbstauthor{L.~F.~Alday and A.~Zhiboedov},
\nbbsttitle{{An Algebraic Approach to the Analytic Bootstrap}},
\nbbstjournal{\doiref{10.1007/JHEP04(2017)157}{JHEP~1704,~157~(2017)}},
\nbbsteprint{\arxivref{1510.08091}{arxiv:1510.08091}}.

\bibitem{Alday:2016njk}
\nbbstauthor{L.~F.~Alday},
\nbbsttitle{{Large Spin Perturbation Theory for Conformal Field Theories}},
\nbbstjournal{\doiref{10.1103/PhysRevLett.119.111601}{Phys.~Rev.~Lett.~119,~111601~(2017)}},
\nbbsteprint{\arxivref{1611.01500}{arxiv:1611.01500}}.

\bibitem{Kaviraj:2015cxa}
\nbbstauthor{A.~Kaviraj, K.~Sen and A.~Sinha},
\nbbsttitle{{Analytic bootstrap at large spin}},
\nbbstjournal{\doiref{10.1007/JHEP11(2015)083}{JHEP~1511,~083~(2015)}},
\nbbsteprint{\arxivref{1502.01437}{arxiv:1502.01437}}.

\bibitem{Kaviraj:2015xsa}
\nbbstauthor{A.~Kaviraj, K.~Sen and A.~Sinha},
\nbbsttitle{{Universal anomalous dimensions at large spin and large twist}},
\nbbstjournal{\doiref{10.1007/JHEP07(2015)026}{JHEP~1507,~026~(2015)}},
\nbbsteprint{\arxivref{1504.00772}{arxiv:1504.00772}}.

\bibitem{Hofman:2016awc}
\nbbstauthor{D.~M.~Hofman, D.~Li, D.~Meltzer, D.~Poland and F.~Rejon-Barrera},
\nbbsttitle{{A Proof of the Conformal Collider Bounds}},
\nbbstjournal{\doiref{10.1007/JHEP06(2016)111}{JHEP~1606,~111~(2016)}},
\nbbsteprint{\arxivref{1603.03771}{arxiv:1603.03771}}.

\bibitem{Li:2015itl}
\nbbstauthor{D.~Li, D.~Meltzer and D.~Poland},
\nbbsttitle{{Conformal Collider Physics from the Lightcone Bootstrap}},
\nbbstjournal{\doiref{10.1007/JHEP02(2016)143}{JHEP~1602,~143~(2016)}},
\nbbsteprint{\arxivref{1511.08025}{arxiv:1511.08025}}.

\bibitem{Li:2015rfa}
\nbbstauthor{D.~Li, D.~Meltzer and D.~Poland},
\nbbsttitle{{Non-Abelian Binding Energies from the Lightcone Bootstrap}},
\nbbstjournal{\doiref{10.1007/JHEP02(2016)149}{JHEP~1602,~149~(2016)}},
\nbbsteprint{\arxivref{1510.07044}{arxiv:1510.07044}}.

\bibitem{Caron-Huot:2017vep}
\nbbstauthor{S.~Caron-Huot},
\nbbsttitle{{Analyticity in Spin in Conformal Theories}},
\nbbstjournal{\doiref{10.1007/JHEP09(2017)078}{JHEP~1709,~078~(2017)}},
\nbbsteprint{\arxivref{1703.00278}{arxiv:1703.00278}}.

\bibitem{Kravchuk:2018htv}
\nbbstauthor{P.~Kravchuk and D.~Simmons-Duffin},
\nbbsttitle{{Light-ray operators in conformal field theory}},
\nbbstjournal{\doiref{10.1007/JHEP11(2018)102}{JHEP~1811,~102~(2018)}},
\nbbsteprint{\arxivref{1805.00098}{arxiv:1805.00098}}.

\bibitem{Cornalba:2006xk}
\nbbstauthor{L.~Cornalba, M.~S.~Costa, J.~Penedones and R.~Schiappa},
\nbbsttitle{{Eikonal Approximation in AdS/CFT: From Shock Waves to Four-Point Functions}},
\nbbstjournal{\doiref{10.1088/1126-6708/2007/08/019}{JHEP~0708,~019~(2007)}},
\nbbsteprint{\arxivref{hep-th/0611122}{hep-th/0611122}}.

\bibitem{Cornalba:2006xm}
\nbbstauthor{L.~Cornalba, M.~S.~Costa, J.~Penedones and R.~Schiappa},
\nbbsttitle{{Eikonal Approximation in AdS/CFT: Conformal Partial Waves and Finite N Four-Point Functions}},
\nbbstjournal{\doiref{10.1016/j.nuclphysb.2007.01.007}{Nucl.~Phys.~B~767,~327~(2007)}},
\nbbsteprint{\arxivref{hep-th/0611123}{hep-th/0611123}}.

\bibitem{Cornalba:2007zb}
\nbbstauthor{L.~Cornalba, M.~S.~Costa and J.~Penedones},
\nbbsttitle{{Eikonal approximation in AdS/CFT: Resumming the gravitational loop expansion}},
\nbbstjournal{\doiref{10.1088/1126-6708/2007/09/037}{JHEP~0709,~037~(2007)}},
\nbbsteprint{\arxivref{0707.0120}{arxiv:0707.0120}}.

\bibitem{Fitzpatrick:2014vua}
\nbbstauthor{A.~L.~Fitzpatrick, J.~Kaplan and M.~T.~Walters},
\nbbsttitle{{Universality of Long-Distance AdS Physics from the CFT Bootstrap}},
\nbbstjournal{\doiref{10.1007/JHEP08(2014)145}{JHEP~1408,~145~(2014)}},
\nbbsteprint{\arxivref{1403.6829}{arxiv:1403.6829}}.

\bibitem{Fitzpatrick:2015qma}
\nbbstauthor{A.~L.~Fitzpatrick, J.~Kaplan, M.~T.~Walters and J.~Wang},
\nbbsttitle{{Eikonalization of Conformal Blocks}},
\nbbstjournal{\doiref{10.1007/JHEP09(2015)019}{JHEP~1509,~019~(2015)}},
\nbbsteprint{\arxivref{1504.01737}{arxiv:1504.01737}}.

\bibitem{Homrich:2022mmd}
\nbbstauthor{A.~Homrich, D.~Simmons-Duffin and P.~Vieira},
\nbbsttitle{{Complex Spin: The Missing Zeroes and Newton's Dark Magic}},
\nbbsteprint{\arxivref{2211.13754}{arxiv:2211.13754}}.

\bibitem{Henriksson:2023cnh}
\nbbstauthor{J.~Henriksson, P.~Kravchuk and B.~Oertel},
\nbbsttitle{{Missing local operators, zeros, and twist-4 trajectories}},
\nbbsteprint{\arxivref{2312.09283}{arxiv:2312.09283}}.

\bibitem{Kehrein:1992fn}
\nbbstauthor{S.~Kehrein, F.~Wegner and Y.~Pismak},
\nbbsttitle{{Conformal symmetry and the spectrum of anomalous dimensions in the N vector model in four epsilon dimensions}},
\nbbstjournal{\doiref{10.1016/0550-3213(93)90124-8}{Nucl.~Phys.~B~402,~669~(1993)}}.

\bibitem{Derkachov:1995zr}
\nbbstauthor{S.~E.~Derkachov and A.~N.~Manashov},
\nbbsttitle{{The Spectrum of the anomalous dimensions of the composite operators in epsilon expansion in the scalar phi**4 field theory}},
\nbbstjournal{\doiref{10.1016/0550-3213(95)00513-R}{Nucl.~Phys.~B~455,~685~(1995)}},
\nbbsteprint{\arxivref{hep-th/9505110}{hep-th/9505110}}.

\bibitem{Kehrein:1995ia}
\nbbstauthor{S.~K.~Kehrein},
\nbbsttitle{{The Spectrum of critical exponents in phi**2 in two-dimensions theory in D = (4-epsilon)-dimensions: Resolution of degeneracies and hierarchical structures}},
\nbbstjournal{\doiref{10.1016/0550-3213(95)00375-3}{Nucl.~Phys.~B~453,~777~(1995)}},
\nbbsteprint{\arxivref{hep-th/9507044}{hep-th/9507044}}.

\bibitem{Derkachov:1997uh}
\nbbstauthor{S.~E.~Derkachov and A.~N.~Manashov},
\nbbsttitle{{Spectrum of anomalous dimensions in scalar field theory}}.

\bibitem{Derkachov:1997qv}
\nbbstauthor{S.~E.~Derkachov, S.~K.~Kehrein and A.~N.~Manashov},
\nbbsttitle{{High-gradient operators in the N-vector model}},
\nbbstjournal{\doiref{10.1016/S0550-3213(97)00131-4}{Nucl.~Phys.~B~493,~660~(1997)}},
\nbbsteprint{\arxivref{cond-mat/9610106}{cond-mat/9610106}}.

\bibitem{Braun:1998id}
\nbbstauthor{V.~M.~Braun, S.~E.~Derkachov and A.~N.~Manashov},
\nbbsttitle{{Integrability of three particle evolution equations in QCD}},
\nbbstjournal{\doiref{10.1103/PhysRevLett.81.2020}{Phys.~Rev.~Lett.~81,~2020~(1998)}},
\nbbsteprint{\arxivref{hep-ph/9805225}{hep-ph/9805225}}.

\bibitem{Braun:1999te}
\nbbstauthor{V.~M.~Braun, S.~E.~Derkachov, G.~P.~Korchemsky and A.~N.~Manashov},
\nbbsttitle{{Baryon distribution amplitudes in QCD}},
\nbbstjournal{\doiref{10.1016/S0550-3213(99)00265-5}{Nucl.~Phys.~B~553,~355~(1999)}},
\nbbsteprint{\arxivref{hep-ph/9902375}{hep-ph/9902375}}.

\bibitem{Belitsky:1999ru}
\nbbstauthor{A.~V.~Belitsky},
\nbbsttitle{{Integrability and WKB solution of twist - three evolution equations}},
\nbbstjournal{\doiref{10.1016/S0550-3213(99)00402-2}{Nucl.~Phys.~B~558,~259~(1999)}},
\nbbsteprint{\arxivref{hep-ph/9903512}{hep-ph/9903512}}.

\bibitem{Korchemsky:2010kj}
\nbbstauthor{G.~P.~Korchemsky},
\nbbsttitle{{Review of AdS/CFT Integrability, Chapter IV.4: Integrability in QCD and N\ensuremath{<}4 SYM}},
\nbbstjournal{\doiref{10.1007/s11005-011-0516-7}{Lett.~Math.~Phys.~99,~425~(2012)}},
\nbbsteprint{\arxivref{1012.4000}{arxiv:1012.4000}}.

\bibitem{Gromov:2023hzc}
\nbbstauthor{N.~Gromov, A.~Hegedus, J.~Julius and N.~Sokolova},
\nbbsttitle{{Fast QSC Solver: tool for systematic study of N=4 Super-Yang-Mills spectrum}},
\nbbsteprint{\arxivref{2306.12379}{arxiv:2306.12379}}.

\bibitem{Dolan:2003hv}
\nbbstauthor{F.~A.~Dolan and H.~Osborn},
\nbbsttitle{{Conformal partial waves and the operator product expansion}},
\nbbstjournal{\doiref{10.1016/j.nuclphysb.2003.11.016}{Nucl.~Phys.~B~678,~491~(2004)}},
\nbbsteprint{\arxivref{hep-th/0309180}{hep-th/0309180}}.

\bibitem{Derkachov:2010zza}
\nbbstauthor{S.~E.~Derkachov and A.~N.~Manashov},
\nbbsttitle{{Anomalous dimensions of composite operators in scalar field theories}},
\nbbstjournal{\doiref{10.1007/s10958-010-0032-9}{J.~Math.~Sci.~168,~837~(2010)}}.

\bibitem{szeg1939orthogonal}
\nbbstauthor{G.~Szego},
\nbbsttitle{Orthogonal polynomials},
\nbbsttitle{volume 23},
American Mathematical Soc. (1939).

\bibitem{dlmf}
\nbbsttitle{{\it NIST Digital Library of Mathematical Functions}},
F.~W.~J. Olver, A.~B. {Olde Daalhuis}, D.~W. Lozier, B.~I. Schneider, R.~F. Boisvert, C.~W. Clark, B.~R. Miller, B.~V. Saunders, H.~S. Cohl, and M.~A. McClain, eds.,
\href{http://dlmf.nist.gov/}{\nbbsturl{http://dlmf.nist.gov/}}.

\bibitem{Alday:2015eya}
\nbbstauthor{L.~F.~Alday, A.~Bissi and T.~Lukowski},
\nbbsttitle{{Large spin systematics in CFT}},
\nbbstjournal{\doiref{10.1007/JHEP11(2015)101}{JHEP~1511,~101~(2015)}},
\nbbsteprint{\arxivref{1502.07707}{arxiv:1502.07707}}.

\bibitem{Gracey2015}
\nbbstauthor{J.~A.~Gracey},
\nbbsttitle{Four loop renormalization of $\phi^{3}$ theory in six dimensions},
\nbbstjournal{\doiref{10.1103/PhysRevD.92.025012}{Phys.~Rev.~D~92,~025012~(2015)}},
\href{https://link.aps.org/doi/10.1103/PhysRevD.92.025012}{\nbbsturl{https://link.aps.org/doi/10.1103/PhysRevD.92.025012}}.

\bibitem{deAlcantaraBonfim:1980pe}
\nbbstauthor{O.~F.~de~Alcantara~Bonfim, J.~E.~Kirkham and A.~J.~McKane},
\nbbsttitle{{Critical Exponents to Order $\epsilon^3$ for $\phi^3$ Models of Critical Phenomena in Six $\epsilon$-dimensions}},
\nbbstjournal{\doiref{10.1088/0305-4470/13/7/006}{J.~Phys.~A~13,~L247~(1980)}},
[Erratum: J.Phys.A 13, 3785 (1980)].

\bibitem{Hasegawa_2017}
\nbbstauthor{C.~Hasegawa and Y.~Nakayama},
\nbbsttitle{$\epsilon$-expansion in critical $\phi^3$-theory on real projective space from conformal field theory},
\nbbstjournal{\doiref{10.1142/s0217732317500456}{Modern~Physics~Letters~A~32,~1750045~(2017)}}.

\bibitem{Gopakumar:2016cpb}
\nbbstauthor{R.~Gopakumar, A.~Kaviraj, K.~Sen and A.~Sinha},
\nbbsttitle{{A Mellin space approach to the conformal bootstrap}},
\nbbstjournal{\doiref{10.1007/JHEP05(2017)027}{JHEP~1705,~027~(2017)}},
\nbbsteprint{\arxivref{1611.08407}{arxiv:1611.08407}}.

\bibitem{Bertucci:2022ptt}
\nbbstauthor{F.~Bertucci, J.~Henriksson and B.~McPeak},
\nbbsttitle{{Analytic bootstrap of mixed correlators in the O(n) CFT}},
\nbbstjournal{\doiref{10.1007/JHEP10(2022)104}{JHEP~2210,~104~(2022)}},
\nbbsteprint{\arxivref{2205.09132}{arxiv:2205.09132}}.

\bibitem{WILSON197475}
\nbbstauthor{K.~G.~Wilson and J.~Kogut},
\nbbsttitle{The renormalization group and the $\epsilon$ expansion},
\nbbstjournal{\doiref{https://doi.org/10.1016/0370-1573(74)90023-4}{Physics~Reports~12,~75~(1974)}},
\href{https://www.sciencedirect.com/science/article/pii/0370157374900234}{\nbbsturl{https://www.sciencedirect.com/science/article/pii/0370157374900234}}.

\bibitem{Derkachov1998}
\nbbstauthor{S.~{\'E}.~Derkachov, J.~A.~Gracey and A.~N.~Manashov},
\nbbsttitle{Four loop anomalous dimensions of gradient operators in $\phi^4$ theory},
\nbbstjournal{\doiref{10.1007/s100529800706}{The~European~Physical~Journal~C~-~Particles~and~Fields~2,~569~(1998)}},
\href{https://doi.org/10.1007/s100529800706}{\nbbsturl{https://doi.org/10.1007/s100529800706}}.

\bibitem{Dey:2017oim}
\nbbstauthor{P.~Dey and A.~Kaviraj},
\nbbsttitle{{Towards a Bootstrap approach to higher orders of epsilon expansion}},
\nbbstjournal{\doiref{10.1007/JHEP02(2018)153}{JHEP~1802,~153~(2018)}},
\nbbsteprint{\arxivref{1711.01173}{arxiv:1711.01173}}.

\bibitem{johan_private}
\nbbstauthor{J.~Henriksson},
\nbbsttitle{{Private communication}}.

\bibitem{PK_JAM_inprep}
\nbbstauthor{P.~Kravchuk and J.~Mann},
\nbbsttitle{{In preparation}}.

\bibitem{Cuomo:2022kio}
\nbbstauthor{G.~Cuomo and Z.~Komargodski},
\nbbsttitle{{Giant Vortices and the Regge Limit}},
\nbbstjournal{\doiref{10.1007/JHEP01(2023)006}{JHEP~2301,~006~(2023)}},
\nbbsteprint{\arxivref{2210.15694}{arxiv:2210.15694}}.

\bibitem{cuomo2024chiral}
\nbbstauthor{G.~Cuomo, Z.~Komargodski and S.~Zhong},
\nbbsttitle{{Chiral Modes of Giant Superfluid Vortices}},
\nbbsteprint{\arxivref{2312.06095}{arxiv:2312.06095}}.

\bibitem{cuomo2023numerical}
\nbbstauthor{G.~Cuomo, J.~M. V.~P.~Lopes, J.~Matos, J.~Oliveira and J.~Penedones},
\nbbsttitle{{Numerical tests of the large charge expansion}},
\nbbsteprint{\arxivref{2305.00499}{arxiv:2305.00499}}.

\bibitem{Jafferis_2018}
\nbbstauthor{D.~Jafferis, B.~Mukhametzhanov and A.~Zhiboedov},
\nbbsttitle{{Conformal Bootstrap At Large Charge}},
\nbbstjournal{\doiref{10.1007/JHEP05(2018)043}{JHEP~1805,~043~(2018)}},
\nbbsteprint{\arxivref{1710.11161}{arxiv:1710.11161}}.

\bibitem{Anous:2016kss}
\nbbstauthor{T.~Anous, T.~Hartman, A.~Rovai and J.~Sonner},
\nbbsttitle{{Black Hole Collapse in the 1/c Expansion}},
\nbbstjournal{\doiref{10.1007/JHEP07(2016)123}{JHEP~1607,~123~(2016)}},
\nbbsteprint{\arxivref{1603.04856}{arxiv:1603.04856}}.

\bibitem{Derkachov:2005hw}
\nbbstauthor{S.~E.~Derkachov},
\nbbsttitle{{Factorization of the R-matrix. I.}},
\nbbsteprint{\arxivref{math/0503396}{math/0503396}}.

\end{thebibliography}

\end{document}